\magnification = 1150
\documentstyle{amsppt}
\define\>{\rangle}
\define\<{\langle}
\define\st{\,|\,}
\define\wt{\text{wt}}

\define\R{\Bbb R}
\define\rwt{\roman{wt}}
\define\({\left(}
\define\){\right)}
\define\lfl{\lfloor}
\define\rfl{\rfloor}
\define\eff{\text{eff}}

\input amstex

\topmatter
 
\title Static- and stationary-complete spacetimes: \\ algebraic and causal structures
\endtitle

\rightheadtext
{Static and Stationary Structures}

\author
Steven G. Harris
\endauthor

\address
Department of Mathematics, Saint Louis University,
St. Louis, MO 63103, USA \endaddress
\email
harrissg\@slu.edu \endemail

\thanks
This research was supported in part by Perimeter Institute for Theoretical Physics, Waterloo, ON.Ê Research at Perimeter Institute is supported by the Government of Canada through Industry Canada and by the Province of Ontario through the Ministry of Economic Development and Innovation.
\endthanks

\thanks
Included in the Perimeter Institute experience was the Kitchener Waterloo Chamber Music Society, whose entertainment was instrumental in providing excellent space for mathematical contemplation.
\endthanks

\thanks
The author wishes to thank Miguel S\'anchez of the University of Granada for helpful discussions.
\endthanks

\endtopmatter

\document

\head 0. Introduction
\endhead 

The purpose of this note is to explore the structure of static-complete and stationary-complete spacetimes, in terms of causal structure, the effect of forming quotients from group actions, and certain algebraic structures previously identified as giving insight into global causal behavior for these spacetimes. 

A spacetime $M$ is {\it stationary-complete}\/ if it has a Killing field $K$ which is everywhere timelike and whose integral curves are all defined on $(-\infty, \infty)$; it is {\it static-complete}\/ if $K^\perp$ is integrable, i.e., there is a {\it static restspace}\/ through every point of $M$, a spacelike hypersurface orthogonal to $K$.  We will be concerned here with spacetimes for which the integral curves are all lines, not circles. 

One of the primary points of interest in stationary spacetimes is what might be termed a decoherence between spatial and temporal organization, as measured by stationary observers.  By this is meant the following fact:  If a stationary observer emits a photon that is constrained to travel along a specific closed track, and this observer measures the elapsed time (in stationary units) it takes for the photon to complete the track and return, then a naive expectation would be that this elapsed time is equal to the length of the track within the space of stationary observers (measured in units that yield speed of light to be one, as measured infinitesimally by each stationary observer); but this is not necessarily the case.  There are two sources to a violation of this naive expectation.  

One source depends on local stationary physics and amounts to a sort of ``wind" blowing through the space of stationary observers; an example is the rotation of the Kerr solution.  This wind produces a delay to the elapsed travel time (relative to the length of the track) if the path of the photon is, on average, counter to the wind; the same path, traveled in reverse, would then yield a shorter elapsed travel time than naively expected.  In general, the deviation in travel time from length of the track varies continuously with varying the track.  Because this depends on the wind blowing through the stationary observer space, this source of deviation is not present in static space times.

But there is another source of deviation in travel time from that naively expected, which has nothing to do with local physics, depending solely on topological properties of the spacetime.  This produces deviations that are constant with slight variations of the track; they depend only on the homotopy class of the path traveled.  Thus, this source of deviation, present in some static spacetimes, vanishes for all simply connected static or stationary spacetimes.  That makes examining this source of deviations from naivet\'e a matter of looking at covering maps and group actions.  

Part of the aim of this note is to tease apart these two sources of travel time deviation.  For this reason, the main focus is spacetimes formed as quotients on static- or stationary-complete chronological spacetimes $M$ by the action of groups $G$ acting isometrically on $M$:  Deviations in photon elapsed travel times are but one example of looking at causal properties of $M$ that are inherited by $M/G$.  Explication will, in part, be by means of algebraic structures defined on the space of Killing orbits, first mentioned in \cite{GHr}, which will be further refined in this note.

One of the motivations for this note is the explication of some somewhat unexpected causal behavior in static-complete spacetimes.  Such spacetimes are not as simple as standard static (which are conformal to a metric product of the Lorentz line with a Riemannian manifold---the space of Killing orbits), but are not far removed from such:  They must be quotients of standard static spacetimes by a group action.  As such, they might perhaps be expected to be fairly simple in causal behavior; but it turns out that there are static-complete spacetimes which exhibit what might be called causally curious behavior, such as being causal but not strongly causal, and being strongly causal but not globally hyperbolic (in spite of having a globally hyperbolic covering space).

In section 1, the basic properties of chronological stationary-complete spacetimes are reviewed and the algebraic operators which define the global causal behavior of static- and stationary-complete spacetimes are explored, particularly addressing quotients by discrete group actions; some of this section is a review of notions introduced in \cite{GHr}, but some important new features are explored after the basic definitions.  In section 2, the effect of group actions is examined with respect to global causal properties, including explication of deviations in photon travel time.  In section 3, a number of examples will be closely examined, both of physical importance (such as cosmic strings in flat, Schwarzschild and other backgrounds or the Kerr solution) and of mathematical curiosity (the causally curious static-complete examples mentioned just above).

\subhead Summary of Highlights
\endsubhead

Here are some of the more important results of this note (given in somewhat different form than occurs in the main part of the note):

The detailed causality of a stationary-complete manifold $M$ is determined by a sort of algebraic invariant $\beta_M$: an operator, called the {\it fundamental cocycle\/}, defined on the loops (or cycles) of the space of stationary observers.  For a given space of stationary observers, two cocycles are called {\it homologous} if they agree on null-homotopic loops.  

\proclaim{Result A (Theorems 1.5 and 1.6)} A stationary-complete spacetime is entirely characterized by its stationary observer space, the length of the Killing field, and the fundamental cocycle.  Changing the fundamental cocycle to a homologous one, while keeping the same stationary observer space and Killing field length, changes only the manner in which the same fundamental group acts on the universal cover to yield the spacetime as a quotient.
\endproclaim

A new function is defined for pairs of points in a stationary-complete spacetime $M$, the {\it interval\/} $I_M$ (making use of the fundamental cocycle); this governs the chronology relation:

\proclaim{Result B (Theorem 2.8)} For $x$ and $y$ in $M$, a stationary-complete spacetime, $x \ll y$ if and only if $I_M(x,y) > 0$.
\endproclaim

Every cocycle has a {\it weight\/}, essentially its norm as an operator (in terms of $L$, the length of the cycles).  The weight of the fundamental cocycle of $M$, $\rwt(\beta_M)$, has a significant bearing on the causality of the spacetime.  

\proclaim{Result C (Corollary 2.17)} Let $M$ be a stationary-complete and globally hyperbolic spacetime  and $M'$ a quotient of $M$ by group action respecting the Killing field.  If $\rwt(\beta_{M'}) < 1$, then $M'$ is also globally hyperbolic.
\endproclaim

\proclaim{Result D (Theorem 2.18 and Corollary 2.20)} Let $M$ be a stationary-complete spacetime.  
\roster
\item If $\rwt(\beta_M) > 1$, then $M$ is chronologically vicious.
\item If $\rwt(\beta_M) = 1$ and some loop realizes this weight, then $M$ is chronological but not causal.
\item If $\rwt(\beta_M) = 1$ and no one loop realizes this weight, and for some sequence of curves $\{c_n\}$ realizing this weight, $\{L(c_n) - \beta_M\<c_n\>\} \to 0$, then $M$ is causal but not future-distinguishing.
\item If $\rwt(\beta_M) = 1$ and no one loop realizes this weight, and for every weight-realizing sequence of loops $\{c_n\}$, $\{L(c_n) - \beta_M\<c_n\>\}$ is bounded away from 0; or, alternatively, if $\rwt(\beta_M) < 1$; then $M$ is future-distinguishing and, indeed, strongly causal, stably causal, and even causally continuous.  Furthermore, this is precisely the condition for $M$ to have a presentation as a standard stationary spacetime.
\endroster
\endproclaim 

\proclaim{Result E (Theorem 2.18)} Let $M$ be a stationary-complete spacetime.  Then $M$ is globally hyperbolic if and only (1) the stationary observer space is complete in the appropriate metric and (2) either  $\rwt(\beta_M) = 1$ and for any weight-realizing sequence of loops $\{c_n\}$, $\{L(c_n) - \beta_M\<c_n\>\}$ grows to infinity, or, alternatively, $\rwt(\beta_M) < 1$.
\endproclaim

\example{Examples} A number of different types of examples are considered in detail, both purely mathematical constructs (showing the variety of behaviors that are possible) and more physical models illustrating how the tools of this paper can be applied.  Among the purely mathematical constructs are a strongly causal spacetime with a quotient that is causal but not strongly causal (Example 3.3) and a globally hyperbolic spacetime with a quotient that is still strongly causal but not globally hyperbolic (Example 3.4b).  The physical models start with flat, Schwarzschild, Kerr, and other backgrounds, then apply quotients by line- and circle-actions---including shifts in time-co\"ordinate, producing non-trivial fundamental cocycle---to show how the tools developed here describe the resulting spacetimes; in particular, we explore cosmic strings with temporal shifts in flat, Rindler, Schwarzschild, Kerr, and other backgrounds (Examples 3.5, 3.6, 3.7b, 3.8, and 3.10).
\endexample

\head 1. Basic Algebraic Structures
\endhead

In \cite{Ge} Geroch established the notion of analyzing a stationary spacetime in terms of the space of stationary orbits:  For $M$ a spacetime with an everywhere-timelike Killing field $K$, this means the integral curves of $K$.  Although, in general, the space of stationary orbits need not be in the least bit nice, it is shown in \cite{Hr1} that if $K$ is complete---and the orbits are lines, not circles---then the leaf space $Q$ of the Killing foliation is a (Hausdorff) manifold; the projection $\pi : M \to Q$ is the same as modding out by the $\Bbb R$ action on $M$ defined by $t\cdot x = \gamma_x(t)$, where $\gamma_x : \Bbb R \to M$ is the integral curve of $K$ with $\gamma_x(0) = x$.  

Actually (see \cite{Hr1}), all that's needed for a complete timelike vector field $K$ on chronological $M$ to result in a Hausdorff manifold for the quotient is that the flow generated by $K$ move a small neighborhood of each point away from (i.e., disjoint from) a sufficiently small neighborhood of any other point.  But with $K$ being Killing, this real action is an isometry, and that equips $Q$ with a Riemannian metric $h$.  Then, as observed by Geroch, we can write the spacetime metric $g$ in $M$ in this form (with signature $(-+\cdots+)$):
$g = -(\Omega \circ \pi)\alpha^2 + \pi^*h$,
where $\Omega : Q \to \Bbb R^+$ gives the length-squared of $K$ and $\alpha$ is the 1-form on $M$ which takes $K$ to 1 and whose kernel is $K^\perp$ (i.e., $\alpha = -\langle-,K\rangle/|K|^2$ and $|K_x|^2 = \Omega(\pi(x))$).  $M$ is static iff $d\alpha = 0$.

For definiteness, let us adopt the following terminology:  For $K$ a timelike Killing field on a spacetime $M$, define the {\it primary Killing form\/} associated with $K$ as $\alpha = -\langle-,K\rangle/|K|^2$, the {\it Killing orbit space\/} as the space $Q$ of stationary orbits, ie., integral curves of $K$ (also called the {\it stationary observer space}), and the {\it Killing projection\/} as $\pi: M \to Q$; defined on $Q$ are the {\it Killing orbit metric\/} $h$ (i.e., the induced Riemannian metric), the {\it Killing field length-squared\/} $\Omega = |K|^2$, and the {\it conformal metric\/} $(1/\Omega)h$.

Although the focus of this paper is chronological spacetimes, it will be helpful not to have to assume that {\it a priori}\/.  Therefore, we will make use of the broader sense referred to above:

\definition{Definition 1.1} A stationary-complete spacetime $M$ will be said to satisfy the {\it observer-manifold condition}\/ (with respect to a specified timelike Killing field)  if for each point $x\in M$ there is a neighborhood $U_x$ of $x$ such that for each point $y\in M$ there is a neighborhood $W_{x,y}$ of $y$ such that for $|t|$ large enough, $(t\cdot U_x) \cap W_{x,y} = \emptyset$.  (Note that this condition implies the Killing orbits are all lines, not circles.)
\enddefinition

By remarks in \cite{Hr1}, stemming from observations by Palais in \cite{P}, the observer-manifold condition suffices to ensure that the space of stationary observers is a (Hausdorff) manifold.  This is not an empty condition, as may be noted by considering a torus spacetime, $M = \Bbb L^2/\Bbb Z^2$ ($\Bbb L^n$ denoting Minkowski space with signature $(- + \cdots +)$) with the two-integer action on $\Bbb L^2$ via $(m,n) \cdot (t,x) = (t+m,x+n)$.  Take as Killing field $K = \partial/\partial t + a(\partial/\partial x)$ where $a$ is any irrational number with $|a| < 1$.  Then the $K$-action is ergodic, with $t\cdot U$, for any open set $U$, filling up the spacetime as $t$ increases; and, indeed, the Killing orbit space is perfectly horrendous.  

But chronology is sufficient to ensure the observer-manifold condition.  In order to have the results of this section stated in terms of this condition, and still be seen to apply to chronological stationary-complete spacetimes, the proof of this claim is provided here.

\proclaim{Proposition 1.2}
Let $M$ be a stationary-complete spacetime.  If $M$ is chronological then it obeys the observer-manifold condition.
\endproclaim
\demo{Proof}
Let $K$ be the specified timelike Killing field.  For any point $x \in M$ let $N_x^\epsilon$ be the image under $\text{exp}_x$ of the ball of radius $\epsilon$ in $K_x^\perp$, for $\epsilon > 0$ sufficiently small; let $B(x,\epsilon) = \{t \cdot z \st z \in N_x^\epsilon \text{ and } |t| < \epsilon\}$.   For some $\epsilon_0(x) > 0$, $\epsilon < \epsilon_0(x)$ implies that for all $z \in N_x^\epsilon$, $\gamma_z$ (the integral curve of $K$ with $\gamma_z(0) = z$) intersects $N_x^\epsilon$ precisely in $z$:  For otherwise, there are sequences of points $\{z_n\}$ and $\{\bar z_n\}$ in $N_x^\epsilon$ (any $\epsilon$ small enough) approaching $x$ with $\gamma_{z_n}(t_n) = \bar z_n$ for some $t_n \neq 0$; we can assume all $t_n > 0$. For $\epsilon$ sufficiently small, $N_x^\epsilon$ is achronal within $B(x,\epsilon)$ (viewed as a spacetime in its own right), so $\gamma_{z_n}$ must exit $B(x,\epsilon)$ before returning to it and encountering $\bar z_n$; hence, $t_n > 2\epsilon$.  For $n$ sufficiently large, this means (since $\gamma_{z_n}$ is not an arbitrary timelike curve but is constrained to be one of the $K$-curves) $\gamma_{z_n}$ enters the future of $x$ before exiting $B(x,\epsilon)$, so $\gamma_{z_n}(s_n) \gg x$ for some $s_n$ with $0 < s_n  < \epsilon$.  Similarly, for $n$ sufficiently large, as $\gamma_{z_n}$ enters   $B(x,\epsilon)$ on its way to $\bar z_n$, it must enter the past of $x$ within $B(x,\epsilon)$, so $\gamma_{z_n}(s'_n) \ll x$ for some $s'_n$ with $t_n - \epsilon < s'_n < t_n$.  As we also have $\epsilon < t_n - \epsilon$, this gives us $x \ll \gamma_{z_n}(s_n) \ll \gamma_{z_n}(s'_n) \ll x$, violating chronology.

Let $\epsilon_x = (1/2)\epsilon_0(x)$, and let $U_x = B(x,\epsilon_x)$.  The entire tube $T_x = \Bbb R \cdot N_x^{\epsilon_x}$ is well behaved, with $(t,z) \mapsto t \cdot z$ providing a diffeomorphism $\Bbb R \times N_x^{\epsilon_x} \to T_x$; indeed, this is true even using $\epsilon_0(x)$ for $\epsilon_x$ instead of half that value.   Now consider any $y \in M$.  If $y \in T_x$, then for some unique $z \in N_x^{\epsilon_x}$ and some unique $t_0$, $y = t_0 \cdot z$.  Let $W_{x,y} = B(t_0\cdot x, \epsilon_x)$; then for $t$ sufficiently large $(t \cdot U_x) \cap W_{x,y}$ is empty, as $t$ pushes $U_x$ well above or below the $t_0$ level.  If $y \in \partial(T_x)$, we do essentially the same thing: $W_{x,y} = B(t_0\cdot x,2\epsilon_x)$, as the tube works equally well at that radius.  Finally, if $y \in M - \text{closure}(T_x)$, let $W_{x,y} = M - \text{closure}(T_x)$, since the $\Bbb R$-action keeps $T_x$ within itself. \qed
\enddemo

The following definitions and constructions are taken from \cite{GHr}.  

Any line bundle such as $\pi : M \to Q$ has a cross-section $z: Q \to M$, i.e., $\pi(z(q)) = q$; this amounts to giving a global choice of starting-time for the clocks carried by the stationary observers.  Any such choice allows the definition of a map $\tau_z: M \to \Bbb R$ by $\tau_z(x) \cdot z(\pi(x)) = x$, i.e., $\tau_z$ gives the elapsed time of an event encountered by a stationary observer from that observer's $z$-starting-time. (In the absence of any confusion, this function will just be called $\tau$.) Note that for any cross-section $z$,  $(d\tau_z)K = 1$.  

Conversely, define a {\it Killing time-function}\/ as any function $\tau : M \to \Bbb R$ satisfying $(d\tau)K = 1$.  Then since $\tau$ necessarily takes on all values on any one $K$-orbit, we can define a cross-section $z_\tau: Q \to M$ by the requirement that $\tau\circ z_\tau = 0$.  Clearly $\tau = \tau_{z_\tau}$.   Also, for any cross-section $z$, $z_{\tau_z} = z$.  

For a Killing time-function $\tau$, note that $d\tau$ and $\alpha$ have the same effect on $K$, and both are invariant under the $K$-action; therefore, there is on $Q$ a unique 1-form $\omega$ (or $\omega_\tau$ for specificity) such that $\alpha - d\tau = \pi^*\omega$; call this the {\it Killing drift-form}\/ associated with the Killing time-function $\tau$ (or, alternatively, associated with the corresponding cross-section $z = z_\tau$).  Thus, we can write the spacetime metric as
$$g = - (\Omega\circ\pi)(d\tau + \pi^*\omega)^2 + \pi^*h,\tag1.1$$
and $M$ is static iff $d\omega = 0$ (i.e., $\omega$ is a closed 1-form).  We have a diffeomorphism $(\tau,\pi) : M \to \Bbb R \times\, Q$, but this is not a metric product nor even conformal to one, so long as $\omega$ is non-zero.  

It should be noted that this formulation, encapsulated in (1.1), is slightly more general than what is sometimes called a {\it standard stationary}\/ spacetime (see, for example \cite{JS}):  That is a spacetime $M$ with maps $\tau: M \to \Bbb R$ and $\pi : M \to Q$, with $Q$ bearing a 1-form $\omega$, a Riemannian metric $g_Q$, and a positive function $\Omega : Q \to \Bbb R$, with the spacetime metric given by
$$g = (\Omega\circ\pi)(-d\tau^2 - d\tau\otimes\pi^*\omega - \pi^*\omega\otimes d\tau + \pi^*g_Q).\tag1.1a$$
This matches up with (1.1) for $h = \Omega(\omega^2 + g_Q$).  But the difference is that in the standard stationary formulation, it is not sufficient that the metric $h$ induced on the stationary observer space $Q$ be Riemannian; rather, it is required that $(1/\Omega)h - \omega^2$ (that is, $g_Q$) be Riemannian, i.e., that $||\omega|| < 1$ for norm calculated using $\bar h = (1/\Omega)h$---the {\it conformal metric}\/ on $Q$, mentioned earlier. (Note that this restriction on the conformal norm of $\omega$ is equivalent to the image of $z$---that is, the $\tau = 0$ hypersurface---being locally spacelike.  Thus, a presentation of the spacetime as standard stationary is the same as having a spacelike cross-section; this observation is the same as Lemma 3.3 in  \cite{JS}.)  As will be seen in section 2, this additional requirement for the standard stationary formulation (i.e., conformal norm of $\omega$ less than 1) is closely related to $M$ being causal, so the extra generality allowed in the present formulation is perhaps not of great use; but this generality makes it possible to treat cases that might not {\it a priori}\/ be known to be causal. 

In essence, $\omega$ (or, rather, $-\omega$) measures, from one stationary observer to the ones infinitesimally close to it, the difference in starting-times for their $z$-clocks, as measured by the universal clock $K$. More precisely:  If $X$ is a vector in $Q$ and $\bar X$ is a lift of $X$ to $M$, perpendicular to $K$, then $-\omega(X) = (d\tau)\bar X$; thus, $-\omega$ measures infinitesimal change in $\tau$ (i.e., in starting-time) in directions perpendicular to the Killing field.  

A good physical interpretation of the drift-form $\omega$ can be gleaned from Proposition 1.3 below.  

At any one point, photons travel at a speed of one $\tau$-unit of time per one conformal unit of length, so a naive expectation is that a closed path of conformal length $L$ in $Q$, traced about by a photon, would lead to the photon coming back with an elapsed $\tau$-time of $L$.  It is the integral of  $\omega$ over such a loop that specifies the extent to which this naive expectation is incorrect.  This is what inspires the term ``drift-form", as one can think of what is being measured by $\int_c \omega$ as being an inherent drift or wind felt in the observer-space $Q$, affecting the transmission of a signal along the path $c$ of observers; however, $\omega$ depends on the choice of cross-section, so we will develop another object, the ``fundamental cocycle", that does not depend on cross-section.

There is a gauge freedom in the choice of the cross-section $z$:  For any map $\eta: Q \to \Bbb R$, we can change $z$ to $z^\eta$, defined by $z^\eta(q) = \eta(q) \cdot z(q)$ (and this encompasses all possible cross-sections).  Then $\tau^\eta = \tau - \eta\circ\pi$ (i.e., $\tau_{z^\eta} = \tau_z - \eta\circ\pi$) and $\omega^\eta = \omega + d\eta$.  Thus, $\omega$ is defined up to an exact 1-form.  In the static case, we can choose the cross-section so that $\omega$ is zero on any simply-connected neighborhood (so the static observers have the same starting-times, as measured by $K$), but whether this can be globalized depends on the global topology of $M$ and whether the closed form $\omega$ interacts with that global topology to prevent its representation as an exact form.  In the general stationary case, things are even more complicated.

The static case has a simple explication of the 1-form $\omega$:  As it is closed and defined up to an exact 1-form, it precisely defines an element $\beta$ of the first de Rham cohomology of $M$, $H^1(M;\Bbb R)$; call it the {\it fundamental cohomology class}\/ of the static spacetime $M$.  This is most easily interpreted as being a real-valued group map $\rho_\omega$ on elements of the fundamental group of $Q$, $G = \pi_1(Q)$:  Choose a base point $q_0$ in $Q$ (better yet, choose a base point $x_0 \in M$, and let $q_0 = \pi(x_0)$).  Then for any element $a \in G$, $a$ is represented by a base-pointed loop $c$ in $Q$ (i.e., $c$ starts and ends at $q_0$); we write $a = [c]$.   We define $\rho_\omega(a) = \int_c\omega$; then $\rho_\omega : G \to \Bbb R$ is a group morphism, as $\rho_\omega(aa') = \int_{c\cdot c'} \omega = \int_c \omega + \int_{c'} \omega$, where $a = [c]$, $a' = [c']$, and $c \cdot c'$ indicates concatenation of curves (first $c'$, then $c$).  As noted above, $\int_c \omega$ has the physical significance of being the difference between conformal length of the loop $c$ and change in observer-time between emission and reception of a photon along that path.  It can be expressed in an explicitly gauge-independent manner as $\beta([c]) =  \int_{\bar c} \alpha$, where $\bar c$ is any loop in $M$ which is a lift of $c$ (the independence among representative loops $c$ is due to Stokes' Theorem, as two homotopic loops form the boundary of a 2-surface---the homotopy between them).  

For the general stationary case, things aren't as neat.  But we can still employ a form of algebraic structure by considering the Abelian group $Z(Q)$ (the {\it cycles}\/ of $Q$) generated by all base-pointed loops $c$ in $Q$ (parametrized with $\dot c$ never 0), subject to these relations: 
\roster
\item same-direction reparametrization is irrelevant: if $c'$ is a reparametrization of $c$ in the same direction, then $c$ and $c'$ represent the same element;
\item reverse-parametrization is inverse: if $c'$ is $c$ with the reverse parametrization, then $c$ and $c'$ represent inverse elements; and
\item concatenation is sum: the concatenation of $c$ and $c'$ represents the same element as their group sum.
\endroster
(This exposition will work for any manifold $Q$, not just the observer space of a stationary spacetime.)  We write that the loop $c$ represents the element $\<c\>$ in $Z(Q)$.  Call a loop {\it simple}\/ if it has no base-pointed sub-loops that are reverse parametrizations of one another. Any cycle $\zeta$ is then represented by an essentially unique base-pointed simple loop $c$ in $Q$; more precisely, for any $\zeta \in Z(Q)$, there is a unique collection of oriented (but unparametrized) base-pointed loops $\{c_i\}$ (with none of the $c_i$ the reversal of another $c_j$) such that $\zeta$ is represented by the concatenation (in any order) of the $\{c_i\}$.  

Define $Z^*(Q) = \text{Hom}(Z(Q),\Bbb R)$ (the {\it cocycles}\/ of $Q$).  Any 1-form $\theta$ on $Q$ defines a cocycle $\{\theta\}$ via integration:  If $\zeta = \<c\>$ is a cycle, then $\{\theta\}(\zeta) = \int_c \theta$, which is independent of the representation for $\zeta$.  Note that if we add any exact form to $\theta$, it doesn't change the cocycle:  For any $\eta : Q \to \Bbb R$, for any loop $c$, $\int_c d\eta = 0$, so $\{\theta + d\eta\} = \{\theta\}$.

(The definition of cycles provides perhaps the minimum identifications needed so that cycles are understood to be platforms for integration of 1-forms.  But does this definition perhaps allow other cocycles than those coming from 1-forms?  It might be provident to  sharpen the notion of cycles so as to capture only 1-forms as cocycles.  For instance, we could employ additional identifications:
\roster
\item[4] Segment interchange:  If loops $c$ and $c'$ are composed of concatenated segments $c = \sigma_2 \cdot \sigma_1$ and $c' = \sigma'_2 \cdot \sigma'_1$, with the join-point of the segments in $c$ the same as that in $c'$, then with $c'' = \sigma'_2 \cdot \sigma_1$ and $c''' = \sigma_2 \cdot \sigma'_1$, $c + c' \sim c'' + c'''$ (i.e., $\sigma_2 \cdot \sigma_1 \cdot \sigma'_2 \cdot \sigma'_1 \sim \sigma'_2 \cdot \sigma_1 \cdot \sigma_2 \cdot \sigma'_1$).
\item Segment reversal: If a loop $c$ is composed of concatenated segments $\sigma_4 \cdot \sigma_3 \cdot \sigma_2 \cdot \sigma_1$ with $\sigma_2$ and $\sigma_3$ the same except for being reverse-oriented, then with $c' = \sigma_4 \cdot \sigma_1$, $c' \sim c$ (i.e., $\sigma_4 \cdot (\sigma_2)^{-1} \cdot \sigma_2 \cdot \sigma_1 \sim \sigma_4 \cdot \sigma_1$).
\endroster
Let $Z'(Q) = \{\zeta - \zeta' \in Z(Q) \st \zeta \sim \zeta'\}$ and let $\tilde Z^*(Q) = \{\beta \in Z^*(Q) \st Z'(Q) \subset \text{Ker}(\beta)\}$; equivalently, $\tilde Z^*(Q) = \text{Hom}(\tilde Z(Q), \Bbb R)$, where $\tilde Z(Q) = Z(Q)/Z'(Q)$.  Then it might be the case that these additional restrictions on cocycles, along with some sort of continuity condition, are sufficient to guarantee that all cocycles are of the form $\{\theta\}$ for some 1-form $\theta$.  However, this possibility will not be further explored in this note.)  

We now define the {\it fundamental cocycle}\/ for a stationary-complete spacetime  satisfying the observer-manifold property $M$ as $\beta_M = \{\omega\}$, where $\omega = \omega_z$ for some cross-section $z$ of the stationary observer line-bundle, the cocycle being independent of the cross-section.\footnote{Observation due to James Hebda: In the language of connections, $\beta_M$ gives the holonomy for curves in the base space of the principle fibre-bundle $\pi: M \to Q$ with connection 1-form $\alpha$.}  The fundamental cocycle precisely gives the difference between the actual and the naively expected travel times for photons along specified paths:

\proclaim{Proposition 1.3}  Let $\pi: M \to Q$ be a stationary-complete spacetime with Killing orbit-space $Q$ and fundamental cocycle $\beta_M$.  If $c$ is any loop in $Q$ from $q$ to $q$ and $\bar c$ any future-null lift of $c$ to $M$, from $x$ to $x'$ (both in $\pi^{-1}(q)$), then $x' = T \cdot x$ for $T = L(c) - \beta_M\<c\>$ (where $L$ is conformal length). 
\endproclaim
\demo{Proof} Let $\tau$ be any Killing time-function on $M$ with corresponding Killing drift-form $\omega$ (so spacetime metric $g = -(\Omega \circ \pi)(d\tau + \pi^*\omega)^2 + \pi^*h)$).  Say $c$ is parametrized as $c : [0,S] \to Q$; we can specify $\bar c : [0,S] \to M$ as $\bar c(s) = t(s) \cdot z(c(s))$ for some function $t : [0,S] \to \Bbb R$, i.e., $\tau(\bar c(s)) = t(s)$, so $(d\tau)\dot{\bar c} = t'$.  Then $\bar c$ is future-null for $t' = -\omega\dot c + \sqrt{h(\dot c, \dot c)/\Omega(c)}$.   The two end-points of $\bar c$, $x$ and $x'$, lie on the same stationary orbit (i.e., $q$) with a separation $T$ (i.e., $x' = T \cdot x$) defined by $T = \int_c t' = -\int_c \omega + \int^S_0 \sqrt{h(\dot c, \dot c)/\Omega(c)}\,ds = - \int_c\omega + \int_0^S||\dot c||\,ds$ (using conformal norm, i.e., from $\bar h$). Then we have $T = L(c) - \beta_M\<c\>$ as desired. \qed
\enddemo

This gives us a physical interpretation of drift-form: If a stationary observer constrains a photon to move along a specific closed track $c$ among stationary observers and measures the time between emission of a photon along that track and its reception, then $\int_c\omega$ is the difference between the conformal length of $c$ and that measured time.

\vskip .1 in

The balance of the material in this section is new.

Cocycles defined by 1-forms work very nicely.  Let $Z_0(Q)$ be the subgroup of cycles generated by null-homotopic loops.  Then we have the following.

\proclaim{Proposition 1.4}Let $Q$ be a manifold and $\theta$ a 1-form on $Q$.  Then
\roster
\item"(a)" $\{\theta\} = 0$ iff $\theta$ is exact.
\item"(b)" $Z_0(Q) \subset \roman{Ker}(\{\theta\})$ iff $\theta$ is closed.
\endroster
\endproclaim
\demo{Proof}
(a) If $\theta = d\eta$ for some function $\eta : Q \to \Bbb R$, then clearly $\int_c \theta = 0$ for any loop $c$.  Conversely, suppose $\int_c \theta = 0$ for all base-pointed loops $c$; this implies that the integral of $\theta$ from base-point $q_0$ to any other point $q$ is path-independent.  Therefore, we can define $\eta: Q \to \Bbb R$ by $\eta(q) = \int_\sigma \theta$, where $\sigma$ is any path from $q_0$ to $q$.  Then $\theta = d\eta$, and we are done.

(b) If $d\theta = 0$, then for any null-homotopic base-pointed loop $c$, $c$ bounds a 2-surface $\Sigma$ (a smooth homotopy from $c$ to base-point).  Then by Stokes' Theorem, $\int_c \theta = \int_{\partial(\Sigma)} \theta = \int_\Sigma d\theta = 0$.  Conversely, suppose $\int_c\theta = 0$ for all null-homotopic base-pointed loops $c$, and suppose $d\theta_{q_1} \neq 0$ for some point $q_1$.  There are two non-linearly-related vectors $X$ and $Y$ at $q_1$ such that $(d\theta)(X,Y) > 0$.  Consider a small disk $\Sigma$ centered at $q_1$, with boundary a loop $\sigma$, such that $T_{q_1}\Sigma = \text{span}\{X,Y\}$; give $\Sigma$ an orientation with the ordering $(X,Y)$ being the positive sense at $q_1$, and let this determine the orientation for $\sigma$.  Let $\sigma'$ be a curve from $q_0$ to $q_1$.  Let $c = (-\sigma')\cdot\sigma\cdot\sigma'$, where $-\sigma'$ denotes $\sigma'$ with reverse parametrization.  Then $c$ is null-homotopic, so $0 = \{\theta\}\<c\> = \int_c \theta = -\int_{\sigma'}\theta + \int_\sigma\theta + \int_{\sigma'}\theta = \int_\sigma \theta = \int_{\partial\Sigma}\theta = \int_\Sigma d\theta$.  But with $\Sigma$ sufficiently small, $\int_\Sigma d\theta > 0$, as $d\theta$ on $T_q\Sigma$ won't vary by much from its value at $q_1$.  This contradiction shows $d\theta = 0$ everywhere. \qed
\enddemo

The cycle and cocycle constructions are, of course, functorial:  For any map $\phi : Q \to Q'$, we have induced maps $\phi_* : Z(Q) \to Z(Q')$ and $\phi^* :Z^*(Q') \to Z^*(Q)$ given by $\phi_*\<c\> = \<\phi\circ c\>$ and $\phi^*\beta' : \zeta \mapsto \beta'(\phi_*\zeta)$.  This allows us to compare cocycles in different stationary spacetimes.  

We also have an easy identification of cocycles under change of base-point:  Let $Z_q(Q)$ and $Z_q^*(Q)$ denote cycles and cocycles for base-point being $q$.  For any other point $q' \in Q$, let $\sigma$ be a curve from $q$ to $q'$.  Then we can define an isomorphism of groups $\phi_\sigma: Z_q(Q) \to Z_{q'}(Q)$ by prepending and appending $\sigma$ to the beginning and end of loops at $q$, i.e., $\phi_\sigma : \<c\> \mapsto \<\sigma\cdot c\cdot(-\sigma)\>$ (first go from $q'$ to $q$ via backwards $\sigma$, then do the loop $c$, then go back to $q'$ via $\sigma$).  Then $\phi_\sigma^* : Z_{q'}^*(Q) \to Z_q^*(Q)$ is also an isomorphism, where $\phi_\sigma^*(\beta) : \zeta \mapsto \beta(\phi_\sigma\zeta)$.

The first important point to be made about the fundamental cocycle is that together with the observer-space and its metric and the length of the Killing field, it provides full information to define the spacetime (that is to say, we don't need the 1-form, but only the cocycle it defines):

\proclaim{Theorem 1.5}
Suppose $M$ and $M'$ are both stationary-complete spacetimes satisfying the observer-manifold condition, with $Q$ and $Q'$ their respective stationary observer spaces (with Killing projections $\pi: M \to Q$ and $\pi': M' \to Q'$) with induced Riemannian metrics $h$ and $h'$ and Killing squared-lengths $\Omega : Q \to \Bbb R$ and $\Omega': Q' \to \Bbb R$.  

If there is a diffeormorphism $\phi: Q \to Q'$ preserving observer-space metric, Killing length, and fundamental cocycles, i.e., 
\roster
\item $\phi^*h' = h$,
\item $\Omega'\circ\phi = \Omega$, and
\item $\phi^*\beta_{M'} = \beta_M$,
\endroster
then $\phi$ is induced by an isometry $\psi : M \to M'$ (i.e., $\pi'\circ\psi = \phi\circ\pi$).
\endproclaim

\demo{Proof}
Choose cross-sections $z$ and $z'$ of the bundles $\pi$ and $\pi'$, yielding representative 1-forms $\omega$ and $\omega'$ and time-functions $\tau$ and $\tau'$.  Let $g$ and $g'$ be the spacetime metrics on $M$ and $M'$, respectively.  Then we have
$$\align
g & = -(\Omega\circ\pi)(d\tau + \pi^*\omega)^2 + \pi^*h \\
g' & = -(\Omega'\circ\pi')(d\tau' + \pi'{}^*\omega')^2 + \pi'{}^*h'
\endalign$$
Define $\psi: M \to M'$ by $(\tau',\pi')\circ\psi = (\tau,\phi\circ\pi)$, i.e., $\psi(x) = \tau(x)\cdot z'(\phi(\pi(x)))$.  We manifestly have $\phi$ induced by $\psi$, and it's clear $\psi$ is a diffeomorphism; we just need to see if $\psi$ is an isometry.  We have
$$\align
\psi^*g' & = -(\Omega'\circ\pi'\circ\psi)(\psi^*d\tau' + \psi^*\pi'{}^*\omega')^2 + \psi^*\pi'{}^*h' \\
& = -(\Omega'\circ\phi\circ\pi)(d\tau + \pi^*\phi^*\omega')^2 + \pi^*\phi^*h' \\
& = -(\Omega\circ\pi)(d\tau + \pi^*\phi^*\omega')^2 + \pi^*h
\endalign$$
Thus, all we need for $\psi$ to be an isometry is that $\phi^*\omega' = \omega$.  However, this is not automatically the case.  

What we know is that $\phi^*\beta_{M'} = \beta_M$, so $\{\phi^*\omega' - \omega\} = 0$; therefore, by Proposition 1.4(a), $\phi^*\omega' - \omega$ is exact, i.e., $\phi^*\omega' = \omega + d\eta$  for some $\eta: Q \to \Bbb R$. But we can replace $z$ by $z^\eta$, resulting in $\tau^\eta = \tau - (\eta\circ\pi)$ and $\omega^\eta = \omega + d\eta$.  This gives us precisely $\phi^*\omega' = \omega^\eta$; thus, changing to this corrected cross-section in $M$ yields the desired isometry. \qed
\enddemo

Cocycles have a considerable amount of freedom, compared to first cohomology classes.  The difference is that cohomology classes are rigidly fixed within a homotopy class of loops (Proposition 1.4(b)), while cocycles can vary within a homotopy class.  (Even if we look only at cocycles obeying conditions (4) and (5), that only restricts them to having an additive kind of condition on sub-loops.)  For a given manifold $Q$, we can classify the possible cocycles by this freedom within a homotopy class, asserting that there is a basic similarity---call it homology---between cocycles which are the same, except for constants which depend only on homotopy class.  Specifically, we will say two cocycles $\beta$ and $\beta'$ on a manifold $Q$ are {\it homologous}\/ if $\beta - \beta'$ vanishes on $Z_0(Q)$; that means there is some $\delta \in H^1(Q;\Bbb R)$ such that for any base-pointed loop $c$ in $Q$, $\beta'\<c\> = \beta\<c\> + \delta[c]$.  This is an important classifying property for stationary spacetimes.

\proclaim{Theorem 1.6} Suppose $M_1$ and $M_2$ are stationary-complete spacetimes satisfying the observer-manifold property, with the same stationary observer-space (including induced metric) and same Killing length-function.  Then $\beta_{M_1}$ and $\beta_{M_2}$ are homologous if and only if $M_1$ and $M_2$ share a covering spacetime respecting the Killing fields.
\endproclaim

\demo{Proof}
We'll do an analysis of group actions on a stationary-complete spacetime, looking at the structures involved.  We start with $M$ and $M'$ stationary-complete and satisfying the observer-manifold condition, with observer-space projections $\pi: M \to Q$ and $\pi': M' \to Q'$, Killing-length-squared maps $\Omega: Q \to \Bbb R$ and $\Omega': Q' \to \Bbb R$ for respective Killing fields $K$ and $K'$, with spacetime metrics $g$ and $g'$ and induced observer-space metrics $h$ and $h'$.  We want to know how it can be that $M' = M/G$ for some group $G$ acting properly discontinuously and isometrically on $M$ (and with $K$ on $M$ inducing $K'$  on $M/G$; if $M$ happens to have more than one Killing field, we need to specify that it is the $K$-structure in $M$ we want mapping to the $K'$-structure in $M'$).  

In the following somewhat technical lemma---needed for the proof of the theorem, but also of some interest in itself, providing insight on the effect of group actions on various structures---and in general in this paper, ``action of a group" is always taken to mean a properly discontinuous and effective action, so that the quotient space is a manifold and the isotropy group is trivial.  Much of the results in this paper could also be formulated for general covering projections, but it is more convenient to restrict to the nicest group actions.

\proclaim{Lemma 1.7}$M'$, with Killing field $K'$, is the quotient of the action by a group $G$ of isometries on $M$, with Killing field $K$ (so $p_M: M \to M'$ is projection by group action, with $K' = {p_M}_*K$), means precisely that  \roster
\item $G$ acts by isometries on $Q$ and $\Omega$ is $G$-invariant,
\item $Q'= Q/G$, and
\item For any Killing drift-form $\omega$ on $Q$, there is a function $\eta : Q \to \Bbb R$ such that $\omega^\eta$ is $G$-invariant.  For a given target-space $M'$, $\eta$ is determined up to an arbitrary $G$-invariant function on $Q$.
\endroster
Then with $p_Q : Q \to Q'$ the projection induced by $p_M$, and taking for base-points $q_0$ in $Q$ and $q'_0 = p_Q(q_0)$ in $Q'$, the following hold:
\roster
\item"(a)" $p_Q^* h' = h$
\item"(b)" $\Omega' \circ p_Q = \Omega$
\item"(c)" $p_Q^* \beta_{M'} = \beta_M$
\item"(d)" $\beta_{M'}\<c'\> = \int_c \omega + \eta(c(1)) - \eta(q_0)$ for any base-pointed loop $c': [0,1] \to Q'$, where $c: [0,1] \to Q$ is the lift of $c'$ starting at base-point.
\item"(e)" $p_Q^*\omega' = \omega^\eta$, where $\omega'$ is the drift-form corresponding to the cross-section $z': Q' \to M'$ defined by $z'(p_Q(q)) = p_M(x)$ for $\tau(x) = \eta(q)$, $\pi(x) = q$ (alternatively: $z' = z_{\tau'}$ for $\tau': M' \to \Bbb R$ obeying $\tau' \circ p_M = \tau^{\eta}$). 
\endroster
\endproclaim
\demo{Proof of Lemma}

First let us suppose that the group $G$ acts on $M$ with $M' = G/M$ and $K' = p_M{}_* K$ for $p_M : M \to M/G = M'$ the projection.  It then follows that the $G$-action on $M$ must preserve $K$ and, hence, the $K$-orbits; thus each element $a \in G$ yields a motion on $Q$, $q \mapsto a \cdot q$, i.e., there is a $G$-action on $Q$.  Since the $G$-action on $M$ is by isometries, it is also by isometries on $Q$; similarly, the properly discontinuous nature of the action on $M$ is inherited by the action on $Q$.  Therefore, $p_M$ defines a projection $p_Q : Q \to Q/G$, and we can identify $Q/G$ with $Q'$ with $h = p_Q^*h'$.  And since the $G$-action must preserve $K$, it also preserves the length of $K$, hence, the map $\Omega$.  Thus, we have conditions (1) and (2).

Let $\tau : M \to \Bbb R$ be any Killing time-function on $M$, with $\omega = \omega_\tau$ the drift-form. 

Now consider any Killing time-function $\tau': M' \to \Bbb R$ on the quotient space, with drift-form $\omega' = \omega_{\tau'}$.  Note that $\bar\tau = \tau' \circ p_M$ is another Killing time-function on $M$ ($(d\bar\tau)K = (d\tau'){p_M}_*K = (d\tau')K' = 1$).  It follows that for some $\eta: Q \to \Bbb R$, $\bar\tau = \tau^\eta$($= \tau - \eta\circ\pi$); accordingly, with $\bar\omega$ the drift-form from $\bar\tau$, we also get $\bar\omega = \omega^\eta$($ = \omega + d\eta$).  Since $\bar\tau$ factors through $p_M$, it is $G$-invariant; it follows that $\bar\omega$ is also (since $G$ acts on $M$ via isometry and preserves the Killing vector $K$, we have $\alpha$ is $G$-invariant; then $\pi^*\bar\omega = \alpha - d\bar\tau$ shows $\bar\omega$ to be $G$-invariant).  This gives us the first sentence in (3).

What freedom is there in choosing $\eta$ to have $\omega^\eta$ be $G$-invariant?  Any drift-form $\tilde{\omega}$ on $Q$ must arise as $\tilde{\omega} = \omega^{\tilde\eta}$ for some $\tilde\eta : Q \to M$; then $\tilde{\omega}$ being $G$-invariant implies $\tilde{\tau} = \tau^{\tilde\eta}$ is as well, which means that $\tilde{\tau}$ factors through $M'$, i.e., $\tilde{\tau} = \tilde\tau' \circ p_M$ for some $\tilde\tau': M' \to \Bbb R$.  Then $\tilde\tau' = {\tau'}^{\eta'}$ for some $\eta': Q' \to \Bbb R$.  Then we have
$$\alignat 2
\bar\tau & = \tau^\eta &   \tilde\tau & = \tau^{\tilde\eta} \\
\tau' \circ p_M & = \tau - \eta\circ\pi &  \tilde\tau' \circ p_M & = \tau - \tilde\eta\circ\pi \\
& &  (\tau' - \eta'\circ\pi') \circ p_M & = \tau - \tilde\eta\circ\pi \\
& &  \tau' \circ p_M - \eta'\circ\pi'\circ p_M & = \tau - \tilde\eta\circ\pi \\
& & \tau - \eta\circ\pi - \eta'\circ p_Q\circ\pi & = \tau - \tilde\eta\circ \pi \\
& & (\tilde\eta - \eta)\circ\pi & = \eta'\circ p_Q\circ\pi \\
& & \quad \tilde\eta - \eta & = \eta'\circ p_Q
\endalignat$$
In other words, $\tilde\eta$ must differ from $\eta$ precisely by something which factors through $Q'$, i.e., is $G$-invariant; further, any such $G$-invariant difference is allowed.  This is the second sentence of (3).

Conversely, suppose we are given the stationary spacetime $M$ (with observer space $Q$) along with a group $G$ and a map $\eta : Q \to \Bbb R$ obeying conditions (1) and (3).  As per condition (2), we define $Q' = Q/G$; by condition (1), we get a Riemannian metric $h'$ on $Q'$ and map $\Omega' : Q' \to \Bbb R$ such that properties (a) and (b) hold.  We will define $M'$ as the quotient of a $G$-action on $M$.  

We will use $(\tau,\pi): M \to \Bbb R \times Q$ to parametrize $M$.  Define, for any $a \in G$, $a\cdot(t,q) = (t + \eta(a\cdot q) - \eta(q), a \cdot q)$.  It is easy to check that this is a group action. Since $G$ acts properly discontinuously on $Q$, the same is true of the action on $M$.  Let us use this convention for any element $a \in G$:  $a_* = (R_a)_*$ and $a^* = (R_a)^*$, where $R_a: Q \to Q$ denotes multiplication by $a$ (or its generalization to $M$).  Then for $X \in T_qQ$, $a_*(t\frac{\partial}{\partial\tau}, X) = ((t + (a^*d\eta\,)X - (d\eta\,)X)\frac{\partial}{\partial\tau}, a_*X)$.  Thus $a^*d\tau = d\tau + a^*\pi^*d\eta - \pi^*d\eta$; we also have $a^*$ and $\pi^*$ commute, and $a^*h = h$.  This gives us (using condition (3)---$\omega + d\eta$ is $G$-invariant---in the last step)
$$\align
a^*g & = -(\Omega\circ\pi)(a^*d\tau + a^*\pi^*\omega)^2 + a^*\pi^*h \\
& = -(\Omega\circ\pi)(d\tau + a^*\pi^*d\eta - \pi^*d\eta + \pi^*a^*\omega)^2 + \pi^*h \\
& = -(\Omega\circ\pi)(d\tau + \pi^*(a^*(d\eta + \omega) -(d\eta + \omega)) + \pi^*\omega))^2 + \pi^*h \\
& = -(\Omega\circ\pi)(d\tau + \pi^*\omega)^2 + \pi^*h \\
& = g.
\endalign$$
This shows us $G$ acts by isometries on $M$, carrying $K$ to $K$. We define $M' = M/G$, with $p_M : M \to M'$ covering $p_Q : Q \to Q'$, and $M'$ is stationary with Killing field $K' = p_M{}_* K$.  

Had we replaced $\eta$ by $\tilde\eta$ such that $\tilde\eta - \eta$ is $G$-invariant---i.e., instead of $\eta$, used $\eta + \eta' \circ p_Q$ for any $\eta': Q' \to \Bbb R$---the $G$-action on $M$ would have been identical, thus yielding the same $M'$.

All that is left to us now is to show that properties (c), (d), and (e) follow from conditions (1), (2), and (3) (i.e., from $M'$ being a quotient of $M$, with corresponding Killing fields).  We'll start by recalling that the Killing time-function $\tau'$ obeys $\tau^\eta = \tau' \circ p_M$.  Then the cross-section $z'$ defined by $\tau'$ is characterized by $\tau' \circ z' = 0$.  We can express this by saying for any $q \in Q$, $z'(p_Q(q)) = p_M(x)$ for some $x \in M$ such that $0 = \tau'(z'(p_Q(q))) = \tau'(p_M(x)) = \tau^\eta(x) = \tau(x) - \eta(\pi(x)) = \tau(x) - \eta(q)$. This establishes the characterization of $z'$ in property (e).

We know that $\omega^\eta$ is $G$-invariant, so there is a unique 1-form $\omega'$ in $Q'$ with $\omega^\eta = {p_Q}^*\omega'$.  We need to show this is the drift-form corresponding to the cross-section $z'$, i.e., using the time-function $\tau'$.  Now, with $\alpha'$ the primary Killing form associated with $K'$, we have $p_M^*\alpha' = \alpha$. Therefore (using $p_M^*d\tau' = d\tau - \pi^*d\eta$), ${p_M}^*(\alpha' - d\tau') = \alpha - d\tau + \pi^*d\eta = \pi^*(d\eta + \omega) = \pi^*p_Q^*\omega' = {p_M}^*\pi'{}^*\omega'$, from which we get $\alpha' - d\tau' = \pi'{}^*\omega'$, precisely as needed.  This proves property (e).

So with $\omega'$ identified by property (e), we can calculate $\beta_{M'}$ for properties (c) and (d).  Let $c': [0,1] \to Q'$ be a loop starting and ending at base-point $q_0' = \pi q_0$ with lift $c: [0,1] \to Q$  starting at $q_0$.  We have $\beta_{M'}\<c'\> = \{\omega'\}\<c'\> = \int_{c'}\omega' = \int_c p_Q^*\omega'  = \int_c (d\eta + \omega) = \int_c\omega + \eta(c(1)) - \eta(q_0)$, which is property (d).  For property (c), we just apply this to starting with a loop $c$ in $Q$: $(p_Q^*\beta_{M'})\<c\> = \beta_{M'}p_Q{}_*\<c\> = \beta_{M'}\<p_Q\circ c\> = \int_c\omega + \eta(q_0) - \eta(q_0) = \int_c\omega = \beta_M\<c\>$ (where we have used that the lift of $p_Q\circ c$ is the loop $c$).  Therefore, $p_Q^*\beta_{M'} = \beta_M$, property (c). \qed
\enddemo

The freedom in defining $\eta$ amounts to saying that $\eta$ is arbitrary on any fundamental region of $Q$, and that it is then fully constrained on the rest of $Q$ by requirement of condition (3).

\vskip .1 in

Continuing with the proof of Theorem 1.6:  First suppose that $M_1$ and $M_2$ (with observer spaces $Q_1$ and $Q_2$ respectively) have a common covering space; we might as well assume this is the universal cover $\tilde M$ of both $M_1$ and $M_2$ (and it has observer space $\tilde Q$).  We have $M_i = \tilde M/G_i$ with $G_i = \pi_1(M_i)$, the fundamental groups; but since $Q_i = \tilde Q/G_i$ and $Q_1 = Q_2$ (call it $Q$) we have $G_1 = G_2$; call it $G$.  By virtue of Lemma 1.7, there are maps $\tilde \eta_i : \tilde Q \to \Bbb R$ with $\tilde \omega + d\tilde \eta_i$ $G$-invariant (where $\tilde\omega$ is a 1-form on $\tilde Q$), and for any base-pointed loop $c$ in $Q$, $\beta_{M_i}\<c\> = \int_{\tilde c}\tilde\omega + \tilde \eta_i(\tilde q) - \tilde \eta_i(\tilde q_0)$, where $\tilde c$ is a lift of $c$ to $\tilde Q$ from $\tilde q_0$ to some $\tilde q$.  Therefore, $(\beta_{M_1} - \beta_{M_2})\<c\> = \tilde\delta(\tilde q) - \tilde\delta(\tilde q_0)$, where $\tilde\delta = \tilde \eta_1 - \tilde \eta_2$.  But condition (3) tells us $d\tilde\delta$ is $G$-invariant, so there is a 1-form $\theta$ on $Q$ with $p_{\tilde Q}{}^* \theta = d\tilde\delta$; and since $d\tilde\delta$ is exact, $\theta$ is closed. We then have $\beta_{M_1} - \beta_{M_2} = \{\theta\}$; and with $\theta$ being closed we know $\{\theta\}$ depends only on the homotopy class of each loop, i.e., $\beta_{M_1}$ and $\beta_{M_2}$ are homologous.  

Now suppose $\beta_{M_1}$ and $\beta_{M_2}$ are homologous.  We can write $\beta_{M_i} = \{\omega_i\}$ for 1-forms $\omega_1$ and $\omega_2$ on $Q$ (the common observer space for $M_1$ and $M_2$); let $\theta = \omega_1 - \omega_2$.  We have $\beta_{M_1} - \beta_{M_2} = \{\theta\}$.  We know that for all null-homotopic base-pointed loops $c$ in $Q$, $\{\theta\}\<c\> = 0$, i.e., $Z_0(Q) \subset \text{Ker}(\theta)$; therefore, by Proposition 1.4(b), $d\theta = 0$.  

For each $i$, let $\tilde M_i$ be the universal cover of $M_i$ with projection $p_{\tilde M_i} : \tilde M_i \to M_i$.  With $\tilde Q_i$ the observer space (and projection $\tilde \pi_i : \tilde M_i \to \tilde Q_i$), we know from Lemma 4.2 of \cite{GHr} that $\tilde Q_i$ is the universal cover of $Q$ and the map $p_{\tilde Q_i}: \tilde Q_i \to Q$ induced by $p_{\tilde M_i}$ is the standard universal covering space projection.  Thus, we can identify both $\tilde Q_i$ with $\tilde Q$, the universal cover of $Q$, and $p_{\tilde Q_i}$ with the standard projection $p_{\tilde Q}: \tilde Q \to Q$; this provides a topological identification of $\tilde M_1$ with $\tilde M_2$ as $\Bbb R$-fibre bundles over $\tilde Q$, $\tilde \pi_i: \tilde M_i \to \tilde Q$.  Our goal is to show a geometric identity between $\tilde M_1$ and $\tilde M_2$, thus providing the same geometric covering space for $M_1$ and $M_2$.  We are assuming the same Killing-length-squared function $\Omega: Q \to \Bbb R$ for $M_1$ and $M_2$, so $\tilde M_1$ and $\tilde M_2$ have the same Killing-length-squared function $\tilde \Omega : \tilde Q \to \Bbb R$.  We just need to examine the 1-forms $\tilde\omega_i$ on $\tilde Q$.
 
From Lemma 1.7(3) we have for each $i$ the map $\tilde \eta_i : \tilde Q \to \Bbb R$  with $d\tilde \eta_i + \tilde\omega_i$ $G$-invariant (as above, there is only one group involved, since $M_1$ and $M_2$ have the same manifold as observer space), where $\tilde \eta_i$ is used to define cross-section  $\tilde z_i : \tilde Q \to \tilde M_i$, yielding the 1-form $\tilde\omega_i$; these maps define $G$-actions via $a \cdot(t,\tilde q) = (t + \tilde \eta_i(\tilde q) - \tilde \eta_i(\tilde q_0), a \cdot \tilde q)$. By Lemma 1.7(e), we have $p_{\tilde Q}^*\omega_i = d\tilde \eta_i + \tilde\omega_i$. Let $\tilde\delta = \tilde \eta_1 - \tilde \eta_2$ and $\tilde\theta = \tilde \omega_1 - \tilde\omega_2$. Then $\tilde\theta = p_{\tilde Q}^*\theta - d\tilde\delta$, so $d\tilde\theta = 0$ (as we have established $d\theta = 0$).  Since $\tilde Q$ is simply connected, there is some $\tilde\eta: \tilde Q \to \Bbb R$ such that $\tilde\theta = d\tilde\eta$. We thus have $\tilde\omega_1 = \tilde\omega_2^{\tilde\eta}$, and so $\tilde M_1$ and $\tilde M_2$ are really the same space (the modified cross-section $\tilde z_2^{\tilde\eta}$ yielding $\tilde\omega_2^{\tilde\eta}$ as the drift-form on $\tilde Q$), though with differing $G$-actions as defined by $\tilde \eta_1$ and $\tilde \eta_2^{\tilde\eta} = \tilde \eta_2 + \tilde\eta$ yielding $M_1$ and $M_2$. \qed 
\enddemo

\remark{Remark 1.8} Suppose we start with a given stationary-complete spacetime $M$ obeying the observer-manifold condition and wish to produce another one, $M'$, over the same observer space $Q$ with homologous fundamental cocycle $\beta_{M'}$ differing from $\beta_M$ by a specified action on homotopy classes of curves in $Q$; say, we want $\beta_{M'} = \beta_M + \theta$ for some $\theta \in H^1(Q;\Bbb R)$ (eliding the distinction between $\theta$ and $\{\theta\}$ for a closed 1-form).  Then the preceding work tells us how to manage this:  

Let $G = \pi_1(Q) = \pi_1(M)$, acting as usual on the universal covers $\tilde M$ of $M$ and $\tilde Q$ of $Q$.  Let $\tilde \eta : \tilde Q \to \Bbb R$ be the map from Lemma 1.7(3); the $G$-action on $\tilde M$, making use of $\tilde \tau: \tilde M \to \Bbb R$ and $\tilde \pi: \tilde M \to \tilde Q$, is $a \cdot (t,\tilde q) = (t + \tilde \eta(a \cdot \tilde q) - \tilde \eta(\tilde q), a \cdot \tilde q)$ (picking base-point $\tilde q_0$ in $\tilde Q$ so that $p_{\tilde Q}\tilde q_0 = q_0$, base-point in $Q$).  We can think of $\theta$ as a closed 1-form on $Q$; then $\tilde\theta = p_{\tilde Q}^*\theta$ is a closed 1-form on $\tilde Q$, hence, exact, so there is some $\tilde\delta: \tilde Q \to \Bbb R$ with $d\tilde\delta = \tilde\theta$.  Define $\tilde \eta' = \tilde \eta + \tilde\delta$, and use this to define a new $G$-action on $\tilde M$: $a \cdot (t,\tilde q) = (t + \tilde \eta'(a \cdot \tilde q) - \tilde \eta'(\tilde q), a \cdot \tilde q)$, that is to say, the $G$-action on $\tilde Q$ followed by  $\Bbb R$-action from $\tilde\delta(a\cdot\tilde q) - \tilde\delta(\tilde q)$; then $M'$ is the quotient of $\tilde M$ by this action, yielding $\beta_{M'} = \beta_M + \theta$.
\endremark

\head 2. Causal Properties and Group Actions
\endhead

In \cite{GHr} the causal structure of a stationary spacetime was explored in terms of the weight of the fundamental cocycle:  For any Riemannian manifold $Q$, for any cycle $\zeta \in Z(Q)$, there is an essentially unique base-pointed loop $c$ in $Q$ with $\<c\> = \zeta$; thus, it makes sense to speak of the length of a cycle, $L(\zeta)$.  Then define the {\it weight}\/ of any cocycle $\beta \in Z^*(Q)$ as
$$\text{wt}(\beta) = \underset{\zeta \neq 0}\to{\sup_{\zeta \in Z(Q)}}\frac{|\beta(\zeta)|}{L(\zeta)}.$$
Note that if $\omega$ is a 1-form with bounded norm in $Q$, then $\text{wt}(\{\omega\}) \le \sup ||\omega||$.  

For $\theta$ a closed 1-form on $Q$, we can express weight in terms of homotopy classes of base-pointed loops:  For $a \in \pi_1(Q)$, let $L(a) = \inf_{[c] = a} L(c)$.  Then
$$\text{wt}(\{\theta\}) = \underset{a \neq e}\to{\sup_{a \in \pi_1(Q)}} \frac{|\{\theta\}(a)|}{L(a)}.$$ 
Another way of thinking of this is to represent $G = \pi_1(Q)$ as a group acting isometrically on $\tilde Q$, the universal cover; in this view, we represent $\theta \in H^1(Q;\Bbb R)$ as the group map $\rho_\theta: G \to \Bbb R$.  Then
$$\text{wt}(\{\theta\}) = \underset{a \neq e}\to{\sup_{a \in G}}\; \frac{|\rho_\theta(a)|}{d(\tilde q_0,a \cdot \tilde q_0)}\,,$$
where $\tilde q_0$ is the base-point in $\tilde Q$.

For $M$ a stationary-complete spacetime satisfying the observer-manifold condition, with observer space $Q$, observer-space metric $h$, and Killing-length-squared function $\Omega : Q \to \Bbb R$, let $\bar h = h/\Omega$, the conformal metric; and for any curve $c$ in $Q$, define $L(c)$, the conformal length of $c$, as its $\bar h$-length.  It is the weight of the fundamental cocycle $\beta_M$, as calculated in the conformal metric, that is of use for causal properties of $M$.  (In this paper, what will be of most importance is whether $\wt(\beta_M)$ is less than, equal to, or greater than 1; but the exact value of $\wt(\beta_M)$ plays an important role in the curvature-based estimates for the behavior of the causality in stationary spacetimes in \cite{GHr}.)

Note that in the standard stationary formulation, as $\beta_M = \{\omega\}$ for $||\omega|| < 1$ in the conformal norm, we always have $\text{wt}(\beta_M) \le 1$.  The statement that $\omega$ has conformal norm less than 1 is not gauge-invariant (since $\omega$ can always be replaced by $\omega^\eta = \omega + d\eta$ for an arbitrary $\eta: Q \to \Bbb R$); but the important consequence that the fundamental cocycle has weight no more than 1 is explicitly gauge-invariant.

It is well to note that weight of the fundamental cocycle---or of any cocycle defined by a 1-form, say, $\theta$---is independent of the choice of base-point in $Q$:  Let $\sigma$ be a curve from $q$ to $q'$, providing the group isomorphism from section 1, $\phi_\sigma^*: Z_{q'}^*(Q) \to Z_q^*(Q)$ (recall $\phi_\sigma: Z_q(Q) \to Z_{q'}(Q)$ is defined as $\phi_\sigma\<c\> = \<\sigma\cdot c\cdot(-\sigma)\>$); let $\{\theta\}_q$ denote the $q$-based cocycle defined by $\theta$ and similarly for $\{\theta\}_{q'}$.  We have $\phi_\sigma^*\{\theta\}_{q'} = \{\theta\}_q$, since $\int_{\sigma\cdot c\cdot(-\sigma)} \theta = \int_c\theta$.  Then, on the one hand,
$$\align
\wt(\{\theta\}_{q'}) & = \sup_{\zeta' \in Z_{q'}(Q)}\(\frac{\{\theta\}_{q'}\zeta'}{L(\zeta')}\)  \\
& = \sup_{\zeta \in Z_q(Q)}\(\frac{\{\theta\}_{q'}(\phi_\sigma\zeta)}{L(\phi_\sigma\zeta)} \)  \\
& = \sup_{\zeta \in Z_q(Q)}\(\frac{\{\theta\}_q\zeta}{L(\phi_\sigma\zeta)} \), 
\endalign$$
while on the other hand,
$$\wt(\{\theta\}_q) = \sup_{\zeta \in Z_q(Q)}\(\frac{\{\theta\}_q\zeta}{L(\zeta)}\).$$
If we knew $\phi_\sigma(\zeta)$ was always at least as long as $\zeta$, we'd have $\wt(\{\theta\}_{q'}) \le \wt(\{\theta\}_{q})$, and by symmetry of $q$ and $q'$, we'd be done.  But since length here is length of the simple loop corresponding to the cycle, that's not necessarily the case:  For instance, if $\zeta = \phi_{-\sigma}\<c'\>$ for $c'$ a simple loop at $q'$, then $\phi_\sigma\zeta$ has $c'$ as its simple loop, and that is shorter than the simple loop for $\zeta$ (i.e., $-\sigma \cdot c' \cdot \sigma$).  But that is the only way that $\phi_\sigma\zeta$ can have a shorter simple loop than $\zeta$, i.e., that $\zeta = \phi_{-\sigma}\zeta'$ with $\zeta'$ having a simple loop $c'$ shorter than $-\sigma \cdot c' \cdot \sigma$.  But then we just consider $n\zeta = \phi_{-\sigma}(n\<c'\>) = \<-\sigma \cdot (c')^n \cdot \sigma\>$:  We have $\{\theta\}_q(n\zeta)/L(n\zeta) = n\{\theta\}_q\zeta/(nL(c') + 2L(\sigma)) = \{\theta\}_q\zeta/(L(c') + (2/n)L(\sigma))$; as the sup for $\wt(\{\theta\}_q)$ includes all these ratios, we see $\wt(\{\theta\}_q) \ge \{\theta\}_q\zeta/L(c') = \{\theta\}_q\zeta/L(\phi_\sigma\zeta)$.  Therefore, we have $\wt(\{\theta\}_{q'}) \le \wt(\{\theta\}_q)$ after all, and we are done.

 (The same result holds for all cocycles in $\tilde Z^*(Q)$.)  

As observed in \cite{GHr}, the causality condition on $M$ is precisely that for any loop $c$ in $Q$, $|\beta_M\<c\>| <  L(c)$ (since the $\tau$-separation of the endpoints of $\bar c$, a null lift of $c$ to a curve in $M$, is $L(c) \pm \beta_M\<c\>$, the sign depending on whether $\bar c$ is future- or past-directed); similarly, the chronology condition is precisely that for all such $c$, $|\beta_M\<c\>| \le L(c)$ .   Therefore we have

\proclaim{Proposition 2.1} Let $M$ be a stationary-complete spacetime obeying the observer-manifold condition.  $M$ is chronological iff $\roman{wt}(\beta_M) \le 1$. If $\roman{wt}(\beta_M) < 1$ then $M$ is causal. \qed
\endproclaim 

It's worth noting that it's possible for $M$ to be causal with wt($\beta_M$) = 1, if $Q$ is sufficiently incomplete (in the conformal metric) that there are no cycles of minimal length for a given evaluation of the fundamental cocycle.  An example is provided in section 3 (example 3.3). 

We can gain an understanding of weight of the fundamental cocycle by noting that this weight approaches 0 if we confine the cycles to smaller and smaller neighborhoods of the base-point:

\proclaim{Proposition 2.2} Let $Q$ be a Riemannian manifold, with $\theta$ a 1-form on $Q$. Let $p$ be any point in $Q$. For all $n$, for a sufficiently small neighborhood $U_n$ of $p$, $\rwt(\{\theta\}_n) \le 1/n$, where $\{\theta\}_n = \{\theta|_{U_n}\}$ is the cocycle formed from the restriction of $\theta$ to $U_n$. 
\endproclaim
\demo{Proof}

Choose a co\"ordinate patch $U$ for $Q$ around $p$, mapping $p$ to the origin; do this so that the pulled back Riemannian metric at $T_pQ$ is the Euclidean metric at $T_0\R^m$ ($m = \text{dim}(Q)$).  For a curve $c$ in $U$, let $\bar c$ be its image in $\R^m$ under the co\"ordinate chart, and similarly for pairing all items between $U$ and $\R^m$.  For any $r > 0$, let $U_r$ be the image in $Q$ of the ball of radius $r$ about the origin in $\R^m$.  For each $q \in U$, let $\bar\delta_q$ be the element of $T_0^*\R^m$ with components from $\bar\theta_{\bar q} - \bar\theta_{\bar p}$ (note $\bar p = 0$), that is to say, $\bar\delta_q = \Sigma_i((\theta_q)_i - (\theta_p)_i)(d\bar x^i)_0$, where $\theta_q = \Sigma_i(\theta_q)_i(dx^i)_q$ (and also $\bar\theta_{\bar q} = \Sigma_i(\theta_q)_i(d\bar x^i)_{\bar q}$, since $\theta(\frac{\partial}{\partial x^i})_q = \bar\theta(\frac{\partial}{\partial \bar x^i})_{\bar q}$).  For each $r$ sufficiently small that this is defined, let $A_r = \sup_{U_r}||\bar\delta_q||$, with norm measured by the Euclidean metric in $\R^m$.  As $\theta$ is continuous, $\lim_{r \to 0} A_r = 0$.

For any loop $c: [0,T] \to U$ at $p$, we have $\{\theta\}\<c\> = \int_c\theta = \int_{\bar c}\bar\theta = \int_0^T \bar\theta_{\bar c(t)}\dot{\bar c}(t) \,dt = \int_0^T\bar\delta_{c(t)}\dot{\bar c}(t)\,dt + \int_0^T\bar\theta_0\dot{\bar c}(t)\,dt = \int_0^T\bar\delta_{c(t)}\dot{\bar c}(t)\,dt + \bar\theta_0\(\int_0^T \dot{\bar c}(t)\,dt\)$.  Note that that last integral is 0, since $\bar c$ is a loop.  Therefore, if $c$ is contained within $U_r$, we have $|\{\theta\}\<c\>| \le A_r\int_0^T||\dot{\bar c}(t)||\,dt$, using Euclidean norm at $T_{\bar c(t)}\R^m$.  For $r$ sufficiently small, the Riemannian metric at $T_qQ$ is no more than twice the Euclidean metric at $T_{\bar q}\R^m$, for all $q \in U_r$; this gives us $|\{\theta\}\<c\>| \le 2A_rL(c)$. Therefore $\wt(\{\theta|_{U_r}\}) \le 2A_r$, and the result follows. \qed
\enddemo 

We can better understand the nature of the weight of a cocycle by looking at a related concept: the efficiency with which a curve interacts with a 1-form.  

\definition{Definition 2.3} For any Riemannian manifold $M$ and a 1-form $\theta$ on $M$, if $c: [a,b] \to M$ is any curve, then define the {\it efficiency\/} of $\sigma$ with respect to $\theta$, $\text{eff}_\theta(\sigma)$, by
$$\text{eff}_\theta(\sigma) = \frac{\int_\sigma \theta}{L(\sigma)}$$
where $L$ denotes length. 
\enddefinition

There is a continuity property of efficiency, with respect to the compact-open topology on curves (equivalently, the $C^0$ topology).  Note that efficiency is not lower-semi-continuous: Given a curve $\sigma^0$ which largely follows an integral curve of the vector field $\theta^\#$ corresponding to $\theta$ (so $\sigma^0$ has fairly high efficiency with respect to $\theta$), and a typical neighborhood $\Cal U$ of $\sigma^0$---say, all curves with image lying within an $\epsilon$-neighborhood of $\sigma^0$---it's easy to find a curve $\sigma \in \Cal U$ with much lower efficiency:  Let $\sigma$ twine tightly around $\sigma^0$, in largely perpendicular directions, so it is largely perpendicular to $\theta^\#$ and captures very little $\langle \dot\sigma,\theta^\#\rangle$ with respect to its length, i.e., it has very low effienciency.  

But efficiency is upper semi-continuous:

\proclaim{Lemma 2.4} Given a 1-form $\theta$ on a Riemannian manifold $M$ and points $p$ and $q$ in $M$, efficiency with respect to $\theta$ is upper semi-continuous on the space $C(p,q)$ of curves from $p$ to $q$ with the compact-open topology.
\endproclaim

\demo{Proof}
Our aim is to show the following:  Given any curve $\sigma^0: [a,b] \to M$ running from $p$ to $q$, for every $\epsilon > 0$, there is some $r > 0$ such that for any curve $\sigma \in C(p,q)$ which lies within the ``cylinder" $B(r) = \{x \st \text{for some }t \in [a,b], d(x,\sigma(t)) < r\}$, $\text{eff}(\sigma) < \text{eff}(\sigma^0) + \epsilon$.

First thing to do is to divide the interval $[a,b]$ into sub-intervals, $a = t_0 <  \dots < t_N = b$, with $I_k = [t_k, t_{k + 1}]$, such that there is a co\"ordinate chart $\phi_k: U_k \to \Bbb R^m$ ($m = \text{dim}(M))$ with $\sigma^0(I_k)$ lying in $U_k$, with  $\phi_k$ taking $\sigma^0(I_k)$ diffeomorphically onto a compact line segment $L_k$ in $\Bbb R^m$.  We will employ a variant of the strategy used in the proof of Proposition 2.2, working first in a single sub-interval, $\sigma^0_k: I_k \to U_k$.  Let $\bar\theta_k$ be the 1-form in $\phi_k(U_k)$ corresponding to $\theta|_{U_k}$.  Using the metric on $\Bbb R^m$ derived from $M$ via $\phi_k$, parametrize $L_k$ with constant speed on $[0,1]$, and for each $t \in [0,1]$, let $D_k^t(r)$ be the $\{m - 1\}$-disk of radius $r$, perpendicular to $L_k$ and centered at $L_k(t)$; we will only consider $r$ sufficiently small that for all $t$, $D_k^t(r)$ lies within $\phi_k(U_k)$.  Let $\bar B_k(r) = \bigcup_{0 \le t \le 1} D_k^t(r)$. Let $D(r)$ be the disk of radius $r$ around the origin in $\Bbb R^{m - 1}$; then by projection parallel to $L_k$ (and by identification of $D_k^0(r)$ with $D(r)$) we have an obvious diffeomorphism $\psi_k = (\pi_k, \tau_k): \bar B_k(r) \to D(r) \times [0,1]$.  Then for each $(x,t) \in D(r) \times [0,1]$, let $\bar\delta_k^t(x)$ be the element of $T^*_{\psi_k^{-1}(x,t)}\Bbb R^m$ giving the co\"ordinate-component  difference between $\bar\theta_k(\psi_k^{-1}(x,t))$ and $\bar\theta_k(L_k(t))$, i.e., $\bar\delta^t_k(x) = \Sigma_i([\bar\theta_k(\psi_k^{-1}(x,t))]_i - [\bar\theta_k(L_k(t))]_i)d\bar x^i$.  In particular, $\bar\delta^t_k(\bold 0) = 0$, and $||\bar\delta^t_k(x)||$ goes to 0, uniformly in $t$, as $||x||$ goes to 0; let us say $||\bar\delta^t_k(x)|| < \epsilon$ for $||x|| < \rho_k(\epsilon)$.  For any $z \in \Bar B_k(r)$, with $x = \pi_k(z)$, $t = \tau_k(z)$, $z_0 = \psi_k^{-1}(\bold 0,t)$, and $F_u^v: T_u^* \Bbb R^m \to T_v^*\Bbb R^m$ the obvious translation via constant co\"ordinate components, we have
$$\bar\theta_k(z) = F_{z_0}^z(\bar\theta_k(z_0)) + \bar\delta^t_k(x).$$
The idea here is to relate $\bar\theta_k$ at any point of $\bar B_k(r)$ to $\bar\theta_k$ at the central line $L_k$ at the same $t$-value, with $\bar\delta^t_k$ giving the difference.

Now consider a curve $\sigma \in C(p,q)$ that lies within $B(r)$.  We need to break $\sigma$ up into segments $\{\sigma_i\}$, each of which remains within a single chart $U_{k_i}$.  We can do this in such a way that each segment has exactly one of four forms:  entering $U_{k_i}$ at $\phi_{k_i}^{-1}(D_{k_i}^{t_{k_i}}(r))$ and exiting at $\phi_{k_i + 1}^{-1}(D_{k_i + 1}^{t_{k_i + 1}}(r))$; doing the reverse; entering at the $k_i$ disk and exiting there (going backwards); and entering at the $k_i + 1$ disk (backwards-pointing) and exiting there.  Call these the forward, backward, front-only, and back-only configurations, respectively (nomenclature here related to thinking of the $k_i$-disk and $(k_i + 1)$-disks as respectively the ``front door" and ``back door" of $U_{k_i}$).

Consider first a forward configuration for $\sigma_i$.  Using barred notation for elements pushed forward by $\phi_{k_i}$ to $\Bbb R^m$, we have 
$$\int_{\sigma_i} \theta = \int_{\bar\sigma_i} \bar\theta_{k_i} = \int_{L_{k_i}} \bar\theta_{k_i} + \int_{\bar\sigma_i} \bar \delta_{k_i}^{\tau_{k_i}}(\pi_{k_i}),$$ 
where by that last integral is meant $\int (\bar\delta^{s(t)}_{k_i}(x(t))) (\dot{\bar\sigma}_i(t)) \,dt$, for $s(t) = \tau_{k_i}(\bar\sigma_i(t))$, $x(t) = \pi_{k_i}(\bar\sigma_i(t))$.  Note that this last integral is bounded by $\epsilon L(\bar\sigma_i)$ so long as $r < \rho_{k_i}(\epsilon)$ ($L$ denoting length in the metric from $M$).  On the other hand, the integral over $L_{k_i}$ is precisely $\int_{\sigma^0} \theta$, restricted to $I_{k_i}$. Thus we can express everything in $M$:
$$\int_{\sigma_i} \theta = \int_{\sigma^0 \text{ on }I_{k_i}} \theta \; + A_i$$
where $|A_i| < (\epsilon/2) L(\sigma_i)$ so long as $r < \rho_{k_i}(\epsilon/2)$.

For a backward configuration for $\sigma_i$, we get the same result, except for a backward parametrization on $L_{k_i}$:

$$\int_{\sigma_i} \theta = -\int_{\sigma^0 \text{ on }I_{k_i}} \theta \; + A_i$$
with the same condition on $A_i$.

For a front-only configuration, the integration of $\bar\theta_{k_i}$ on $L_{k_i}$ goes forward and backward along the line segment in equal amounts, yielding 0; and the same for a back-only configuration.  Either of these configurations thus yields
$$\int_{\sigma_i} \theta = A_i$$
with the same condition on $A_i$.

Adding up all the different segments $\{\sigma_i\}$, we find that, since $\sigma$ goes from $p$ to $q$ overall, the forward and backward parametrizations on the various $\{I_{k_i}\}$ must add up to a simple forward-parametrization from $p$ to $q$:
$$\int_\sigma \theta = \int_{\sigma^0} \theta + \Sigma_i A_i$$
where $|\Sigma_i A_i| < (\epsilon/2) \Sigma_i L(\sigma_i) = (\epsilon/2) L(\sigma)$ so long as $r < \min\{\rho_{k_i}(\epsilon/2)\}$.

We thus have
$$\text{eff}_\theta(\sigma) < \frac{L(\sigma^0)}{L(\sigma)}\text{eff}_\theta(\sigma^0) + \epsilon/2$$
when $r < \min\{\rho_{k_i}(\epsilon/2)\}$.  

Finally, for $r$ sufficiently small, $\sigma$ cannot be much shorter than $\sigma^0$:  For any $\epsilon > 0$, $r$ sufficiently small that $L(\sigma) > L(\sigma^0)/(1 + \epsilon/(2\;\text{eff}_\theta(\sigma^0))$ and $r < \min\{\rho_{k_i}(\epsilon/2)\}$,
$$\text{eff}_\theta(\sigma) < \text{eff}_\theta(\sigma^0) + \epsilon$$
\qed\enddemo

Whether a spacetime has a presentation as standard static is determined by the fundamental cocycle:

\proclaim{Proposition 2.5} Let $M$ be a stationary-complete spacetime obeying the observer-manifold condition with Killing projection $\pi: M \to Q$.  Then $M$ has a presentation as a standard static spacetime  iff $\beta_M = 0$. 
\endproclaim 

\demo{Proof}  To say that $M$ has a standard static presentation is to say we can regard $M$ as $\Bbb R^1 \times Q$ with metric $g = -(\Omega \circ \pi)(d\tau)^2 + \pi^*h$, where $h$ is a Riemannian metric on $Q$ and $\Omega$ a positive scalar function on $Q$; more properly, there is a Killing time-function $\tau : M \to \Bbb R$ and $(\tau,\pi): M \cong \Bbb R \times Q$ carries the metric $g$ on $M$ to $-(\Omega \circ \pi)(dt)^2 + \pi^*h$ on $\Bbb R \times Q$.

If $M$ has a standard static presentation, then we see that $\alpha = d\tau$ (where $\alpha = -\<-,K\>/|K|^2$), so we have the corresponding drift-form $\omega = 0$ (as $\alpha = d\tau + \omega$).  Thus, $\beta_M = \{\omega\} = 0$.  

On the other hand, suppose $\beta_M = 0$.  Then, for any Killing time-function $\tau$ on $M$, we have $\beta_M = \{\omega\}$ (with $\omega = \omega_M$), so $\omega$ gives rise to the zero-cocyle: for any loop $c$ in $Q$, $\int_c \omega = 0$.  In other words, integration of $\omega$ along curves in $Q$ is path-independent; accordingly, there is a function $\eta: Q \to \Bbb R$ with $\omega = d\eta$ (i.e., $\omega$ is exact).  We then have $\omega^{-\eta} = 0$.  It then follows that $\alpha = d\tau^{-\eta}$, and $(\tau^{-\eta},\pi): M \to \Bbb R \times Q$ is a standard static presentation for $M$. \qed
\enddemo

We need to examine in detail the chronology relation in these spaces.  First we'll look at static spacetimes, then take a different path altogether to look at stationary spacetimes, then see how the latter specializes to the result we'll find first for the former.

First note that when $M$ is static and simply connected, we can always find an expression for the metric conformal to a product metric:  For any Killing time-function $\tau : M \to \Bbb R$ with associated drift-form $\omega$ on $Q$, we get a splitting $(\tau,\pi): M \cong \Bbb R \times Q$; but this may not be the optimal splitting. Since $\omega$ is closed and $Q$ is simply connected, there is some $\eta: Q \to \Bbb R$ with $\omega = d\eta$.  Then $\omega^{-\eta} = 0$, so $\tau^{-\eta} = \tau + \eta\circ\pi$ yields a splitting with $g = -(\Omega\circ\pi)(d\tau^{-\eta})^2 + \pi^*h$; in other words, $(M, (1/\Omega\circ\pi)g)$ is isometric, via $(\tau^{-\eta},\pi)$, to $\Bbb L^1 \times (Q,(1/\Omega)h)$.  So for a static-complete, simply connected spacetime, using the appropriate splitting (call it the {\it product time-function}\/), we get, with $q_i = \pi(x_i)$, 
$$x_1 \ll x_2 \iff \tau(x_2) - \tau(x_1) > d(q_1, q_2),\tag2.1$$
using the conformal metric for the distance function on $Q$.

This formula for simply connected static-complete leads directly to a slightly more complex result for general static-complete.

\proclaim{Theorem 2.6}
Let $\pi : M \to Q$ be a static-complete spacetime, with static observer-space $Q$, satisfying the observer-manifold condition; let $G = \pi_1(M)$ be the fundamental group.  Interpret the fundamental cohomology class of $M$ as a real representation $\rho : G \to \Bbb R$.  Let $\tilde\pi: \tilde M\to \tilde Q$ be the universal cover, with covering maps $p_{\tilde M}: \tilde M \to M$ and $p_{\tilde Q} : \tilde Q \to Q$.  Let $\tilde\tau: \tilde M \to \Bbb R$ be the product time-function on $\tilde M$.

For any $x_1$ and $x_2$ in $M$, let $\tilde x_i$ be any pre-image of $x_i$ under $p_{\tilde M}$, each $i$, and let $\tilde q_i = \tilde\pi(\tilde x_i)$.  Then 
$$x_1 \ll x_2 \iff \tilde\tau(\tilde x_2) - \tilde\tau(\tilde x_1) > {\inf_{a \in G}}\{\tilde d(\tilde q_1, a \cdot \tilde q_2) - \rho(a)\},$$ 
where $\tilde d$ is the conformal distance in $\tilde Q$.
\endproclaim

\demo{Proof}  From Proposition 1.1 in \cite{Hr2}, $x_1 \ll x_2$ if and only some pre-image of $x_2$ (under $p_M$) lies in the future of any one specific pre-image of $x_1$, that is to say, if and only if for some $a \in G$, $\tilde x_1 \ll a\cdot \tilde x_2$.  The points $\tilde x_i$ are representable as $(\tilde t_i, \tilde q_i)$ where $\tilde t_i = \tilde\tau(\tilde x_i)$.  Then $a \cdot \tilde x_2 = (\tilde t_2 + \rho(a), a \cdot \tilde q_2)$.  So from (2.1) we have $\tilde x_1 \ll a\cdot \tilde x_2 \iff \tilde t_2 - \tilde t_1 > \tilde d(\tilde q_1, a \cdot \tilde q_2) - \rho(a)$.  Therefore, $x_1 \ll x_2 \iff \exists\; a \in G \text{ such that } \tilde t_2 - \tilde t_1 > \tilde d(\tilde q_1, a \cdot \tilde q_2) - \rho(a) \iff \tilde t_2 - \tilde t_ 1 > \inf_a \{\tilde d(\tilde q_1, a \cdot \tilde q_2) - \rho(a)\}$. \qed
\enddemo

(Applying this for $x_1 = x_2$, we can read off the requirement that $M$ be chronological: For $x \not\ll x$, we need ${\inf_{a \in G}}\{\tilde d(\tilde q, a \cdot \tilde q) - \rho(a)\} \ge 0$ ($\tilde q$ being any pre-image of $q = \pi(x)$ in $\tilde Q$).  Now, this inf is always $\le 0$, since we get 0 from choosing $a = e$.  Thus, the requirement for chronology in $M$ is that for all $\tilde q \in \tilde Q$, for all $a \neq e$, $\tilde d(\tilde q, a \cdot \tilde q) - \rho(a) \ge 0$, which is equivalent to $\text{wt}(\beta_M) \le 1$, as stated in Proposition 2.1.)

For general stationary-complete, there is no simplicity in looking at the universal cover, though we will still investigate the effect of covering maps.  We will examine an invariant for stationary-complete spacetimes that precisely gives the chronology relation.

Let $\pi: M \to Q$ be a stationary-complete spacetime satisfying the observer-manifold condition, with $\tau: M \to \Bbb R$ a Killing time-function and associated drift-form $\omega$ on $Q$.  For any curve $c:[s_1,s_2] \to Q$, let  
$$L_\omega(c) = \int_{s_1}^{s_2} (|\dot c| - \omega\dot c)\, ds = L(c) - \int_c\omega,$$
where we use the conformal metric in the integral, with $L$ the conformal length.  Note that $L_\omega$ is not necessarily positive, though it will be in the case of a standard stationary presentation.  Also $L_\omega$ is independent of parametrization, so long as the same orientation is maintained; in other words, $L_\omega$ is naturally defined on oriented but unparametrized curves. Note also that $L_\omega$ is additive on curves under the operation of concatenation.  

For any points $q_1$ and $q_2$ in $Q$, define
$$d_\omega(q_1,q_2) = \inf\{L_\omega(c) \st c \text{ goes from } q_1 \text{ to } q_2\}.$$ 
As with $L_\omega$, we have that $d_\omega$ is not necessarily non-negative---nor even finite---though it is non-negative for standard stationary presentations.  Nor is it symmetric.  But it does obey the triangle inequality, as $L_\omega$ is additive.  For a standard stationary presentation, this is a Finsler metric, as detailed in \cite{CJS} and \cite{FHS}.  Note that the triangle property implies $d_\omega$ is locally Lipschitz in each argument:  Changing $q_2$ to $q'_2$ changes $d_\omega$ by no more than $(1 + A)d(q_2,q'_2)$ where $d$ is the conformal distance function in $Q$ and $A$ is a bound on the conformal norm of $\omega$ in a neighborhood containing $q_2$ and $q'_2$.  This implies that if $d_\omega$ is $-\infty$ on one pair of points, it's $-\infty$ for all pairs.  From the triangle inequality (applied to $d_\omega(q,q) + d_\omega(q,q')$), if $d_\omega$ is finite, then for all $q$, $d_\omega(q,q) \ge  0$.

In the case of a static-complete spacetime $\pi: M \to Q$, we can express $d_\omega$ in terms of the representation $\rho: G \to \Bbb R$ as described in Theorem 2.6 (noting that as $M$ is a line bundle over $Q$, we can interpret $G$ equally as $\pi_1(M)$ or $\pi_1(Q)$): A loop $c$ in $Q$ from $q$ to $q$ corresponds to a curve $\tilde c$ in $\tilde Q$ from $\tilde q$ to $a \cdot \tilde q$ for some $a \in G$ with $\tilde q$ any pre-image of $q$ under $p_Q: \tilde Q \to Q$, where $p_Q \circ \tilde c = c$.  Define $\tilde d^\rho$ on $\tilde Q$ by
$$\tilde d^\rho(\tilde q_1, \tilde q_2) = \inf_{a \in G}\(\tilde d(\tilde q_1, a \cdot \tilde q_2) - \rho(a)\)$$
where $\tilde d$ denotes the (conformal) metric in $\tilde Q$.  Note that as $\tilde M$ is static, $\beta_M$ is constant on homotopy classes of loops, so to figure $L_\omega$, we need consider only distance-minimizing curves within a given homotopy class (or, in case of an incomplete metric, nearly-distance-minimizing).  Thus we have
$$d_\omega(q_1,q_2) = \tilde d^\rho(\tilde q_1,\tilde q_2)$$
for any choice of pre-images $\tilde q_i$ of $q_i$.

Finally, define the {\it interval}\/ between points $x_1$ and $x_2$ in $M$ by
$$I_M(x_1,x_2) = \tau(x_2) - \tau(x_1) - d_\omega(\pi(x_1),\pi(x_2)).$$
This interval is independent of the choice of splitting:

\proclaim{Lemma 2.7}
$I_M(x_1,x_2)$ does not change with change of Killing time-function.
\endproclaim

\demo{Proof}
Let $q_i = \pi(x_i)$. For any $\eta : Q \to \Bbb R$, consider a curve $c$ in $Q$ from $q_1$ to $q_2$.  We have $L_{\omega^\eta}(c) = L(c) - \int_c \omega - \int_c d\eta = L_\omega(c) - (\eta(q_2) - \eta(q_1))$.  Therefore, $d_{\omega^\eta}(q_1,q_2) = d_\omega(q_1,q_2) - (\eta(q_2) - \eta(q_1))$.

Now note that $\tau^\eta(x_1) - \tau^\eta(x_2) = \tau(x_1) - \tau(x_2) - (\eta(q_2) - \eta(q_1))$.  The result follows.  \qed
\enddemo

The interval gives precisely the chronology relation:

\proclaim{Theorem 2.8}
Let $\pi: M \to Q$ be a stationary-complete spacetime satisfying the observer-manifold condition.  Then for any points $x_1$ and $x_2$ in $M$,
$$x_1 \ll x_2 \iff I_M(x_1,x_2) > 0\tag{a}$$
$$I_M(x_1,x_2) \ge 0 \iff I^+(x_2) \subset I^+(x_1) \iff I^-(x_1) \subset I^-(x_2)\tag{b}$$
\endproclaim
\demo{Proof}
(a) We have $x_1 \ll x_2$ if and only if there is a timelike curve from $x_1$ to $x_2$; given a splitting from a Killing time-function $\tau: M \to \Bbb R$, this can be represented as $\bar c(s) = (s, c(s))$ for $c: [t_1,t_2] \to Q$ with $c(t_i) = q_i$, where $t_i = \tau(x_i)$ and $q_i = \pi(x_i)$, so long as $\dot {\bar c}$ is future-timelike, i.e., $1 + \omega \dot c > |\dot c|$, using the conformal metric for the norm.  In other words, $x_1 \ll x_2$ if and only if there is a curve $c$ from $q_1$ to $q_2$ with $L_\omega(c) < t_2 - t_1$, i.e., if $d_\omega(q_1,q_2) < t_2 - t_1$; and that is precisely saying $I(x_1,x_2) > 0$. 

(b) $I^+(x_2) \subset I^+(x_1)$ means that everything that is in the future of $x_2$ is also in the future of $x_1$.  Thus we have from part (a):
$$\align
I^+(x_2) \subset I^+(x_1) & \iff x_1 \ll \frac1n \cdot x_2 \quad\text{for all }n > 0 \\
& \iff \tau\(\frac1n\cdot x_2\) - \tau(x_1) > d_\omega(\pi x_1, \pi x_2) \quad\text{for all }n > 0 \\
& \iff \frac1n + \tau(x_2) - \tau(x_1) > d_\omega(\pi x_1, \pi x_2) \quad\text{for all }n > 0 \\
& \iff I_M(x_1, x_2) > -\frac1n \quad\text{for all }n > 0 \\
& \iff I_M(x_1, x_2) \ge 0
\endalign$$
and the other part follows time-symmetrically.
\qed
\enddemo

Note that if $I_M(x_1,x_2) = \infty$ for one pair of points, that's true for all pairs, and $M$ is chronologically vicious: the future (or past) of any point is the whole spacetime.  $I_M$ obeys the reverse triangle inequality; applying that to $I_M(x,x) + I_M(x,x')$, we see that if $I_M < \infty$, then for all $x$, $I_M(x,x) \le 0$.  We also have this characterization of weight of the fundamental cocycle:

\proclaim{Proposition 2.9}
Let $M$ be a stationary-complete spacetime obeying the observer-manifold property.  Then the following are equivalent:
\roster
\item $\roman{wt}(\beta_M) > 1$ 
\item For some $x \in M$, $I_M(x,x) > 0$.
\item $I_M = \infty$
\item $M$ is chronologically vicious.
\endroster
\endproclaim

\demo{Proof}
Suppose (1) is true; represent $\beta_M$ as $\{\omega\}$ for some splitting. Then there is some loop $c_0$ at the base-point $q_0 = \pi(x_0)$ such that $\{\omega\}\<c_0\> > (1 +\epsilon)L(c_0)$ for some $\epsilon > 0$ ($L$ being conformal length). We have $I_M(x_0,x_0) = -d_\omega(q_0,q_0) = -\inf\{L_\omega(c) \st c \text{ is loop at }x_0\} = -\inf\{L(c) - \{\omega\}\<c\>\} = \sup\{\{\omega\}\<c\> - L(c)\} \ge \{\omega\}\<c_0\> - L(c_0) > \epsilon L(c_0) > 0$, so (2) is true.
 
Suppose (2) is true.  Then for some loop $c$ at $q = \pi(x)$, $\{\omega\}\<c\> - L(c) > \epsilon$ for some $\epsilon > 0$.  Let $c^n$ be the $n$-fold concatenation of $c$; then $I_M(x,x) \ge \{\omega\}\<c^n\> - L(c^n) = \{\omega\}(n\<c\>) - nL(c) = n\{\omega\}\<c\> - nL(c) > n\epsilon$.  As this is true for any $n$, $I_M(x,x) = \infty$; by the arguments above, $I_M = \infty$.

Suppose (3) is true.  By the arguments above, (4) is proved.

Suppose (4) is true.  Then, in particular, $1 \cdot x_0 \ll x_0$, so $I_M(1 \cdot x_0, x_0) > 0$, i.e., $-1 - d_\omega(q_0,q_0) > 0$, so $d_\omega(q_0,q_0) < -1$.  That means there is a loop $c_0$ at $q_0$ with $L(c_0) - \{\omega\}\<c_0\> < -1$; thus, $\text{wt}(\beta_M) = \text{wt}(\{\omega\}) \ge \{\omega\}\<c_0\>/L(c_0) > (1 + L(c_0))/L(c_0) > 1$, proving (1). \qed
\enddemo

We need to consider how to calculate $I_M$ in the case that $M$ is static and is given in the convenient form of a universal cover, a fundamental group, and a real representation of the fundamental group.  Let us first explore how to express group actions in general for stationary-complete spacetimes.

\proclaim{Proposition 2.10} Let $\pi: M \to Q$ and $\pi': M' \to Q'$ be  stationary-complete spacetime satisfying the observer-manifold condition, connected by a group action from a group $G$ of isometries on $M$, i.e., $M' = M/G$ in the sense of Lemma 1.7, with projections $p_M: M \to M'$ and $p_Q: Q \to Q'$.  For any Killing time-function $\tau$ on $M$, if the corresponding drift-form $\omega$ is $G$-invariant, then the $G$-action on $M$ can be realized, under the identification $(\tau,\pi): M \cong \Bbb R \times Q$, as $a \cdot (t,q) = (t + \rho(q), a \cdot q)$ for a group morphism $\rho : G \to \Bbb R$.
\endproclaim

\demo{Proof}  We know from Lemma 1.7 that there is a $G$-action on $Q$ that commutes (via $\pi$) with that on $M$.  The general picture we get is this, for $q \in Q$ and $a \in G$:
$$a \cdot (t,q) = (t + \rho_q(a), a \cdot q)$$
for some function $\rho_q: G \to \Bbb R$; but want to have $\rho$ independent of $q$.  Evidently, we have 
$$\rho_q(a) = \tau(a\cdot x) - \tau(x)$$ 
where $(\tau,\pi)(x) = (t,q)$.  Then for any $b \in G$, we also have
$$\rho_{a \cdot q}(b) = \tau(b \cdot a \cdot x) - \tau(a \cdot x)$$
and putting the last two together yields
$$\align
\rho_{a \cdot q}(b) + \rho_q(a) & = \tau((ba) \cdot x) - \tau(x) \\
& = \rho_q(ba)
\endalign$$
Thus, if $\rho_q$ is independent of $q$, then $\rho$ is a group morphism.

So consider any vector $X = (d/dt)q_t$ in $T_qQ$ ($q_0= q$), and, for fixed $a \in G$, look at how $\rho_{q_t}(a)$ varies along the curve $\{q_t\}$.  Let $\bar X$ be the lift of $X$ to $T_xM$ which is perpendicular to $K$, and $\{x_t\}$ the lift of $\{q_t\}$, a $K$-perpendicular curve.  Let $R_a: M \to M$ be the action of $a$ on $M$.
$$\align
\frac{d}{dt}\rho_{q_t}(a) & = \frac{d}{dt}\(\tau(a \cdot x_t) - \tau(x_t)\) \\
& = (d\tau)_{a \cdot x}(R_a{}_*\bar X) - (d\tau)_x(\bar X) \\
& = (R_a^*d\tau - d\tau)\bar X
\endalign$$
Thus, $\rho_q$ is, indeed, independent of $q$ if $d\tau$ is $G$-invariant.  As $\alpha = d\tau + \pi^*\omega$ and $\alpha$ is $G$-invariant, this is the same as having $\omega$ $G$-invariant. \qed

Note that for static-complete spacetimes, this provides a very pretty picture:  If $\pi: M \to Q$ is static-complete (satisfying observer-manifold condition), then let $\tilde \pi: \tilde M \to \tilde Q$ be the universal cover of $M$, and let $G = \pi_1(M)$, the fundamental group of $M$.  Let $\tilde\tau$ be a Killing time-function on $\tilde M$ with corresponding drift-form $\tilde\omega$.  Since $M$ is static, $d\tilde\omega = 0$; and since $\tilde Q$ is simply connected, $d\tilde\omega$ is exact: for some $\tilde\eta: \tilde Q \to \Bbb R$, $\tilde\omega = d\tilde\eta$.  Then $\tilde\omega^{-\tilde\eta} = 0$, so $\beta_{\tilde M} = \{d\tilde\omega^{-\tilde\eta}\} = 0$; by Proposition 2.5, we know $(\tilde\tau^{-\tilde\eta},\tilde\pi)$ is a standard-static presentation of $\tilde M$.  Finally, since $\tilde\omega^{-\tilde\eta}$ is 0, it is plainly $G$-invariant, so Proposition 2.10 yields for us a group-morphism $\rho: G \to \Bbb R$ so that $M$ can be presented as $(\Bbb R \times \tilde Q)/G$ with group action $a \cdot (t,q) = (a + \rho(a), a \cdot q)$.  In fact, $\rho : \pi_1(M) \to \Bbb R$ identifies an element of $H^1(Q;\Bbb R)$; and that is precisely the same as the drift-form $\omega$, representing an element of de Rham cohomology.

(If we wish, we can employ Lemma 1.7 to obtain a $G$-invariant drift-form for $M$ for any group action $M \to M'$.  But as that yields $\tau = p_M \circ \tau'$, it produces not just $d\tau$, but $\tau$ as $G$-invariant, and that just yields viewing $M \to M'$ as $\Bbb R \times Q \to \Bbb R \times Q'$ with $G$-action only on the second component.  It doesn't, in general, produce a nice presentation for $M$.)

\enddemo

Now to consider $I_M$ for $M$ static: A direct calculation of $I_M$ in terms of a cross-section $z$, using the identification of $H^1(Q;\R)$ with first de Rham cohomology is not so obvious.  But Theorems 2.6 and 2.8 provide an easy short-cut:

\proclaim{Proposition 2.11}
Let $\pi : M \to Q$ be a static-complete spacetime satisfying the observer-manifold condition.  Suppose $M$ is presented as $M = \tilde M /G$ where $\tilde M$ is the universal cover of $M$, $G$ is the fundamental group of $Q$, and the $G$-action on $\tilde M$ is given in terms of a real representation $\rho : G \to \R$; that is to say, for $\tilde\tau: \tilde M \to \R$ the product time-function for $\tilde \pi : \tilde M \to \tilde Q$ (covering $\pi$ via $p_{\tilde M} : \tilde M \to M$ and $p_{\tilde Q} : \tilde Q \to Q$), with $\tilde M$ identified as $\R \times \tilde Q$ via $(\tilde\tau,\tilde\pi)$, for $a \in G$, the $G$-action on $\tilde M$ is specified by $a \cdot (t, \tilde q) = (t + \rho(a), a \cdot \tilde q)$. 

For any $x_1$ and $x_2$ in $M$, let $\tilde x_i$ be any pre-image of $x_i$ under $p_{\tilde M}$, each $i$, and let $\tilde q_i = \tilde\pi(\tilde x_i)$.  Then 
$$I_M(x_1,x_2) =  \tilde\tau(\tilde x_2) - \tilde\tau(\tilde x_1) - {\inf_{a \in G}}\{\tilde d(\tilde q_1, a \cdot \tilde q_2) - \rho(a)\},$$ 
where $\tilde d$ is the conformal distance in $\tilde Q$.  In other words, using $[\;]$ to indicate identity under the $G$-action:
$$I_M([t_1,\tilde q_1],[t_2,\tilde q_2]) = t_2 - t_1 + \sup_{a\in G}\{\rho(a) - \tilde d(\tilde q_1,a \cdot\tilde q_2)\}.$$
\endproclaim

\demo{Proof} The first formulation for $I_M$ follows directly from Theorems 2.6 and 2.8, and the second formulation is clearly equivalent to the first. \qed
\enddemo

Now we will consider how the interval changes under a covering map.  Recall that a covering map $p : X \to Y$ is characterized by $Y$ having, for each point $y$, a neighborhood $U$ such that $p^{-1}(U)$ is a disjoint union of open sets $\{V_\alpha\}$ (over some indexing set) such that for any $\alpha$, the restriction of $p$ to $V_\alpha$ is a homeomorphism onto $U$; and that there is a set of global homeomorphisms $D$ on $X$, called deck transformations, such for any $\delta \in D$, $\delta$ carries each $V_\alpha$ to some $V_\beta$, and any two such pre-images of $U$ are related by one such deck transformation.  The indexing set for the $\{V_\alpha\}$ is thus the set of deck transformations, which can be identified with the cosets $\pi_1(Y)/p_*(\pi_1(X))$.  (By a covering map of stationary spacetimes, let us understand an isometry that preserves specified Killing fields.  Note this induces a covering map on the orbit spaces, with deck transformations on the spacetime level inducing deck transformations on the orbit-space level.)

\proclaim{Theorem 2.12}
Let $p_M: M \to M'$ be a covering map of stationary-complete spacetimes $\pi: M \to Q$ and $\pi' : M' \to Q'$, each satisfying the observer-manifold condition ($p_Q: Q \to Q'$ the induced covering map on the observer spaces).  Let $D_M$ be the deck transformations for $p_M$ and $D_Q$ the associated deck transformations for $p_Q$.  For any two points $x'_1$ and $x'_2$ in $M'$, pick pre-images $x_i$ of $x'_i$ in $M$.  Then
$$I_{M'}(x'_1,x'_2) = \sup_{\delta \in D_M}I_M(x_1, \delta(x_2)).$$
\endproclaim

\demo{Proof}
Let $\tau' : M' \to \Bbb R$ be a Killing time-function for $M'$; then $\tau = \tau' \circ p_M$ is a Killing time-function for $M$.  With $\omega$ and $\omega'$ the accompanying drift-forms on $Q$ and $Q'$, we have $\omega = p_Q^*\omega'$ (with $\alpha$ and $\alpha'$ the  primary Killing forms in $M$ and $M'$, we have $\pi^*\omega = \alpha - d\tau = p_M^*(\alpha' - d\tau') = p_M^*\pi'{}^*\omega' = \pi^*p_Q^*\omega'$).  Therefore, for any curve $c$ in $Q$, $L_\omega(c) = L_{\omega'}(p_Q\circ c)$ (using Lemma 1.7(a,b) to relate $\Omega'$ and $h'$ to $\Omega$ and $h$ via $p_Q$).  Thus, for any $q_1$ and $q_2$ in $Q$, we have
$$d_\omega(q_1,q_2) = \inf\{L_{\omega'}(p_Q\circ c) \st c \text{ goes from } q_1 \text{ to } q_2\}.\tag2.2$$

Let $q'_i = p_Q(q_i)$; then for any such curve $c$ above, $p_Q\circ c$ is a curve from $q'_1$ to $q'_2$; but not all curves $c'$ from $q'_1$ to $q'_2$ arise in such a manner.  In general, a curve $c'$ from $q'_1$ to $q'_2$ can be lifted to a curve $c$ in $Q$ starting at $q_1$, but ending at $\delta(q_2)$ for some deck transformation $\delta$, depending on the homotopy class of $c'$ (different homotopy classes give rise to the same deck transformation if they lie in the same coset in $\pi_1(Q')/p_Q{}_*(\pi_1(Q))$).  Therefore, (2.2) can be rewritten as
$$\align
d_\omega(q_1,q_2) = \inf & \{L_{\omega'}(c') \st c' \text{ goes from } q'_1 \text{ to } q'_2\ \text{ and} \\
& \text{ is in the correct coset of homotopy classes}\}.\tag2.3
\endalign$$
Now let $q_1'$ and $q_2'$ be any points in $Q'$ with any choice of pre-images $q_i$ for $q'_i$; then (2.3) leads us to
$$\align
d_{\omega'}(q'_1,q'_2) & = \inf\{L_{\omega'}(c') \st c' \text{ goes from } q'_1 \text{ to } q'_2\} \\ 
& = \inf_{\delta \in D_Q}\,\inf\{L_\omega(c) \st c \text{ goes from } q_1 \text{ to } \delta(q_2)\} \\
& = \inf_{\delta \in D_Q} d_\omega(q_1,\delta(q_2)).\tag2.4
\endalign$$

From (2.4) it follows that 
$$\align
I_{M'}(x'_1,x'_2) & = \tau'(x'_2) - \tau'(x'_1) - \inf_{\delta \in D_Q} d_\omega(q_1,\delta(q_2)) \\
& = \tau(x_2) - \tau(x_1) - \inf_{\delta \in D_Q} d_\omega(q_1,\delta(q_2)) \\
& = \sup_{\delta \in D_M} I_M(x_1,\delta(x_2)).
\endalign$$ \qed
\enddemo

We will now see how to employ $I_M$ in characterizing further causal properties.  Recall that a spacetime $M$ is {\it future-distinguishing}\/ if for all $x$ and $y$ in $M$, $x \neq y$ implies $I^+(x) \neq I^+(y)$ (and dually for {\it past-distinguishing}); {\it distinguishing}\/ if it is both future- and past-distinguishing; {\it strongly causal}\/ if every point has arbitrarily small neighborhoods $U$ such that the chronology relation in $U$ as a spacetime in its own right is precisely the chronology relation of $M$, restricted to $U$; {\it stably causal}\/ if all metrics with light cones sufficiently close to that of the metric of $M$, in the sense of being topologically close in the tangent space, are causal metrics; and {\it causally continuous}\/ if $I^+$ and $I^-$ are outer-continuous, i.e., if $I^+(x)$ omits some compact set $K$, then so does $I^+(y)$ for all $y$ in some neighborhood of $x$.

It is a remarkable fact that a chronological stationary-complete spacetime which is distinguishing is also strongly causal, stably causal, and even causally continuous, this being a chain of nominally increasingly stronger conditions (see Proposition 3.1 in \cite{JS}, citing Proposition 3.21 and Theorem 3.25 of \cite{BEEs}, calling upon \cite{HwSs} and \cite{D}).  In fact, just future-distinguishing can be appended to this list:

\proclaim{Proposition 2.13} A stationary-complete spacetime $M$ satisfying the observer-manifold property is future-distinguishing iff it is distinguishing, hence, strongly causal, stably causal, and causally continuous.
\endproclaim
\demo{Proof}
By Theorem 2.8(b), if $I^+(x_1) = I^+(x_2)$, then $I_M(x_1,x_2) = 0$ and also $I_M(x_2, x_1) = 0$, from which $I^-(x_1) = I^-(x_2)$.  Hence, if $M$ is future-distinguishing, it is distinguishing. \qed\enddemo

For measuring how close a stationary-complete manifold comes to being globally hyperbolic, it will useful to define another type of causality property. 

\definition{Definitions 2.14} Let $\pi: M \to Q$ be a stationary-complete spacetime satisfying the observer-manifold condition.  We will say $M$ is {\it causally bounded}\/ if for all $x$ and $x'$ in $M$, $\pi\(I^+(x) \cap I^-(x')\)$ is bounded in the conformal metric on $Q$ (i.e., in $\bar h = (1/\Omega)h$, where $g = -(\Omega \circ \pi)\alpha^2 + \pi^*h$).  $M$ is {\it spatially complete} if $Q$ is complete in the conformal metric.
\enddefinition

Here is the relation to global hyperbolicity:

\proclaim{Proposition 2.15} Let $\pi: M \to Q$ be a stationary-complete spacetime satisfying the observer-manifold condition.  $M$ is globally hyperbolic if and only if $M$ is
\roster
\item future-distinguishing,
\item causally bounded, and
\item spatially complete.
\endroster
\endproclaim

\demo{Proof}
Suppose $M$ is globally hyperbolic.  Then it is strongly causal, hence, future-distinguishing.  For any $x$ and $x'$ in $M$, since $I^+(x) \cap I^-(x')$ is relatively compact, so is $\pi\(I^+(x) \cap I^-(x')\)$; accordingly, it must be bounded in any Riemannian metric.  

To show spatial completeness, consider any curve $c: [0,L) \to Q$ which is unit-speed in the conformal metric; pick a point $x$ in the fiber of $\pi$ above $c(0)$ and let $\bar c: [0, L) \to M$ be the lift of $c$ starting at $x$ with $\dot{\bar c}$ everywhere perpendicular to the Killing field. Let $\Pi = \pi^{-1}(c)$, a timelike 2-surface in $M$; we can parametrize $\Pi$ as $t \cdot \bar c(s)$ for $(t,s) \in \Bbb R \times [0,L)$. Then the induced conformal metric on $\Pi$ is $-dt^2 + ds^2$ (the conformal metric being $\tilde g = - \alpha^2 + \tilde h$ with $\alpha(K) = 1$); in other words, $\Pi$ is just a strip of Minkowski 2-space.  That makes it easy to calculate within $\Pi$, and we clearly have that all of $\bar c$ is contained within $I_\Pi^+(-2L \cdot x) \cap I_\Pi^-(2L \cdot x)$; therefore, the same is true in $M$: all of $\bar c$ is contained within $I^+(-2L \cdot x) \cap I^-(2L \cdot x)$.  Then global hyperbolicity of $M$ tells us that $\bar c$ has a limit point $\bar c(L)$; and then $\pi(\bar c(L))$ is a limit point of $c$ at $L$.

Suppose $M$ has properties (1), (2), and (3).  By Proposition 2.13, since $M$ is future-distinguishing, it is strongly causal.  Pick any $x \ll y$ in $M$, and let $A = I^+(x) \cap I^-(y)$; we need to show that $A$ is relatively compact.  As $M$ is causally bounded and spatially complete, $\pi(A)$ is relatively compact in $Q$.  Pick a Killing time-function $\tau: M \to \Bbb R$; then, as $(\tau,\pi): M \cong \Bbb R \times Q$ is a diffeomorhphism, all we need to show is that $\tau(A)$ is bounded in $\Bbb R$.

First note that there is some $T > 0$ such that $T \cdot x \gg y$:  If we consider any curve $c$ from $\pi(x)$ to $\pi(y)$, then, as shown above, $\pi^{-1}(c)$ is (in the conformal metric) a strip of $\Bbb L^2$; in particular, $\pi^{-1}(\pi(x))$ enters the future of $y$.  Let $x' = T \cdot x$ and $q = \pi(x) = \pi(x')$. We will now see that $\tau$ is bounded on $I^+(x) \cap I^-(x')$.  For consider any timelike curve $\gamma$ from $x$ to $x'$; then $\pi \circ \gamma = \sigma$ is a loop in $Q$ at $q$, and we can take $\sigma$ to be unit-speed in $Q$.  Again, we can consider $\pi^{-1}(\sigma)$ as a strip of $\Bbb L^2$, that is to say, for $L$ the conformal length of $\sigma$, we have a map $\psi: \Bbb R^1 \times [0,L] \to M$, $\psi(t,s) = t \cdot \bar\sigma(s)$, where $\bar\sigma: [0,L] \to M$ is the lift of $\sigma$, starting at $x$, which is everywhere perpendicular to $K$; then $\psi$ is a local isometry from $(\Bbb R^1 \times [0,L], -dt^2 + ds^2)$ onto its image with the conformal metric.  We have $\gamma = \psi \circ \delta$, where $\delta(s) = (\bar t(s),s)$ for some function $\bar t$ with $\bar t(0) = 0$, $\bar t(L) = T$, and $\bar t' > 1$ (since $\gamma$ is timelike); it follows that $L < T$.  

Note that the $dt$ in the Minkowski strip obeys $dt = \psi^*\alpha = \psi^*(d\tau + \pi^*\omega)$.  We have $(dt)\dot\delta = \bar t' = \alpha(\dot\gamma) = d\tau(\dot\gamma) + \omega(\dot\sigma)$.  For any $s_0$, let $\gamma_{s_0}$ denote the restriction of $\gamma$ to $[0,s_0]$; and similarly for $\sigma_{s_0}$ and $\delta_{s_0}$.  Then the change in $\tau$ over $\gamma$, from $x =\gamma(0)$ to $\gamma(s_0)$ is $\Delta_{s_0}(\tau) = \tau(\gamma(s_0)) - \tau(\gamma(0)) = \int_{\gamma_{s_0}} d\tau = \int_{\delta_{s_0}} dt - \int_{\sigma_{s_0}} \omega = \bar t(s_0) - \int_{\sigma_{s_0}} \omega$.  As $\pi(A)$ is relatively compact, there is an upper bound $B$ to $||\omega||$ (in the conformal metric); and we know that $\sigma$ has length no greater than $T$ (in the conformal metric).  It follows that $|\Delta(\tau)| < T + BT$. \qed
\enddemo

\proclaim{Theorem 2.16} Let $M$ be a stationary-complete spacetime acted on by a group $G$ of isometries, with quotient $M' = M/G$.  Then  $M'$ is  globally hyperbolic iff 
\roster
\item $M'$ is causal,
\item $M$ is globally hyperbolic, and 
\item for $x \in M$, for all $y \in M$, the $G$-orbit of $y$ has only finite intersection with $I^+(x)$.
\endroster
\endproclaim

\demo{Proof} From Proposition 1.4 in \cite{Hr2}, we know that if $M'$ is globally hyperbolic, then so is $M$, and in $M$, the $G$-orbit of a point has only finite intersection with the future of any point.  That same proposition yields the converse, so long as we know that $M$ has a fundamental neighborhood system for each point, consisting of neighborhoods, each of whose $G$-orbits is well behaved (no timelike relations between components of the orbits).  But we want to obtain the converse without making such a strong assumption; instead we will use Proposition 2.15.

Let $\pi: M \to Q$ and $\pi': M' \to Q'$ be the projections to observer spaces.  Let $p_M: M \to M'$ be the quotient projection, with induced quotient projection $p_Q: Q \to Q'$.  

Spatial completeness is inherited by $M'$ from $M$ automatically:  If $M$ is spatially complete, then, since $p_Q$ is a covering projection and a local isometry of the conformal metrics, so is $M'$. 

Causal boundedness passes from $M$ to $M'$ as a result of the assumption on $G$-orbits:  For any $x$ and $y$ in $M$, let $x' = p_M(x)$ and $y' = p_M(y)$, and let $A' = \pi'\(I^+_{M'}(x') \cap I^-_{M'}(y')\)$.  For any $a \in G$, let $A_a = \pi\(I^+_M(x) \cap I^-_M(a \cdot y)\)$.  It is straight-forward to see that $A' = p_Q\(\bigcup_{a \in G} A_a\)$ (Proposition 1.1 in \cite{Hr2}: $p_M(x) \ll p_M(z)$ if and only if for some $a \in G$, $x \ll a \cdot z$).  If $M$ is causally bounded then each $A_a$ is relatively compact; and if $G \cdot y$ has only finite intersection with $I^+_M(x)$, then there are only finitely many such $A_a$ which are non-empty, and $A'$ is relatively compact.

For $M'$ to inherit future distinguishing from $M$, we will need both that $M$ is globally hyperbolic and also causality for $M'$. In this presentation, we will crucially employ both the $G$- and $\Bbb R$-actions on $M$; so as to prevent any confusion let us denote $G$-action with $\cdot$ and (for this proof and the next one only) $\Bbb R$-action with $*$.  

We know $M$ is future-distinguishing; we wish to show the same is true of $M'$.  So let $x$ and $y$ be any points in $M$, with $x' = p_M(x)$, $y' = p_M(y)$.  Suppose first that $I^+_{M'}(y') \subset I^+_{M'}(x')$.  That means precisely that $I^+_M(y) \subset \bigcup_{a \in G} a \cdot I^+_M(x)$; and that in turn is equivalent to saying that for all positive integers $n$, $(1/n)*y \gg a_n \cdot x$ for some $a_n \in G$.  This gives us $x \ll a_n^{-1}\cdot (1/n)*y$; and, further, $(-2)*x \ll (-2)*a_n^{-1}\cdot(1/n)*y = a_n^{-1}\cdot(1/n - 2)*y \ll a_n^{-1} \cdot y$, for all $n$.  But since $(-2)*x$ can have only finitely many $a_n^{-1} \cdot y$ in its future (and the $G$-action is effective), infinitely many of those $a_n$ coincide; say, for all $k$, $a_{n_k} = a$ for some $a \in G$.  Then for all $k$, $(1/n_k)*y \gg a \cdot x$, and it follows that $I^+_M(y) \subset I^+_M(a \cdot x)$.  Dually, if $I^+_{M'}(x') \subset I^+_{M'}(y')$, then for some $b \in G$, $I_M^+(x) \subset I_M^+(b \cdot y)$.  Thus, if $I^+_{M'}(x') = I^+_{M'}(y')$, then for some $a$ and $b$ in $G$, $I_M^+(x) \subset I_M^+(b \cdot y) = b \cdot I^+_M(y) \subset b \cdot I^+_M(a \cdot x) = I^+_M((ba)\cdot x)$.  Since $M$ is globally hyperbolic, this gives us a causal curve $\gamma$ from $(ba)\cdot x$ to $x$ (the space of causal curves between $(ba)\cdot x$ and, say, $1*x$ includes causal curves curves from $(ba)\cdot x$ to $(1/n)*x$ to $1*x$, and these have a causal limit curve from $(ba)\cdot x$ to $x$ to $1*x$).  If $(ba)\cdot x$ is different from $x$, then $\gamma$ must be a nondegenerate curve (i.e., not just a point), and $\pi_M \circ \gamma$ is a closed causal curve in $M'$, violating causality of $M'$.  Therefore, $b = a^{-1}$; this gives us $I^+_M(y) \subset I^+_M(a \cdot x)$ and also $I^+_M(x) \subset I^+_M(a^{-1}\cdot y)$, i.e., $I^+_M(a \cdot x) \subset I^+_M(y)$.  Thus we have $I^+_M(a \cdot x) = I^+_M(y)$, and by $M$ being future distinguishing, we have $a \cdot x = y$.  And that implies $x' = y'$. \qed

\enddemo

The three conditions of Theorem 2.16 are mutually independent.  For instance, an example which has (2) and (3) satisfied but for which $M'$ is not causal:  Let $M = \Bbb L^1 \times \Bbb S^1$, where $\Bbb S^1$ is the circle realized as $\Bbb R/\Bbb Z$.  Then let $\Bbb Z_2 = \{0,1\}$ act on $M$ via $1 \cdot (t,[x]) = (-t, [x + 1/2])$; and $M' = M/\Bbb Z_2$.  Then $M$ is globally hyperbolic, $\Bbb Z_2$ acts effectively via isometry, and all $\Bbb Z_2$-orbits are finite; but $M'$ is not causal: In $M$, we have the timelike curve $\gamma(s) = (2s-1, [s/2])$, which projects to a timelike curve $\gamma'$ in $M'$; but $\gamma'(1) = \gamma'(0)$.  An example showing $M'$ can be causal and $M$ globally hyperbolic, but with infinite intersection with a $G$-orbit of a point in $M$ and the future of another point, is provided in Example 3.4b.  (And, of course, if $M$ fails to be globally hyperbolic, so does $M'$, even with $M'$ causal and a very simple---or even non-existent---group action on $M$.)

\proclaim{Corollary 2.17} Let $M$ be a globally hyperbolic stationary-complete spacetime acted on by a group $G$ of isometries, with quotient $M' = M/G$.  Suppose $\rwt(\beta_{M'}) < 1$; then $M'$ is also globally hyperbolic.
\endproclaim

\demo{Proof} By Proposition 2.1, $M'$ is causal.  We only need to show that the $G$-orbit of each point in $M$ has finite intersection with the future of any point.  We will make use of the $G$-invariant Killing time-function $\tau$ and drift-form $\omega$ given to us from Lemma 1.7.  We also again use $*$ to expess the $\Bbb R$-action.

Consider any $x$ and $y$ in $M$, with $p = \pi(x), q = \pi(y)$.  For any $a \in G$, we have 
$$\align
x \ll a \cdot y & \iff I_M(x,a \cdot y) > 0 \\
& \iff \tau(a \cdot y) - \tau(x) > d_\omega(p,a \cdot q) \\
& \iff \tau(y) - \tau(x) > d_\omega(p,a \cdot q)
\endalign$$
(so we could say the $G$-orbit of $y$ intersects the future of $x$ only to the extent that the $G$-orbit of $q$ resides within a fixed $d_\omega$-distance of $p$).  

It is worthwhile to note the relation between $d_\omega$ and $d_{\omega'}$:  For $q_i \in Q$ and $q'_i = p_Q(q_i)$ ($i = 1,2$), we have
$$\align
d_{\omega'}(q'_1,q'_2) & = \inf\{L_{\omega'}(c') \st c' \text{ goes from } q'_1 \text{ to } q'_2\} \\
& = \inf\{L_{\omega'}(c') \st c' = p_Q \circ c,\; c \text{ goes from } q_1 \text{ to } b \cdot q_2 \text{ for some } b \in G\} \\
& = \inf_{b \in G}\(\inf\{L_{\omega'}(c') \st c' = p_Q \circ c,\; c \text{ goes from } q_1 \text{ to } b \cdot q_2\}\) \\
& = \inf_{b \in G}\(\inf\{L_{\omega}(c) \st c \text{ goes from } q_1 \text{ to } b \cdot q_2\}\) \\
& = \inf_{b \in G}\(d_\omega(q_1, b \cdot q_2)\)
\endalign$$
using the fact that $p_Q^*\omega' = \omega$, from Lemma 1.7(e).

Note that with $\rwt(\beta_{M'}) = w < 1$,  for any loop $c'$ in $Q'$, $|\int_{c'}\omega| = |\beta_M\<c'\>| \le wL(c')$, so
$$\align
L_{\omega'}(c') & = L(c') - \int_{c'}\omega \\
& \ge (1 - w)L(c').
\endalign$$ 
Furthermore, for any $q \in Q$ and $a \in G$, for any curve $c$ from $q$ to $a \cdot q$, $c' = p_Q \circ c$ is a loop in $Q'$, so we then have
$$\align
L_\omega(c) & = L_{\omega'}(c') \\
& \ge (1 - w)L(c') \\
& = (1 - w)L(c)
\endalign$$
From this it follows that for any $a \in G$,
$$\align
d_\omega(q, a \cdot q) & = \inf\{L_\omega(c) \st c \text{ goes from } q \text{ to } a \cdot q\} \\
& \ge (1 - w)\inf\{L(c) \st c \text{ goes from } q \text{ to } a \cdot q\} \\
& = (1 - w) d(q, a \cdot q)
\endalign$$ 

Now consider any collection $\{a_n\}$ in $G$ such that for all $n$, $a_n \cdot y \gg x$.  We have for some $S < 0$, $S * y \ll x$, so for all $n$, $S*y \ll a_n \cdot y$.  Thus, we have (recalling $\tau$ is $G$-invariant), for all $n$, $I_M(S*y,a_n \cdot y) > 0$, so
$$\align
d_\omega(q,a_n \cdot q) & < \tau(a_n \cdot y) - \tau(S * y) \\
& = -S
\endalign$$
from which it then follows, for all $n$,
$$\align
d(q, a_n \cdot q) & \le d_\omega(q, a_n \cdot q)/(1 - w) \\
& \le -S/(1 - w).
\endalign$$
But it's not possible for an infinite number of elements of the $G$-orbit of $q$ to be within any fixed distance of $q$, since $G$ acts properly discontinuously.  Therefore, $\{a_n\}$  must be a finite collection. \qed

\enddemo

Recall that for a stationary spacetime $M$ with the observer-manifold property, $M$ is chronological if and only if $\text{wt}(\beta_M) \le 1$.  The following theorem more carefully distinguishes sub-cases within that.  It breaks up all stationary-complete spacetimes into six mutually exclusive conditions involving the weight of the fundamental cocycle and various other properties of the behavior of that cocycle on loops; these cocycle categories are shown to  devolve into four mutually exclusive causality categories, corresponding to cocycle categories (1), (2), (3), and \{(4), (5), (6)\}; (5) and (6) have the same causal category, which subsumes that of (4).  Thus, this theorem completely characterizes the global causal properties of stationary-complete spacetimes in terms of the fundamental cocyle and related phenomena.  

\vskip .1 in

(Note that for a loop $c$, $L_\omega(c)$ is independent of the choice of cross-section, as changing cross-section changes $\omega$ by an exact 1-form.  As usual $L$ denotes conformal length of a curve. Recall $L_\omega(c) = L(c) - \int_c \omega = L(c) - \beta_M(c)$, $\text{eff}_\omega(c) = \beta_M(c)/L(c)$, and $\text{wt}(\beta_M) = \sup_{\text{loops }c}(\text{eff}_\omega(c))$.) 

\proclaim{Theorem 2.18}  Let $\pi: M \to Q$ be a stationary-complete spacetime satisfying the observer-manifold condition.  There are only these mutually exclusive possibilities:
\roster
\item If 
$$
\roman{wt}(\beta_M) > 1,$$
then $M$ is chronologically vicious. 

\item If  
$$\align
 & \roman{wt}(\beta_M)   = 1 \text{ and}  \\
 & \text{there is a loop } c \text{ in } Q \text { with } L(c)  = \beta_M\<c\>, 
\endalign$$
then $M$ is chronological but not causal.

\item If 
$$\align
& \roman{wt}(\beta_M)  = 1; \\ 
& \text{for every loop } c \text{ in } Q,\; L(c) > \beta_M\<c\>; \text{ and} \\ 
& \text{there is a sequence of base-pointed loops } \{c_n\} \text{ in } Q \text{ with} \\  
& \quad\quad\quad \{\roman{eff}_\omega(c_n)\} \to 1 \text{ and } \{L_\omega(c_n)\} \to 0;
\endalign$$ 
then $M$ is causal but not future- or past-distinguishing (in particular, not strongly causal).

\item If 
$$\align
& \roman{wt}(\beta_M) = 1; \\ 
& \text{for every loop } c \text{ in } Q,\; L(c) > \beta_M\<c\>; \\
& \text{for every sequence of base-pointed loops } \{c_n\} \text{ in } Q   \text{ with } \{\roman{eff}_\omega(c_n)\} \to 1, \\
& \quad\quad\quad \{L_\omega(c_n)\} \text{ is bounded away from } 0  \\
& \quad\quad\quad (\text{from which it follows that } \{L(c_n)\} \to \infty); \text{ and } \\
& \text{there is such a sequence } \{c_n\} \text{ with } \{L_\omega(c_n)\}  \text{ bounded above};
\endalign$$  
then $M$ is strongly causal (and causally continuous) but not spatially complete or not causally bounded (hence, not globally hyperbolic).  

\item If 
$$\align
& \roman{wt}(\beta_M) = 1; \\ 
& \text{for every loop } c \text{ in } Q,\; L(c) > \beta_M\<c\>; \text{ and}\\
& \text{for every sequence of base-pointed loops } \{c_n\} \text{ in } Q  \text{ with }\{\roman{eff}_\omega(c_n)\} \to 1, \\
& \quad\quad\quad \{L_\omega(c_n)\} \to \infty; \\
\endalign$$
then $M$ is strongly causal (and causally continuous) and causally bounded (hence, $M$ is globally hyperbolic iff it is spatially complete).  

\item If 
$$\roman{wt}(\beta_M) < 1,$$ 
then $M$ is strongly causal (and causally continuous) and causally bounded (hence, $M$ is globally hyperbolic iff it is spatially complete).
\endroster
\endproclaim

\demo{Proof} The conclusion from case (1) follows from Proposition 2.9.  Proposition 2.1 shows us that cases (2) through (6) at least imply that $M$ is chronological. 

A key idea here is to consider any loop $c$ in $Q$, say from $q$ to $q$.  Let $\bar c$ be the future-directed null lift of $c$ starting at some choice of pre-image $x$ of $q$.  Proposition 1.3 shows that the future endpoint of $\bar c$ is $T \cdot x$ where $T = L(c) - \beta_M\<c\>$. Note that for any splitting, yielding a drift-form $\omega$, $T = L(c) - \int_c \omega = L_\omega(c)$.

Suppose we have case (2): $\text{wt}(\beta_M) = 1$ and $L(c) = \beta_M\<c\>$ for some loop $c$ in $Q$.  Then we have $\bar c$ (as above) is a closed null curve in $M$.  This establishes the full result in case (2).

At this point we pause to use Theorem 2.8(b) to characterize future distinguishing in a chronological stationary-complete spacetime:  $M$ fails to be future distinguishing iff there are points $x \neq y$ with $I^+(x) = I^+(y)$, i.e., $I_M(x,y) \ge 0$ and $I_M(y,x) \ge 0$; given a splitting, this means $\tau(y) - \tau(x) \ge d_\omega(x,y)$ and $\tau(x) - \tau(y) \ge d_\omega(x,y)$, i.e., for every $n$, there is a curve $\sigma_n$ from $\pi(x)$ to $\pi(y)$ and a curve $\sigma'_n$ from $\pi(y)$ to $\pi(x)$ with $L_\omega(\sigma_n) < \tau(y) - \tau(x) + 1/(2n)$ and $L_\omega(\sigma'_n) < \tau(x) - \tau(y) + 1/(2n)$.  

Note that we can assume $\pi(x) \neq \pi(y)$, since if $x$ and $y$ have the same projection to $Q$, then either $x \ll y$ or $y \ll x$, and $I^+(x) = I^+(y)$ would imply a failure of chronology in $M$.  Thus $M$ fails to be future-distinguishing implies there is a point $x$ and for all $n$ there is a loop $c_n$ from $\pi(x)$ to $\pi(x)$ with $L_\omega(c_n) < 1/n$, with those loops not converging on $\pi(x)$ (let $c_n = \sigma'_n \cdot \sigma_n$, the concatenation; then $L_\omega(c_n) = L_\omega(\sigma'_n) + L_\omega(\sigma_n)$, and $c_n$ contains both $\pi(x)$ and $\pi(y)$).  

Furthermore, that condition (i.e., $L_\omega(c_n) < 1/n$ and $\{c_n\}$ doesn't converge to $\pi(x)$) implies failure of future distinguishing: Since the loops $\{c_n\}$ don't converge on $p = \pi(x)$, there is some point $q \neq p$ in $Q$ and a subsequence $c_{n_k}$ each containing a point $q_{n_k}$ with $\{q_{n_k}\}$ approaching $q$ (there is some sphere of positive radius around $p$ with infinitely many of the loops intersecting that sphere); let $\sigma_k$ be the portion of $c_{n_k}$ from $p$ to $q_k$  and $\sigma'_k$ the balance of the loop, so $c_{n_k} = \sigma'_k \cdot \sigma_k$.  Let $y^T \in \pi^{-1}(q)$ be such that $\tau(y^T) = T$. Then 
$$\align
I_M(x,y^T) & = \tau(y^T) - \tau(x) - d_\omega(p,q) \\
& = \tau(y^T) - \tau(x) - \lim_{k \to \infty} d_\omega(p,q_k) \\
& \ge T -\tau(x) - \liminf_{k \to \infty} L_\omega(\sigma_k)
\endalign$$ and, similarly, 
$$I_M(y^T,x) \ge \tau(x) - T - \liminf_{k \to \infty} L_\omega(\sigma'_k).$$  

Note that we cannot have any subsequence with $\{L_\omega(\sigma'_{k_i})\} \to -\infty$, for if there were, consider curves $\rho_i$ from $q$ to $q_{k_i}$, contracting to the point $q$, and any curve $\rho$ from $p$ to $q$: each $\gamma_i = \sigma'_{k_i} \cdot \rho_i \cdot \rho$ is a loop at $p$, and $\{L_\omega(\rho_i)\} \to 0$, so $\{L_\omega(\gamma_i)\} = \{L_\omega(\sigma'_{k_i}) + L_\omega(\rho_i) + L_\omega(\rho)\} \to -\infty$, but $L_\omega(\gamma_i) < 0$ violates chronology in $M$ at $x$.  And likewise there is no subsequence with $\{L_\omega(\sigma_{k_i})\} \to -\infty$.  For all $k$ we have $L_\omega(\sigma_k) + L_\omega(\sigma'_k) = L_\omega(c_{n_k}) < 1/n_k$; therefore we also cannot have any subsequence with $\{L_\omega(\sigma_{k_i})\} \to \infty$ or $\{L_\omega(\sigma'_{k_i})\} \to \infty$.  Thus, both $\{L_\omega(\sigma_{k_i})\}$ and $\{L_\omega(\sigma'_{k_i})\}$ are bounded above and below, and it follows there is a subsequence with $\{L_\omega(\sigma_{k_i})\} \to l$ and $\{L_\omega(\sigma'_{k_i})\} \to l'$ for some finite numbers $l$ and $l'$ obeying $l + l' \le 0$.  Let $T$ be chosen so that $\tau(x) + l \le T \le \tau(x) - l'$ (possible because $l + l' \le 0$); then $I_M(x,y^T) \ge 0$ and $I_M(y^T,x) \ge 0$, and by Proposition 2.8(b), $I^+(x) = I^+(y^T)$.    
Thus, we have shown:

\proclaim{Lemma 2.19}Chronological $M$ is future-distinguishing iff there is no sequence of loops in $Q$, all passing through one point but not converging to that point, for which $L_\omega$ of that sequence goes to 0. \qed
\endproclaim 

Suppose now we have case (3): $\text{wt}(\beta_M) = 1$ and $L(c) > \beta_M\<c\>$ for every loop $c$ in $Q$; and there is a ``weight-realizing" sequence of loops $\{c_n\}$ with $\{L_\omega(c_n)\} \to 0$.  By the discussion above, any future-directed null lift $\bar c$ of any such loop has the endpoints of $\bar c$ separated by an amount $T = L(c) - \beta_M\<c\> > 0$ along the Killing orbit.  This shows $M$ cannot have a closed null curve, hence, no closed causal curve.  We know the loops $\{c_n\}$ cannot collapse to a point, for then Proposition 2.2 would imply $\{\text{eff}_\omega(c_n)\} \to 0$.  It follows from Lemma 2.19 that $M$ is not future-distinguishing.

Now suppose we have case (4):  $\text{wt}(\beta_M) = 1$ and $L(c) > \beta_M\<c\>$ for every loop $c$ in $Q$; all weight-realizing sequences of loops have $L_\omega$ bounded away from 0, and at least one such sequence has $L_\omega$ bounded above.  First we want to apply Lemma 2.19, so we consider any sequence of loops $\{c_n\}$ in $Q$, each containing a point $p$, with $\{L_\omega(c_n)\} \to 0$; if we can show that $\{c_n\}$ must always collapse to $p$, then Lemma 2.19 implies $M$ is future-distinguishing.  Since $\{L_\omega(c_n)\}$ is not bounded away from 0---nor is that true for any subsequence---we know that $\{c_n\}$ does not realize $\text{wt}(\beta_M)$, nor does any subsequence; in other words, for some $w < 1$, for all $n$, $\text{eff}_  \omega(c_n) \le w$. Thus, $L_\omega(c_n) = L(c_n) - \int_{c_n}\omega \ge (1 - w)L(c_n)$ (since $|\int_{c_n}\omega|/L(c_n) \le w$).  Therefore, $\{L_\omega(c_n)\} \to 0$ tells us $\{L(c_n)\} \to 0$ also, from which we know that $\{c_n\}$ collapses to $p$.  So we know is $M$ is future-distinguishing, and, by Proposition 2.13, strongly causal (and causally continuous).

However, given that there is a sequence of loops $\{c_n\}$ in $Q$ through a point $p$ with $\{\text{eff}_\omega(c_n)\} \to 1$ and $L_\omega(c_n) < A$ for all $n$, $M$ cannot be globally hyperbolic:  For pick a point $x \in \pi^{-1}(p)$, and let $y = A \cdot x$; for each $n$ let $\bar c_n$ be the future-null lift of $c_n$ starting at $x$---necessarily terminating at some point $T_n \cdot x$ with $T_n < A$---and let $\delta_n$ be the extension of $\bar c_n$ to $y$ by concatenation with $\{t \cdot x \st T_n \le t \le A\}$.  If $M$ were globally hyperbolic, then $\{\delta_n\}$ would have a limit curve $\delta$, future causal from $x$ to $y$.  But then $c = \pi\circ\delta$ would be a limit loop of the loops $\{c_n\}$, and by Lemma 2.4, $\text{eff}_\omega(c) = 1$, contrary to our hypothesis.  Thus, $M$ cannot be globally hyperbolic; and by Proposition 2.15 it follows $M$ cannot be both causally bounded and spatially complete.

Note that for any sequence with $\{\text{eff}_\omega(c_n)\} \to 1$ and $L_\omega(c_n) \ge \epsilon > 0$ for all $n$, it's always true that $\{L(c_n)\} \to \infty$: We have $\{\text{eff}_\omega(c_n)\} = (\int_{c_n}\omega)/L(c_n) = (L(c_n) - L_\omega(c_n))/L(c_n) = 1 - L_\omega(c_n)/L(c_n)$, so $\{L_\omega(c_n)/L(c_n)\} \to 0$; $L_\omega(c_n)$ bounded away from 0 thus implies $L(c_n)$ goes to infinity.

Now consider case (5):  $\text{wt}(\beta_M) = 1$ and $L(c) > \beta_M\<c\>$ for every loop $c$ in $Q$; all weight-realizing sequences of loops have $L_\omega$ bounded away from 0; and for any such sequence $\{c_n\}$, $\{L_\omega(c_n)\} \to \infty$ .  Just as in case (4), we know $M$ is future-distinguishing, hence, strongly causal; we need to show it is causally bounded.  Consider any $x \ll y$ in $M$; we want to show $I^+(x) \cap I^-(y)$ has bounded projection to $Q$.  For some $T > 0$, $T \cdot x \gg y$, and it suffices to show the same for $I^+(x) \cap I^-(T \cdot x)$.  Then all we need to look at are loops $c$ in $Q$, based at $p = \pi(x)$, as any point in $\pi(I^+(x) \cap I^-(T \cdot x))$ occurs as an element of such a loop.  So consider a sequence of such loops $\{c_n\}$ reaching out as far as possible, i.e., with points $p_n \in c_n$ such that $d(p_n,p)$ approaches the maximum possible; we need to see if this max distance is finite.  This amounts to showing that $\{L(c_n)\}$ must be bounded above.  Note that we have $L_\omega(c_n) \le T$ for all $n$. Now, if $\{\text{eff}_\omega(c_n)\} \to 1$, then by assumption $\{L_\omega(c_n)\}$ cannot be bounded above.  So it follows there is some $w < 1$ such that for all $n$, $\text{eff}_\omega(c_n) \le w$.  Then, as above, $T \ge L_\omega(c_n) \ge (1 - w)L(c_n)$, and we have $L(c_n) \le T/(1 - w)$, all $n$.

Finally, we have case (6):  $\text{wt}(\beta_M) < 1$.  Just as immediately above we have $M$ is causally bounded (choosing $w = \text{wt}(\beta_M)$).  To show $M$ is future-distinguishing (hence, strongly causal), consider a sequence of $p$-based loops $\{c_n\}$ in $Q$ with $\{L_\omega(c_n)\} \to 0$.  By the same argument, $\{L(c_n)\} \to 0$ also, and Lemma 2.19 yields the result. \qed
\enddemo

One easy corollary of Theorem 2.18 is a direct measure of whether or not a stationary-complete spacetime has a presentation as standard stationary (this is essentially the content of Theorem 1.2 of \cite{JS}):  

\proclaim{Corollary 2.20} Let $\pi: M \to Q$ be a stationary-complete spacetime satisfying the observer-manifold condition.  Then the following are equivalent: 
\roster
\item $M$ has a presentation as standard stationary
\item $M$ is stably causal
\item either $\rwt(\beta_M) < 1$ or, alternatively, $\rwt(\beta_M) = 1$ and for every sequence of $\{c_n\}$ of base-pointed loops in $Q$ with $\{\roman{eff}_\omega(c_n)\} \to 1$, $\{L_\omega(c_n)\}$ is bounded away from 0.
\endroster
\endproclaim

\demo{Proof} Theorem 2.18 gives us that statements \therosteritem2 and \therosteritem3 are equivalent (i.e., falling within categories (4), (5), or (6) of that theorem).

If $M$ has a standard stationary presentation (see equation (1.1a)), then the corresponding time function $\tau: M \to \R$ is a global time function in the sense of being strictly increasing along every causal curve; this is because $\nabla\tau$ is perpendicular to the $\tau = \text{constant}$ slices, and those are all spacelike in a standard stationary presentation.  That is equivalent to being stably causal (see, for instance, \cite{BEEs}, p. 64, citing \cite{Hw}). 

On the other hand, if $M$ is stably causal, there is a continuous global time function $T_0: M \to \R$.  Moreover, by \cite{BrS}, Theorem 1.2, where there is a continuous gobal time function, there is a smooth one $T : M \to \R$ with timelike gradient.  Let $t_0$ be any number in the range of $T$; we need to see that $\Sigma = T^{-1}(t_0)$ defines a spacelike cross-section of $\pi: M \to \R$.  

Since $T$ has a non-vanishing gradient, $\Sigma$ is a smooth hypersurface; since $\nabla T$ is timelike, $\Sigma$ is spacelike; and since $T$ is increasing on timelike curves, $\Sigma$ intersects each Killing orbit at most once.  All that remains is show that $Q_0 = \{ q \in Q \st \Sigma \text{ intersects } \pi^{-1}(q)\}$ is actually all of $Q$.  By continuity of $T$, $Q_0$ is closed.  Suppose $q \in Q_0$, with $T(x) = t_0$ for $\pi(x) = q$; since $\Sigma$ is spacelike, for all points $q'$ nearby to $q$, $\pi^{-1}(q')$ intersects $\Sigma$ also (since $\Sigma$ is transverse to the fibres of $\pi$).  Therefore, $Q_0$ is also open, so $Q_0 = Q$.  Then $\Sigma$ defines a cross-section $z: Q \to M$ by $z(q) = x$ for $T(x) = t_0$.  This is a standard stationary presentation of $M$.\qed

\remark{Remark 2.21} Category (4) in Theorem 2.18 has two (non-exclusive) possibilities: spatial incompleteness and causal unboundedness.  If, in the context of Theorem 1.5, we consider the Killing orbit space and the Killing length function as fixed, but allow ourselves the freedom to modify the fundamental co-cycle (in the sense of seeing how changes to the fundamental cocyle affect the causal structure of the spacetime)---for instance, in the static case, by ramping the values of the representation of the fundamental group up and down---then spatial completeness or incompleteness is unaffected by changes solely to the fundamental cocycle.  

Assuming the fundamental cocycle is selected to have weight 1 and the spacetime is category (4), the issue of causal boundedness can be investigated with the given sequence of loops $\{c_n\}$ with $\{\text{eff}(c_n)\} \to 1$ and $L_\omega(c_n) < A$ for some finite $A$: With $\bar c_n$ the null lift of $c_n$ to the spacetime $M$, all starting at the same point $x$, then we have each $\bar c_n$ contained in $I^+(x) \cap I^-(A \cdot x)$, so the boundedness of the loops $\{c_n\}$ is of issue in discovering causal unboundedness:  Specifically, if the loops (which we know to be unbounded in length) actually travel unbounded distances from the base-point, then $M$ is causally unbounded.  If that behavior is impossible for any such sequence of loops in $Q$, then $M$ is causally bounded.

Alternatively, we can consider the fundamental cocycle as fixed and consider modifying only the Killing length through a conformal change in spacetime metric.  There is always some conformal change in metric which will make $Q$ complete; in the case of category (4) this will necessarily render the spacetime causally unbounded.  Contrariwise, there is always a conformal change in metric that will bring all the points of any sequence of loops to within a bounded distance of the base-point.  Thus, we can see spatial completeness and causal boundedness as trading off with one another under various choices of conformal factor. 
\endremark

\enddemo 

\head Section 3: Examples
\endhead

Here is presented a large range of examples of static- and stationary-complete spacetimes, along with presentations of fundamental cocycles and the like.  Several purposes are served here: showing simple examples of how these ideas play out; exhibiting examples that show various hypotheses in the theorems really are needed and really are independent; and presenting physically plausible models of spacetimes with non-trivial behaviors in terms of fundamental cocycle behavior.  (Some of these last are generated by taking existing physical spacetimes with a circle factor, unwrapping around that factor, and rewrapping with temporal lapse.)

There are more static than purely stationary examples, as the former are simpler to deal with; for instance, to calculate $I_M$, one need only work with $\tilde d^\rho$ in the universal cover (as exemplified in the material before Lemma 2.7), and that is a good deal easier than trying to find minimizing curves for $L_\omega$.

\subhead Static Examples
\endsubhead

Throughout the static examples, the following notation will be used:  For a group $G$ operating isometrically on a (typically simply connected) space $\tilde Q$, and for $\rho: G \to \Bbb R$ a representation of $G$, $(\Bbb L^1 \times \tilde Q)/G^\rho$ (typically denoted $M$) indicates modding out by the $G$-action on $\Bbb L^1 \times \tilde Q$ specified by $a \cdot (t, \tilde q) = (t + \rho(a), a \cdot \tilde q)$.  Let $\tilde M = \Bbb L^1 \times \tilde Q$,  $M = \tilde M/G^\rho$, and $Q = \tilde Q/G$.  We have commuting maps $\tilde \pi: \tilde M \to \tilde Q$, $\pi: M \to Q$, $p_M : \tilde M \to M$, and $p_Q: \tilde Q \to Q$.  With $\tilde K$ the obvious unit-length Killing field $\partial/\partial t$ on $\tilde M$, $K = p_M{}_* \tilde K$ is a Killing field on $M$, and $\tilde \pi$ and $\pi$ are the respective Killing projections.  The notation $[t, \tilde q]$ will often be used for $p_M(t, \tilde q)$ and likewise $[\tilde q]$ for $p_Q(\tilde q)$.

Then, as in the material at the beginning of Section 2,  
$$\rwt(\beta_M) = \underset{a \neq e}\to{\sup_{a \in G}}\; \frac{|\rho(a)|}{d(\tilde q,a \cdot \tilde q)}$$
for any choice of base-point $\tilde q \in \tilde Q$.  We will also want to calculate the interval using 
$$\tilde d^\rho(\tilde q, \tilde q') = \sup_{a \in G} \(d(\tilde q, a \cdot \tilde q') - \rho(a)\)$$
though that is typically rather messy and not always capable of a closed form.

\subsubhead Flat Examples: Not Physically Significant
\endsubsubhead

\example{Example 3.1: Minkowski cylinders}

The simplest example of a non-standard static-complete spacetime is a refastening of the $1 + 1$ Minkowski cylinder, $\Bbb L^1 \times \Bbb S^1$.  We start with Minkowsi 2-space, $\Bbb L^2 = \Bbb L^1 \times \Bbb R^1$ (i.e., $\tilde Q = \Bbb R^1$).  We then apply a $\Bbb Z$-action to $\Bbb R^1$, $m \cdot x = x + m$, and choose a representation $\rho: \Bbb Z \to \Bbb R$, determined by a constant $\lambda$ as $\rho(m) = \lambda m$.  Then the static spacetime we obtain is $\pi: M = (\Bbb L^ 1 \times \Bbb R^1)/\Bbb Z^\rho \to Q = \Bbb R^1/\Bbb Z = \Bbb S^1$.  We easily find $\rwt(\beta_M) = |\lambda|$, so $M$ is causal iff $|\lambda| < 1$, in which case it is globally hyperbolic; if $\lambda = 1$ (or -1), then $M$ is in category (2) of Theorem 2.18, as there is a closed causal loop, $[t(s),x(s)] = [s,s]$.  

In sum: $M$ is in category (1), (2), or (6) according as $|\lambda|$ is $> 1$, $= 1$, or $< 1$, respectively.

Calculating $\tilde d^\rho$ always involves finding an infimum over a discrete group.  Especially useful is the floor function on reals, denoted by $\lfl x \rfl$; this indicates the unique integer $n$ such that $x = n + \delta$ for $0 \le \delta < 1$.  In this example we have
$$\tilde d^\rho(x,x') = -\lambda\lfl x - x'\rfl + \min\{\delta, 1 - \delta - \lambda\}$$
(with $\delta = x - x' - \lfl x - x' \rfl$) telling us that $[t,x] \ll [t',x']$ iff $t' - t > -\lambda\lfl x - x'\rfl + \min\{\delta, 1 - \delta - \lambda\}$.  Since $\tilde M = \Bbb L^2$ is globally hyperbolic, so is $M$ iff $|\lambda| < 1$.

One striking aspect of this example is that it has a plethora of alternate interpretations as a static-complete spacetime, due to the existence in $Q$ of a Killing vector $X$, which we may as well take to be $p_Q{}_*(\partial/\partial x)$.  Then for any $a$ with $|a| < 1$, $K_a = K + aX$ is a timelike Killing field on $M$.  We have the primary Killing form $\alpha_a = (dt - a\,dx)/(1 - a^2)$.  The $K_a$-Killing action is $s\cdot [t,x] = [t + s, x + as]$, so the $K_a$-Killing orbits have the form $l_x = \{[s, x + as] \st s \in \Bbb R\}$.  But note that for any integer $m$, $[s, x + (1 - a\lambda)m + as] = [s, x + m + a(s - a\lambda m)] = [s - \lambda m, x + a(s - \lambda m)] = [s', x + as']$ for $s' = s - \lambda m$, so $l_{x + (1 - a\lambda)m} = l_x$; thus, the space of $K_a$-Killing orbits is $Q_a = \Bbb R/\Bbb Z$ with the action $m \cdot x = x + (1 - a\lambda)m$, and we have the Killing projection $\pi_a: M \to Q_a$, $\pi_a[t,x] = [x - at]_a$ (with $[\;\;]_a$denoting an element of $Q_a$).  The Killing orbit metric (using $u$ for the co\"ordinate in $Q_a$) is $h_a = (du)^2/(1 - a^2)$, the Killing squared-length is $\Omega_a = 1 - a^2$, and the conformal metric is $(du)^2/(1 - a^2)^2$.  The representation of $\Bbb Z$ is $\rho_a(m) = ((\lambda- a)/(1- a^2))m$; this can be calculated, for instance, by using the Killing time-function $\tau_a: M \to \Bbb R$ given by $\tau_a(t,[x]) = (t - \lambda x)/(1 - a\lambda)$, yielding Killing drift-form $\omega_a = (\lambda - a)/((1 - a^2)(1 - a\lambda))du$ and making use of the loop $c_a: [0,1] \to Q_a$, $c_a(s) = [(1 - a\lambda)s]_a$.  Then using the  conformal length of $c_a$ is $(1 - a\lambda)/(1 - a^2)$, we get the weight of $\beta_{M}$ using $K_a$ as $\rwt(\omega_a) = |\lambda - a|/(1 - a\lambda)$.

In particular, taking $a = \lambda$ yields $\rwt(\beta_M) = 0$:  With respect to the Killing field $K_\lambda$, $M$ is the standard static spacetime $\Bbb L^1 \times Q_\lambda$.  In other words, our rewrapping of the standard cylinder with circumference 1 has resulted in nothing other than a cylinder with circumference $1 - \lambda^2$, albeit with respect to a different Killing field.  More physically:  It was the choice of a non-ideal set of static observers (or clocks) that led to the conclusion that there was any globally anomalous behavior in $M$; choosing the optimal collection of clocks reveals a perfectly ordinary global behavior.  

In general, though, we do not have the freedom from a spacelike Killing field to shift from one timelike Killing field to another.

\vskip .1 in

We can do a similar operation on $\Bbb L^1 \times \Bbb R^k$, using the $\Bbb Z$-action on $\Bbb R^n$ given by  $m \cdot \vec x = (x_1 + m, x_2, \dots, x_k)$.  The results are very similar, though it's more complicated to calculate $\tilde d^\rho$, and it doesn't have as pleasant a formulation.  Since it involves a formula that reappears, it helps to set that up first:

The basic issue is to find the infimum of $\sqrt{b^2 + (x+ c)^2} - ax$.  This infimum is $|b|\sqrt{1 - a^2} + ac$, occurring at $x = a|b|/\sqrt{1 - a^2} - c$, when $x$ is allowed to take on all real values; but things are more complicated when $x$ can take on only a discrete set of values.  So let us define
$$\align
Z(a,b,c) & = \inf_{m \in \Bbb Z} {\(\sqrt{b^2 + (m + c)^2} - am\)} \\
& = -aN + \min\left\{\sqrt{b^2 + (N + c)^2}, \;\sqrt{b^2 + (N + 1 + c)^2} - a\right\} \\
& \quad\quad \text{where }\; N \le \frac{a}{\sqrt{1 - a^2}}|b| - c < N + 1 \;\text{ and } N \in \Bbb Z
\endalign$$
or, alternatively expressed,
$$\align
Z(a,b,c) & = -aN + \min\left\{\sqrt{b^2 + \(\frac{a}{\sqrt{1 - a^2}}|b| - \delta\)^2}, \right. \\
& \left. \quad\quad\quad\quad\quad\quad\quad \sqrt{b^2 + \(\frac{a}{\sqrt{1 - a^2}}|b| + 1 - \delta\)^2} - a\right\} \\
& \quad\quad \text{where }\; \frac{a}{\sqrt{1 - a^2}}|b| - c = N + \delta \;\text{ and } N \in \Bbb Z,\; 0 \le \delta < 1.
\endalign$$ 

Then for the given $\Bbb Z$-action on $\Bbb R^k$ and the same real representation $\rho$ of $\Bbb Z$, we have
$$\tilde d^\rho(x,x') = Z\!\(\lambda,\,||(x' - x)^\perp||,\, x'_1 - x_1\)$$
where $y^\perp = (y_2, ..., y_k)$.

\vskip .1 in

A related notion is to have a torus for $Q$, i.e., let $\Bbb Z^2$ act on $\Bbb R^2$ via $(m,n) \cdot (x,y) = (x + am, y + bn)$, with representation $\rho(m,n) = \lambda m + \mu n$.  We get $\rwt(\beta_M) = \max\{|\lambda/a|, |\mu/b|\}$; categories are as before.  It is a deal more complicated to calculate $\tilde d^\rho$, though, and a closed form doesn't appear easy to obtain:
$$\tilde d^\rho((x,y),(x',y')) = \inf_{m,n}\(\sqrt{(x - x' - am)^2 + (y - y' - bn)^2} - \lambda m - \mu n\)$$

\endexample

\example{Example 3.2: Minkowski glides}

Next most complex action would be $\Bbb Z$ acting via glides instead of translations on $\Bbb R^2$, producing a M\"obius strip for $Q$: $m \cdot (x,y) = (x + m, (-1)^m y)$.  With representation $\rho(m) = \lambda m$, again we have $\rwt(\beta_M) = |\lambda|$.  Calculation of $\tilde d^\rho$ proceeds just as in the Minkowski ``cylinder" over $\Bbb R^k$ (with $k = 2$), with slightly different expressions for even and odd integers:
$$\align
\tilde d^\rho((x,y),(x',y'))  = \min & \left\{2Z\!\(\lambda,\frac12(y'-y),\frac12(x'-x)\), \right. \\
& \;\; \left. 2Z\! \(\lambda,\,\frac12(y'+y),\,\frac12(x'-x + 1)\) - \lambda \right\}
\endalign$$
\endexample

\example{Example 3.3: Infinite connectivity: Causal with wt($\beta_M$) = 1}  

In this example we will make use of a $\Bbb Z$-action, but not on the universal cover.  Instead, the cover will be $\bar Q$ formed by putting slits in the plane $\Bbb R^2$:  For every even number $2n$, remove all vertical segments  $\{2n\}\times[4k, 4k + 3]$  (all integers $k$); and for every odd number $2n + 1$, remove all vertical segments $\{2n + 1\}\times[4k - 2, 4k + 1]$ (all integers $k$). ÊThe idea is to provide windows of access between adjacent slits in any one column, arranged to be not aligned with the windows in the next column over, so that travel between two columns must be on a diagonal whose slope has magnitude greater than 1.
With the standard Euclidean metric on the plane this has an action by the integers $\Bbb Z$ of $m \cdot (x, y) = (x + 2m, y)$; let $Q = \bar Q / \Bbb Z$.  

Let $\bar M =  \Bbb L^1 \times \bar Q $.  Define $\rho: \Bbb Z \to \Bbb R$ by $\rho(m) = 2\sqrt2m$ and let $\Bbb Z$ act on $\bar M$ by the usual $m \cdot (t,p) = (t + \rho(m), m \cdot p) = (t + 2\sqrt2m, x + 2m, y)$; $M = \bar M/\Bbb Z^\rho$.  The curves in $\bar Q$ that come close to being loops in $Q$ (approximately, from the bottom of one window to the top of a window one column over, then back up to the top of the window of the next column) have lengths approaching $2\sqrt2$; thus, $\rwt(\beta_M) = 1$.  But no curves actually realize this weight; thus, $M$ lies in category (3), (4), or (5) of Theorem 2.18.  To discover which, we need to look at a Killing drift-form.

Define a cross-section $z: Q \to M$ by $z[x,y] = [\sqrt2x, x,y]$; the associated Killing time-function $\tau: M \to \Bbb R$ is given by $\tau[t,x,y] = t - \sqrt2x$.  We have the primary Killing form $\alpha = dt$, so $\alpha - d\tau = \sqrt2dx$; this yields the Killing drift-form $\omega = \sqrt2dx$.  Pick a base-point $\bar q_0 = (.5,2.5)$ in $\bar Q$, with corresponding base-point $q_0 = [.5,2.5]$ for $Q$; then we can find curves $\bar c_n$ in $\bar Q$ from $(0,3 + 1/n)$ to $(1, 2 - 1/n)$ to $(2, 3 + 1/n)$, passing through $\bar q_0$, which project to base-pointed loops $c_n$.  Then we have $\{\text{eff}(c_n)\} \to 1$ and $\{L_\omega(c_n)\} \to 0$.  Thus $M$ is in category (3) of Theorem 2.18: While $\bar M$ is strongly causal, $M$ is not; it is causal but not future-distinguishing.

We will look at $\bar d^\rho$ in $\bar Q$ (instead of $\tilde d^\rho$ in $\tilde Q$).  Calculating this precisely is something of a chore.  For a large selection of points, though, it has an easy formula of
$$\bar d^\rho([x,y],[x',y']) = \min\left\{\sqrt{\delta^2 + (y - y')^2}, \sqrt{(2 - \delta)^2 + (y - y')^2}\right\}$$
where $|x - x'| = 2N + \delta$, $N \in \Bbb Z$ and $0 \le \delta < 2$.
\vskip .1 in
We could also do a very similar example with finite connectivity (i.e., finite number of generators of the fundamental group) by taking a quotient of $M$ by the $\Bbb Z$-action, $k\cdot[t,x,y] = [t,x,y + 4k]$.

\endexample

\example{Example 3.4a: Infinite connectivity: Variable results}

In this example, we start with a manifold $Q$ of infinite connectivity, $\Bbb R^2$ with a countable collection of holes in it:  for each $n > 0$, remove $p_n = (n,0)$.  With $G = \pi_1(Q)$ we have $Q = \tilde Q/G$; we need to show how to construct this concretely.

We can realize $G$ in this manner:  Let $* = (0,0)$ be the base-point.  Every element of $G$ can be characterized as a finite sequence of loops, $l_1, \dots, l_n$, with each $l_i$ issuing from $*$ and going around exactly one of the holes  exactly once, either clockwise or counterclockwise.  Thus we can represent an element of $G$ as a finite ordered list of non-zero integers $(k_1,\dots,k_n)$ with $k_i$ representing a loop around $p_{|k_i|}$: counterclockwise for $k_i > 0$, clockwise for $k_i < 0$.  The group operation is by concatenation of lists, $(k_1,\dots,k_n) \cdot (j_1,\dots,j_m) = (j_1,\dots,j_m,k_1,\dots,k_n)$; the identity element is $(\;)$, the empty list.  There are these relations among the elements of $G$: If $l_i$ and $l_{i + 1}$ go around the same hole in opposite directions, they cancel.  Thus, for $g = (k_1, \dots, k_n)$, if $k_i = - k_{i+1}$, then $g = (k_1,\dots,k_{i - 1}, k_{i + 2},\dots,k_n)$.  This yields a presentation of $G$ with an infinite set of generators, $\{(n) \st n  > 0\}$, with relators as just given (including $(n)^{-1} = (-n)$).

We then form the universal cover $\tilde Q$ as follows:  Denote by $S_n$ the segment from $p_n$ to $p_{n+1}$, $S_n = (n,n+1)\times\{0\}$.  Let $N$ be the result of deleting all the segments $S_n$ from $Q$, i.e., $N = \Bbb R^2 - (1,\infty)\times\{0\}$. Consider a family of copies of $N$, indexed by the elements of $G$: $\Cal N = \{N_g \st g \in G\}$; we can think of this as $N \times G$.  Then we can realize $\tilde Q$ via identifications placed on the topological space $\bigcup \Cal N$:  For each $g \in G$ and positive integer $n$, $N_g$ is joined to $N_{(n)g}$ along $S_n$, with the lower side of $S_n$ in $N_g$ joined to the upper side of $S_n$ in $N_{(n)g}$; and similarly for $N_g$ joined to $N_{(-n)g}$ along $S_n$, but with upper and lower sides reversed.

A representation of $G$ is characterized entirely by the numbers $\{\lambda_n \st n  > 0\}$ with $\rho((n)) = \lambda_n$; then $\rho(k_1,\dots,k_n) = \Sigma_i(\pm_i)\lambda_{|k_i|}$ with $\pm_i$ the same sign as $k_i$.  Any choice of $\{\lambda_n\}$ yields a representation.  A cycle representing the group element $(n)$ ($n > 0$) is a loop from $(0,0)$ that passes through $S_n$, from below to above, and then back down through $S_{n-1}$, going around $(n,0)$ counterclockwise but no other $p_k$; clearly we can find a sequence $\{c^n_k \st k \ge 1\}$ of these approaching (though not reaching) a length of $2n$; and so we have $|\beta_M\<c^n_k\>|/L(c^n_k) = |\rho((n))|/L(c^n_k)$ having a supremum of $|\lambda_n|/(2n)$.  For figuring the weight of $\beta_M$, these are the only group elements we need to consider. This gives us the following possibilities for $M = (\Bbb L^1\times \tilde Q)/G^\rho$:

If $\sup_n |\lambda_n|/(2n)$ is achieved for some $n = n_0$, then $\rwt(\beta_M) = |\lambda_{n_0}|/(2n_0)$, and $M$ is in category (1), (3), or (6) of Theorem 2.18, depending on whether this weight is $> 1$, $= 1$, or $< 1$, respectively.  (For weight = 1, it can't be category (2), as $L(c^{n_0}_k) > 2n_0$ for all $k$; we have $L_\omega(c^{n_0}_k) = L(c^{n_0}_k) - 2n_0$, and this goes to 0 as $k \to \infty$, yielding category (3).)

Otherwise, $\rwt(\beta_M) = \limsup_n |\lambda_n|/(2n)$ and $M$ is again in categories (1) or (6) for the weight $> 1$ or $< 1$; for weight $= 1$, several possibilities obtain:  We can take $c_n$, generating $((n))$, essentially to have length $2n$ (for instance, $c_n = c^n_{k_n}$ with $k_n$ chosen large enough that $L(c^n_{k_n}) < 2n + 1/n$).  Then, in essence, $L_\omega(c_n) = 2n - \lambda_n$.  If we restrict $n$ to $\{n_i\}$ generating the limit-supremum (i.e., $|\lambda_{n_i}|/(2n_i) \to 1$), then these are the possibilities:
\roster
\item $\liminf_i (2n_i - \lambda_{n_i}) = 0$: category (3) 
\item $\liminf_i (2n_i - \lambda_{n_i}) > 0$ and $\limsup_i (2n_i - \lambda_{n_i})$ finite: category (4), causally unbounded (since the $\{c_{n_i}\}$ extend indefinitely far)
\item $\liminf_i (2n_i - \lambda_{n_i}) > 0$ and $\limsup_i (2n_i - \lambda_{n_i}) = \infty$: category (5)
\endroster
Examples: $\lambda_n = 2n - 1/n$ for the first, $\lambda_n = 2n - 1$ for the second, and $\lambda_n = 2n - \sqrt n$ for the third.

\vskip .1 in

Calculating $\tilde d^\rho$ is not difficult, but it has numerous cases.  Here is one such:

Suppose $y > 0$ and $y' > 0$. Then for $n > 0$,
$$\tilde d^\rho((x,y,g),(x',y',(n)\cdot g)) = \sqrt{(x -1)^2 + y^2} + n - 1 + \sqrt{(x' - n)^2 + y'{}^2} - \lambda_n$$

\endexample

\example{Example 3.4b: Infinite complexity with global hyperbolicity: variable results}

A slight modification of the previous example produces a globally hyperbolic $\tilde M$:  Revise the metric on a disk of, say, size $1/(2n)$ around $p_n$ so as to produce a complete metric; this can be thought of as erecting a half-infinite cylinder of diameter $1/(2n)$ around each hole, embedding the one finite end of the cylinder in the plane.  This has no practical effect on calculating $\rwt(\beta_M)$; there is a slight change in weight in case $\sup_n |\lambda_n|/(2n)$ is obtained at some $n = n_0$, but otherwise no change at all.  Thus we have the following examples:  

With $\lambda_n = 2n - 1/n$, we have a globally hyperbolic $\tilde M$ with a quotient $M$ which, while causal, is not even future-distinguishing.  

If we take $\lambda_n = 2n - 1$, then we have a globally hyperbolic $\tilde M$ with a quotient $M$ which is causally continuous, yet not globally hyperbolic due to being not causally bounded (we know $Q$ is complete, so it's not spatial completeness which fails: it must be causal boundedness).  The essence of this is that we have infinitely many homotopy classes of future-causal curves between a specific pair of points in $M$.  This answers a question raised in \cite{Hr2} (with regard to Proposition 1.4 in that paper):  If $\tilde M$ is globally hyperbolic and $M = \tilde M/G$ is strongly causal, does it necessarily follow that $M$ is globally hyperbolic, or must one add the condition that in $\tilde M$, the future of any point has only finite intersection with the $G$-orbit of each other point?  This example shows that global hyerpoblicity of $M$ does not follow without additional hypothesis such as that suggested.

Finally, if we take $\lambda_n = 2n - \sqrt n$, we have an example of a globally hyperbolic $M$ with $\rwt(\beta_M) = 1$. 
\endexample

\subsubhead Physically Significant Examples
\endsubsubhead

Numerous classical spacetimes have a spacelike $\Bbb S^1$-symmetry, so that if $M$ is the spacetime with the stationary worldsheet of the axis of symmetry removed, then $M$ can be expressed as $N \times \Bbb S^1$, and the $\Bbb S^1$-action of $\phi \cdot (x,\theta) = (x, \theta +\phi)$ is isometric.  Such a spacetime can be ``rewrapped" via a covering space, $\bar M = N \times \Bbb R^1$ (thinking of $\Bbb S^1$ as covered by $\Bbb R^1$ via the $\Bbb Z$-action $m \cdot s = s + m2\pi$):  For any $\gamma$ we can obtain the spacetime $M' = \bar M/\Bbb Z$ via the action $m \cdot (x, s) = (x, s + m\gamma)$.  For $\gamma$ other than $2\pi$ this is a spacetime locally identical to $M$ but globally distinct; typically it has a ``cone singularity".   A slightly more complex but similar technique is used several times in the following examples.

Also, anything with a spacelike $\Bbb R^1$-symmetry has a $\Bbb Z$-quotient with $\Bbb S^1$-symmetry.  This allows further play as above. 

\example{Example 3.5: Flat cosmic string}

An ``ideal" cosmic string (for general concept see, for instance, \cite{Go}) can be treated as a standard static spacetime with the static observer space given as a planar cone-singularity cross $\Bbb R^1$.  By a planar cone-singularity is meant the following:  Delete from $\Bbb R^2$ a wedge of angle $\delta$, i.e, two rays starting at the origin, with angle $\delta$ between them, and all points between the two rays.  Then glue the two edges together, leaving the origin omitted; this (when crossed with $\Bbb L^2$) is the 0-diameter cosmic string of mass $\delta$.

More generally, the diameter can be some non-zero $r_0 > 0$, and there can be ``negative mass": a wedge of angle larger than $\delta$ can be glued to the two edges.  We will look only at the positive-diameter case, as that lends itself to modifications as per the methods in this note, while remaining causal.

Here is another way to conceptualize the classic cosmic string: With $D(a)$ denoting the closed disk of radius $a$ about the origin in $\Bbb R^2$,  let $N = \Bbb R^2 - D(r_0)$, and let $\tilde N$ be the universal cover of $N$: a helicoid, formed by gluing together copies of $\Bbb R^2 - D(r_0) - (r_0,\infty)\times\{0\}$ edge-to-edge.  We can parametrize $\tilde N$ by polar co\"ordinates $(r,\theta)$ taking all values in $(r_0, \infty)\times\Bbb R$.  We take $\tilde Q = \tilde N \times \Bbb R^1$.  Let $\Bbb Z$ act on $\tilde Q$ (in cylindrical co\"ordinates) via $m \cdot(r,\theta,z) = (r,\theta + m\gamma, z)$ for some constant $\gamma >  0$ ($\gamma = 2\pi - \delta$ for comparison with previous model).  Then $(\Bbb L^1 \times \tilde Q)/\Bbb Z$, with no $\Bbb Z$-action on the $\Bbb L^1$ factor, is the standard cosmic string.

But we can add in a representation of $\Bbb Z$, $\rho(m) = \lambda m$.  This gives us, for $M = (\Bbb L^1 \times \tilde Q)/\Bbb Z^\rho$, $\rwt(\beta_M) = |\lambda|/{r_0\gamma}$ (to see this, pick, for instance,  base-point $(2r_0,0,0)$ and loops $c_n$ going around $D(r_0)$ an angle of $n\gamma$ radians  almost at radius $r_0$, having lengths approximately $2r_0 + n\gamma r_0$; $\beta_M\<c_n\>/L(c_n)$ is about $n\lambda/(r_0(2 + n\gamma))$).  We have $M$ in category (1), (4), or (6), depending on whether $|\lambda|$ is $>$, $=$, $<$ $r_0\gamma$, respectively.  To see that it is category (4) when weight = 1, note that with loops $c_n$ as expressed above, $\{L_\omega(c_n)\} \to 2r_0$; and no matter what the chosen sequence of loops with efficiency limit of 1, there will be that residual length from base-point to the ``circles" of essentially radius $r_0$, not canceled by $\beta_M\<c_n\>$. 

In all cases, $M$ is clearly spatially incomplete; with $|\lambda| = r_0\gamma$ (or, of course, smaller) it is causally bounded:  We have $Q = (r_0,\infty) \times \Bbb S^1 \times \Bbb R^1$ with metric $dr^2 + r^2d\theta^2 + dz^2$ and $\Bbb S^1$ having total length $\gamma$.   To have base-pointed loops $\{c_n\}$ traveling unboundedly far from base point, they must increase indefinitely in the first or third factors; but it is only travel in the middle factor that increases $\beta_M\<c_n\>$, and that increases only by (just less than) the same amount that $L(c_n)$ increases.  Therefore, it's not possible for $\{L_\omega(c_n) = L(c_n) - \beta\<c_n\>\}$ to remain bounded while $\{c_n\}$ reaches unboundedly far from base point.

Calculating $\tilde d^\rho$ involves distance in the punctured helicoid between two points; a near-geodesic between points separated by the $r_0$-disk lies close to being tangent from the first point to the disk, then around the disk the correct amount, then tangent to the disk for going to the second point.  This yields
$$\tilde d^\rho((r, \theta, z), (r', \theta', z')) = \inf_m
\(\sqrt{(z' - z)^2 + R_m^2}- m\lambda\)$$
where $R_m = R((r,\theta),(r',\theta' + m\gamma))$ is distance in the punctured helicoid:  For $p = (r,\theta)$ and $p' = (r',\theta')$,
$$R(p,p') = \left\{ 
\aligned
& |\theta' - \theta| < \Delta\!: \sqrt{r'{}^2 - 2r'r\cos(\theta'-\theta) + r^2}  \\
& |\theta' - \theta| \ge \Delta\!: \sqrt{r'{}^2 - r_0^2} + \sqrt{r^2 - r_0^2} +  r_0\(|\theta' - \theta| - \Delta\)
\endaligned
\right.$$
with $\Delta =  \cos^{-1}\!\dfrac{r_0}{r'} + \cos^{-1}\!\dfrac{r_0}r$.

\vskip .1 in

We can also put in a glide along the $z$-axis, i.e., using as $\Bbb Z$-action $m \cdot (r, \theta, z) = (r, \theta + m\gamma, z + ma)$ for some constant $a$, yielding $\rwt(\beta_M) = |\lambda|/\sqrt{a^2 + (r_0\gamma)^2}$.  But the same thing can be accomplished just with changing the Killing field to $\partial/\partial t + b(\partial/\partial z)$ for appropriate $b$ and adjustment of the representation.

\vskip .1 in 

We can create a cylindrical version (with finite-length cosmic string) with a second, different $\Bbb Z$-action, $n \cdot (r,\theta, z) = (r, \theta, z + na)$, leading to a $\Bbb Z^2$-action overall, $(m,n) \cdot (r,\theta,z) = (r,\theta + m\gamma, z + na)$ and real representation $\rho(m,n) = \lambda m + \mu n$.  We get $\rwt(\beta_M) = \max\left\{|\lambda|/(r_0\gamma),|\mu|/a \right\}$.  As in the double-$\Bbb Z$ action on Minkowski space, it's not apparent how to get a closed form for $\tilde d^\rho$. 

\endexample

\example{Example 3.6: Rindler cosmic string}

This example uses accelerated static observers in a flat spacetime: $(3 + 1)$-dimensional Rindler space (for example, see \cite{MiTW}).

We'll begin with standard rectangular coordinates in $\Bbb L^4$, in which the metric $g$ is 
$$ds^2 =  - dt^2 + dx^2 + dy^2 + dz^2.$$
We generate a Killing field via boosts in $(t,x)$ co\"ordinates, i.e., the isometric $\Bbb R$-action $s \cdot(t,x,y,z) = (t\cosh s + x\sinh s, x\cosh s + t\sinh s,y,z)$; this yields the field $K_p = x(\partial/\partial t) + t(\partial/\partial x)$ (for $p = (t,x,y,z)$).  Then $K$ is timelike in
$M_0 = \{p \in \Bbb L^4 \st |t| < x\}$ (the $(3 + 1)$ Rindler wedge).    Define $\xi$ and $\tau$ on $M_0$ by $\xi(p) = \sqrt{x^2 - t^2}$, $\tau(p) = \tanh^{-1}(t/x)$; $\tau$ corresponds to motion by the $K$-action, and for any $p \in M_0$,  $p = \tau(p) \cdot (0, \xi(p),y,z)$.  (Note that ($\tau$ = constant)-surfaces are perpendicular to $K$, so $M_0$ is static with respect to $K$.)  Thus the space $Q_0$ of Killing orbits is $\Bbb R^+\times\Bbb R^2$ (with $\Bbb R^+ = (0,\infty)$); we have the Killing projection $\pi: M_0 \to Q_0$ with $\pi(t,x,y,z) = (\xi,y,z)$.  The primary Killing form is $\alpha = (x\,dt - t\,dx)/(x^2 - t^2)$, and the Killing length-squared is $\Omega(\xi,y,z) = \xi^2$ (reinterpreting $\xi$ as a function---first co\"ordinate---on $Q_0$; that reinterpretation yields $\pi^*d\xi = d\xi$, with 3-function on the left, 4-function on the right).  The Killing orbit-space metric  is $ h = d\xi^2 + dy^2 + dz^2$ (as can be seen from $g + (\Omega\circ\pi)\alpha^2 = \(\dfrac{tdt - xdx}{\sqrt{x^2 - t^2}}\)^2  + dy^2 + dz^2 = d\xi^2 + dy^2 + dz^2$).  Then the conformal metric is 
$$\bar h = \frac{d\xi^2 + dy^2 + dz^2}{\xi^2}$$
yielding $Q_0$ as the Poincar\'e half-space, i.e., $\Bbb H^3$, hyperbolic 3-space.  Alternatively, it can be analyzed, using $\sigma = \ln\xi$, as the warped-product metric $d\sigma^2 + e^{-2\sigma}(dy^2 + dz^2)$, i.e.,  $Q_0 = \Bbb R \times_{e^{-2\sigma}} \Bbb R^2$.

There are many discrete isometric actions on $\Bbb H^3$, and these can be used to obtain variations on the Minkowski cylinder.  Some of these actions are radically different from Minkowski translations in that if we employ a non-zero temporal shift, we get the weight of the fundamental cocycle is infinite (example: $m \cdot (\xi,y,z) = (\xi,y + m,z)$; consider curves $c_{m,\Xi}$ from $(\xi_0,y_0,z_0)$ to $(\xi_0 + \Xi,y_0,z_0)$ to $(\xi_0 + \Xi,y_0 + m,z_0)$ to $(\xi_0,y_0 + m,z_0)$, first with fixed $\Xi$ and $m \to \infty$, then with $\Xi \to \infty$).  But others behave much like that in Example 3.1 (such as $m \cdot (\xi,y,z) = (e^m\xi, e^my, e^mz)$).  However, the resultant spacetimes are likely of more mathematical than physical interest.

Of somewhat more physical concern is considering what Rindler observers would take to be a cosmic string.  To do this, we proceed much as in Example 3.5 (though with some change due to the non-constant length of the Rindler Killing field):  For some $a > 0$, let $P_a = \{(\xi,y,z) \in Q_0 \st y^2 + z^2 > a^2\xi^2\}$; let $\tilde P_a$ be the universal cover of $P_a$. Define $(r,\phi)$ as polar co\"ordinates in the $(y,z)$-plane.  We then select $\gamma > 0$ and let $\Bbb Z$ act on $\tilde P_a$ with $m \cdot(\xi,r,\phi) = (\xi,r,\phi + m\gamma)$; then $Q_a = \tilde P_a/\Bbb Z$ is our cosmic string orbit-space.  

With a representation $\mu$ of $\Bbb Z$ via $\mu(m) = \lambda m$, we end up with a spacetime $M_a = (\Bbb L^1 \times \tilde P_a)/\Bbb Z^\mu$, and we have $\rwt(\beta_{M_a}) = |\lambda|/(a\gamma)$.   $M_a$ is in category (1), (4), or (6); much as for the previous  comsic string, it is spatially incomplete and (for $|\lambda| \le a\gamma$) causally bounded.  The formula for $\tilde d^\mu$ is rather difficult: It's modestly complex to work in $\tilde P_0$, which is the warped product $\Bbb R \times_{e^{-2\sigma}} H^2$, where $H^2$ is the helicoid (universal cover of $\Bbb R^2 - \{(0,0)\}$); but for $\tilde P_a$, one must figure distances going around the deleted cone, and that is much more of a challenge. 

\vskip .1 in

This spacetime represents what is seen as a cosmic string by the Rindler accelerated family of static observers.  Like the classic cosmic string, it exhibits a cone-singularity type of behavior in a geometrically flat background; and, as in the classic cosmic string, we deleted from the orbit space  a cylinder of (presumably) small radius---but that is as measured by the Rindler conformal metric.  To get a sense of what inertial observers see, look at $\pi^{-1}(P_a) = \{(t,x,y,z) |\; |t| < x \text{ and } y^2 + z^2 > a^2(x^2 - t^2)\}$: At a given $t$-level, this is the region between one of the sheets of a hyperboloid of two sheets and the plane tangent to that sheet at its apex.  Thus, the deleted region grows unboundedly (in terms of $(y,z)$-cross-section) with increasing $x$; however, no one observer sees it grow without bound, as any inertial observer has a finite lifetime.  But the Rindler static observers see the deleted region as having constant diameter $a$, as they use the conformal metric in the orbit space (that being what yields instantaneous speed of light as 1 when measured by static clocks).  

\endexample

\example{Example 3.7a: Melvin's compactified magnetic universe}

In \cite{M}, Melvin describes a static spacetime with cylindrical symmetry, representing, as he puts it, ``a parallel bundle of magnetic or electric flux held together by its own gravitational pull".  The manifold is $\Bbb R^4$ with metric (using cylindrical co\"ordinates $r, \theta, z$)
$$ds^2 = F(r)^2(-dt^2 + dr^2 + dz^2) + F(r)^{-2}r^2d\theta^2$$
where $F(r) = 1 + (b_0r/2)^2$; the constant $b_0$ is the value of the magnetic field on the axis.   In terms used in this paper, then, we have conformal factor $\Omega = F(r)^2$, static orbit space $\tilde Q = \Bbb R^3$, and conformal orbit-space metric 
$$\bar h = dz^2 + dr^2 + (r/F(r)^2)^2d\theta^2.$$

We can impose a $\Bbb Z$-action on the axis of cylindrical symmetry, compactifying in that direction: $\Bbb Z$ acts on $\tilde Q$ by $m \cdot (r, \theta, z) = (r, \theta, z + ma)$ for some $a > 0$.  With a representation $\rho(m) = \lambda m$, we have $M = \tilde M/\Bbb Z^\rho$ with $\rwt(\beta_M) = |\lambda|/a$, much as in Example 3.1; it is in category (1), (2), or (6), depending on $|\lambda|$.  The chief difference is that distance in $\tilde Q$ is not easy to calculate; for instance, a circle in a plane perpendicular to the $z$-axis and center on the axis, of radius $r$, has length $2\pi r/F(r)^2$, which has a maximum (of $\pi\sqrt3/b_0$) at $r = 2/(\sqrt3b_0)$, approaching 0 both as $r \to 0$ and $r \to \infty$.  This makes calculating $\tilde d^\rho$ difficult.  

\endexample

\example{Example 3.7b: Melvin's magnetic cosmic strings}

On the other hand, we can play cosmic string games with Melvin's magnetic universe:  With $\tilde Q$ as Example 3.7a, delete points with $r \le r_0$ for some $r_0 > 0$; but we also must delete points with $r \ge r_{\text{max}}$ for some $r_{\text{max}} > r_0$; call the remaining space $P$.  Then we let $\Bbb Z$ act on $\tilde P$, the universal cover, by $m \cdot (r, \theta, z) = (r, \theta + m\gamma, z)$ for some constant $\gamma > 0$.  With $\rho(m) = \lambda m$, we have 
$$\wt(\beta_M) = \max\left\{\frac{|\lambda|F(r_0)^2}{r_0\gamma}, \frac{|\lambda|F(r_{\text{max}})^2}{r_{\text{max}}\gamma}\right\}.$$
This has the same properties as the flat cosmic string---categories (1), (4), or (6), spatial incompleteness, and (for categories (4) or (6)) causal boundedness---irrespective of which radial value yields the weight of $\beta_M$.  Again, $\tilde d^\rho$ is made complicated by the distance function in $\tilde Q$.

\endexample

\example{Example 3.8: Schwarzschild and other cosmic strings}  

Schwarzschild space (see, for example, \cite{HwEl}) is static in the exterior region; for mass-parameter $m$, we can take the manifold to be $\Bbb R^1 \times (2m,\infty) \times \Bbb S^2$ with metric
$$ds^2 = - \(1 - \frac{2m}r\)dt^2 + \(1 - \frac{2m}r\)^{-1}dr^2 + r^2\(d\theta^2 + \sin^2\theta\,d\phi^2\) $$
We get Killing length-squared $\Omega = 1 - 2m/r$ and Killing orbit space $Q_0 = (2m,\infty) \times \Bbb S^2$ with conformal metric 
$$\bar h = \frac1{\(1 - \frac{2m}r\)^2}\,dr^2 + \dfrac{r^2}{1 - \frac{2m}r}\(d\theta^2 + \sin^2\theta\,d\phi^2\).$$
We can form a cosmic string out of this as before:  Form $Q_a$ for some $a > 0$ by deleting the polar caps with $\sin\theta \le a$, let $\tilde Q_a$ be the universal cover $(2m,\infty) \times (\theta_a, \pi - \theta_a) \times \Bbb R^1$ (with $\theta_a = \sin^{-1}\!a$), let $\Bbb Z$ act on $\tilde Q_a$ with $n \cdot(r,\theta,\phi) = (r,\theta,\phi + n\gamma)$ for some $\gamma > 0$, and choose a representation $\rho(n) = n\lambda$.  Then with $\tilde M_a$ being $\Bbb R^1 \times \tilde Q_a$ (with the Schwarzschild spacetime metric), we have our Schwarzschild cosmic string $M_a = \tilde M_a/\Bbb Z^\rho$.  To find the weight of the fundamental cocycle we need to know where the shortest cycles of non-trivial homotopy class lie; that is at $r = 3m$ (just select the $r$ that minimizes the conformal length of a $\phi$-curve of constant $r$ and $\theta = 0$).  We get $\rwt(\beta_{M_a}) = |\lambda|/(3\sqrt3ma\gamma)$.  Behaving similarly to the other cosmic strings, this is in category (1), (4), or (6), is spacelike incomplete, and is  causally bounded when in category (4).  

There's no barrier to extending from $M_a$ to an analogue of interior Schwarzschild, though that won't be static.

\vskip .1 in

We could work this slightly differently: deleting a smaller portion of $\tilde Q_0$ to get $\tilde Q_a$.  The important thing is to have a positive infimum for the length of a cycle that represents a homotopy class in $Q_a$.  We could do this with $\tilde Q'_a = \{(r,\theta,\phi) \in (2m,\infty) \times [0,\pi] \times \Bbb R \st \sin\theta > (a/r)\sqrt{1 - \frac{2m}r}\}$.  Then $M'_a$ has all the same properties as $M_a$, which it contains.  The advantage of $M'_a$ is that the cosmic string has a constant circumference of $a\gamma$; we have $\rwt(\beta_{M'_a}) = |\lambda|/(a\gamma)$.

What sort of physical process would give rise to an otherwise spherically symmetric geometry having a cone singularity in some particular direction, is far from clear.

\vskip .1 in

We can do very similar things with Reissner--Nordstr\"om \cite{HwEl} or Schwarz\-schild--de Sitter \cite{R}.
\endexample

\subhead{Stationary Examples}
\endsubhead

To find the classification of a stationary spacetime $M$ is a fairly straight-forward calculation of the Killing drift-form $\omega$ from knowing the fundamental Killing form $\alpha$ and the projection $\pi$ to the orbit space $Q$.  To create a new stationary stationary spacetime with a specified addition to the fundamental cocycle, we will use the prescription in Remark 1.8, working in the universal cover $\tilde \pi: \tilde M \to \tilde Q$, acted on by the fundamental group $G = \pi_1(M)$.

One considerable complication for stationary spacetimes is that the fundamental cocycle is not constant on homotopy classes of cycles, so in principle one must consider all possible cycles.  However, there is much simplification for spacetimes with circular symmetry and a drift-form operating along the direction of symmetry:

\proclaim{Lemma 3.9} Let $Q$ be a stationary orbit space with a Riemannian metric $h$, where $Q$ is an open subset of $(\rho_{\text{min}},\rho_{\text{max}}) \times \Bbb S^2$ with $(\rho_{\text{min}},\rho_{\text{max}})$ parametrized by $\rho$ and $\Bbb S^2$ parametrized by $(\theta,\phi) \in [0,\pi]\times[0,\gamma]$ (with $\phi = 0$ identified with $\phi = \gamma$ and with $\theta = 0$ and $\theta = \pi$ being the poles), $Q$ is a $\phi$-invariant subset, and $h$ is invariant in $\phi$.  (Most common is to have $\gamma = 2\pi$.) Let $\omega$ be a drift-form on $Q$ having the form
$$\omega = \omega_\phi(\rho,\theta)\,d\phi.$$
Then the weight of the cocycle $\{\omega\}$ is found by looking only at loops of constant $\rho$ and $\theta$; that is to say (with $h_{\phi,\phi} = h(\frac{\partial}{\partial\phi}, \frac{\partial}{\partial\phi})$),
$$\roman{wt}(\{\omega\}) = \sup_{\rho,\theta}\left\{\frac{|\omega_\phi(\rho,\theta)|}{\sqrt{h_{\phi,\phi}(\rho,\theta)}}\right\}.$$
\endproclaim

\demo{Proof}
For any loop $c$ in $Q$, recall $\eff_\omega(c) = (\int_c \omega)/ L(c)$, where $L$ denotes length using $h$; $\text{wt}(\{\omega\})$ is the supremum of $|\eff_\omega(c)|$ over all loops.  Since everything is $\phi$-invariant, the only way to maximize $\eff_\omega(c)$ is by traveling in only one $\phi$-direction, that is to say, if $c$ reverses $\phi$-direction, $|\int_c \omega|$ loses value.  Furthermore, there is nothing to be gained from $c$ going multiple times around the sphere. Consequently, we can assume $c$ is parametrized by $\phi$ with $c(s) = (\rho(s),\theta(s),s)$ (in $(\rho,\theta,\phi)$-co\"ordinates) for $s \in [0,\gamma]$.

Let $r = \omega_\phi/\sqrt{h_{\phi,\phi}}$, defined on $(\rho_{\text{min}},\rho_{\text{max}}) \times (0,\pi)$, and let $\{(\rho_n,\theta_n)\}$ be a sequence of points yielding $\sup(r)$ on this domain.  Let $c_n$ be the loop $c_n(s) = (\rho_n,\theta_n,s)$.  Then we  have 
$$\align
\text{wt}(\{\omega\}) & \ge \sup_n \eff_\theta(c_n) \\
& = \sup_n \frac{\gamma \omega_\phi(\rho_n,\theta_n)}{\gamma\sqrt{h_{\phi,\phi}(\rho_n,\theta_n)}} \\
& = \lim_n r(\rho_n,\theta_n) \\
& = \sup(r).
\endalign$$

Now consider any loop $c(s) = (\rho(s),\theta(s),s)$.  In $\(\int_c \omega\)/ L(c)$, we have $L(c) \ge \int_c \sqrt{h_{\phi,\phi}(\rho(s),\theta(s))}$, since that integral just omits the $d\rho$ and $d\theta$ terms in evaluating $|\dot c(s)|$.  We also have, for all $s$, $\omega_\phi(\rho(s),\theta(s)) \le \sqrt{h_{\phi,\phi}(\rho(s),\theta(s))} \sup(r)$, whence $\int_c \omega \le \(\int_c \sqrt{h_{\phi,\phi}(\rho(s),\theta(s))}\)\sup(r)$.  Thus we have
$$\align
\eff_\theta(c) & \le \frac{\(\int_c \sqrt{h_{\phi,\phi}(\rho(s),\theta(s))}\)\sup(r)}{\int_c \sqrt{h_{\phi,\phi}(\rho(s),\theta(s))}} \\
& = \sup(r)
\endalign$$
from which it follows that $\text{wt}(\{\omega\}) \le \sup(r)$. \qed
\enddemo

However, this is no short-cut to finding minimizing curves for $d_\omega$, so $I_M$ is not explicitly presented for these examples.

\example{Example 3.10 Kerr and Kerr cosmic strings}
\endexample

The Kerr spacetime represents a spinning black hole in vacuum; see \cite{HwEl} or \cite{O}.  The metric for a model of mass $m$ and angular momentum $a$, on the manifold of $(t,r,\theta,\phi) \in \Bbb R \times \Bbb R \times [0,\pi] \times \Bbb S^1$ (last factor is a unit circle) is
$$\align
ds^2  = & \(-1 + \frac{2mr}{\rho^2}\)dt^2  + \frac{\rho^2}\Delta dr^2 + \rho^2 d\theta^2 + \(r^2 + a^2 + \frac{2mra^2\sin^2\theta}{\rho^2}\)\sin^2\theta\,d\phi^2 \\
\quad\quad\quad & - \frac{2mra\sin^2\theta}{\rho^2}\(d\phi\,dt + dt\,d\phi\)
\endalign$$
where $\rho^2 = r^2 + a^2\cos^2\theta$ and $\Delta = r^2 - 2mr + a^2$.

We will be concerned here only with ``slow Kerr", $0 < a< m$, and only with the portion external to the horizon at $r_+ = m + \sqrt{m^2 - a^2}$; we must further restrict to that portion of external slow Kerr where the Killing field $\partial/\partial t$ is timelike.  The easiest way to do that is to restrict to $r > 2m$; there is a portion between $r = r_+$ and $r = 2m$ where $\partial/\partial t$ is still timelike, but it's a bit complex in that we get only polar caps of the sphere $[0,\pi] \times \Bbb S^1$.

So we take the orbit space to be $Q = (2m,\infty)\times\Bbb S^2$, and this tells us what $M = \pi^{-1}(Q)$ is.  We easily find $\alpha = -\<-,\frac{\partial}{\partial t}\>/(1 - \frac{2mr}{\rho^2}) = dt + \frac{2mra \sin^2\theta}{\rho^2-2mr}\,d\phi$, and setting $g = -(1 - \frac{2mr}{\rho^2})\alpha^2 + \pi^*h$ yields
$$h = \frac{\rho^2}\Delta \,dr^2 + \rho^2\,d\theta^2  + \(r^2 + a^2 + \frac{2mra^2\sin^2\theta}{\rho^2 - 2mr}\)\sin^2\theta\,d\phi^2$$
with the conformal metric $\bar h = \frac{\rho^2}{\rho^2-2mr}h$.  The drift-form is $$\omega = \frac{2mra\sin^2\theta}{\rho^2 - 2mr} \,d\phi.$$
We employ Lemma 3.9 to find $\text{wt}(\beta_M)$:  The supremum occurs at $\theta = \pi/2$ (the equator) and as $r \to 2m$, giving us $\text{wt}(\beta_M) = 1$:  $M$ is category (4), spatially incomplete, and causally bounded.

To generalize Kerr to a Kerr cosmic string, we'll need to cut our manifold $M$ apart.  More specifically, we consider $\tilde Q = (2m,\infty) \times (0,\pi) \times \Bbb R^1$ (deleting the poles before taking the universal cover) and a $\Bbb Z$-action on $\tilde Q$, $n \cdot (r, \theta, \phi) = (r, \theta, \phi + n\gamma)$ for some $\gamma > 0$; then $Q^\gamma = \tilde Q/\Bbb Z$.  We will seek a manifold $M'$ (using the language of Remark 1.8) with $\omega' - \omega = (\lambda/\gamma)d\phi$ for some $\lambda$ (i.e., so $\beta_{M'} = \beta_M + \zeta$, with $\zeta = (\lambda/\gamma)d\phi$ filling the role of $\theta$ in Remark 1.8---that letter being a co\"ordinate for Kerr).  By Remark 1.8, we first need to identify $\tilde\eta: \tilde Q \to \Bbb R$ such that $\omega^{\tilde\eta} = \omega + d\tilde\eta$ is $\Bbb Z$-invariant; but $\omega = \frac{2mra\sin^2\theta}{\rho^2-2mr}\,d\phi$ is itself $\Bbb Z$-invariant, so $\tilde\eta = 0$.  Then we seek $\tilde\delta$ so that $d\tilde\delta = \tilde\zeta = (\lambda/\gamma)d\phi$; clearly, $\tilde\delta = (\lambda/\gamma)\phi$, defined on $\tilde Q$.  With $\tilde\eta' = \tilde\eta + \tilde\delta$, this defines the $\Bbb Z$-action on $\tilde M = Å \tilde Q \times \Bbb R$: 
$$\align
n\cdot(t,r,\theta,\phi) & = (t + \tilde\eta'(r,\theta,\phi + n\gamma) - \tilde\eta'(r,\theta,\phi), r,\theta,\phi + n\gamma) \\
& = ( t + (\lambda/\gamma)(\phi + n\gamma) - (\lambda/\gamma)\phi, r,\theta, \phi + n\gamma) \\
& = (t + n\lambda, r,\theta,\phi + n\gamma).
\endalign$$
Call this $\Bbb Z$-action $\Bbb Z^\lambda$.  Then we have $M^\gamma = \tilde M/\Bbb Z^\lambda$ as our new spacetime, with orbit space $\tilde Q^\gamma = \tilde Q/\Bbb Z$ and action $n\cdot(r,\theta,\phi) = (r,\theta,\phi + n\gamma)$; the specific representation on $\Bbb Z$ is $\mu(n) = n\lambda$ (where we refrain from using $\rho$ for the representation, as that is a reserved letter in Kerr).  

However, we have the same problem here as with other cosmic strings given a time-shift: If we let $c_{r,\theta} : [0,\gamma] \to Q^\gamma$ be the loop $c_{r,\theta}(s) = [r,\theta,s]$, then we have this expression for $R(r,\theta) = \eff_{\omega'}(c_{r,\theta}) = \beta_{M^\gamma}\<c_{r,\theta}\>/L(c_{r,\theta})$:
$$R(r,\theta) = \frac{\lambda + \frac{2mra\gamma\sin^2\theta}{r^2 + a^2 -2mr - a^2\sin^2\theta}}{\gamma\sin\theta\sqrt{r^2 + a^2 - 2mr}\(\frac{r^2 + a^2 - a^2\sin^2\theta}{r^2 + a^2 -2mr - a^2\sin^2\theta}\)}$$
so we have $\text{wt}(\beta_{M^\gamma}) = \infty$ as $R(r,\theta)$ is unbounded as $\theta \to 0$.  Perhaps the simplest resolution is simply to restrict $Q^\gamma$ to those values of $r$ and $\theta$ which have $R(r,\theta) < B$ for some chosen $B > 0$ (amounting to an adaptive removal of spherical caps, unlike the constant-size removal examined in the Schwarzschild case, Example 3.8).  Call this restricted quotient space $Q_B^\gamma$, with $M_B^\gamma$ the corresponding subset of $M^\gamma$; we then have $\text{wt}(\beta_{M^\gamma_B}) = B$.  The set $Q_B^\gamma$ is formed by deleting polar caps, $\theta \le \theta_B(r)$ and $\theta \ge \pi - \theta_B(r)$ for some function $\theta_B$.  To find an equation for $\theta_B$ first note that the equation for $R(r, \theta)$ simplifies if we substitute $y = \frac{2mra\gamma\sin^2\theta}{r^2 + a^2 -2mr - a^2\sin^2\theta}$; then setting this to $B$ yields
$$ \sqrt{\frac{2mr\frac\gamma a +y}y}\frac{\lambda + y}{(r^2 + a^2)\frac\gamma a + y} = B$$
which yields a cubic equation for $y$.  When $B = 1$, this is a quadratic equation yielding $y = \frac{4mra\gamma\lambda^2}{A + \sqrt{A^2 + C}}$, where $A = (r^2 + a^2)^2\gamma^2 - a^2\lambda^2 - 2mra\gamma\lambda$ and $C = 16mra^2\gamma\lambda^2((r^2 + a^2)\gamma - a\lambda - mr\gamma)$; if we look in the regime $r \gg a$ this comes down to approximately $\sin(\theta_1) = \frac\lambda\gamma \frac{\sqrt{r^2 - 2mr}}{r^2}$, the same result as choosing the adapative approach for Schwarzschild with $B = 1 $.  For other values of $B$ near 1, a differential analysis leads, roughly, to $\sin(\theta_B) = (1 + (1 - B^2)/4) \sin(\theta_1)$ (for $r \gg m$, $r \gg a$). Depending on $B$, this gives us $M^\gamma_B$ in category (1), (4), or (6); for $B = 1$ (category (4)), it is spatially incomplete and causally bounded.

\example{Example 3.11 Bonnor and Weyl metrics for Lynden-Bell--Katz Toroidal Solenoid}

In \cite{LK}, Lynden-Bell and Katz present a spacetime representing current flow on a shell torus, circulating around the circular cross-section, thus a limit of toroidal solenoid with infinitely many windings.  For the interior of the torus, they use a stationary metric due to Bonnor in \cite{Bo}; for Bonnor this is the exterior (vacuum) geometry of an infinite cylindrical light beam.  For the exterior of the torus, Lynden-Bell and Katz graft on a static vacuum metric due to Weyl (\cite {BaW}).

The Bonnor metric is given in cylindrical co\"ordinates for radial co\"ordinate $r > a$ as
$$ds^2 = -F(r)(dt - (1 - F(r)^{-1})\,dz)^2 + dr^2 + r^2\,d\phi^2 + F(r)^{-1}\,dz^2$$
where $F(r) = B\ln(r/a) + C$ for constants $B$ and $C$.  Though this is used only for the interior of the torus in the Lynden-Bell--Katz solenoid, we can also examine this spacetime in its own right (though ignoring the portion interior to the cylinder $r = a$, where Bonnor locates the cylindrical light beam).  This spacetime $M_{\text{Bon}}$ has  Killing orbit space $Q_{\text{Bon}}$ as that portion of $\Bbb R^3$ outside the cylinder of radius $a$, i.e., $r > a$, with conformal metric $\bar h = F(r)^{-1}\,dr^2 + F(r)^{-1}r^2\,d\phi^2 + F(r)^{-2}\,dz^2$; clearly, the Killing drift-form is $\omega = (1 - F(r)^{-1})\,dz$. (The drift-form is proportional to the magnetic vector potential, recast as a 1-form.  Thus, in both Bonnor's spacetime and in the Lynden-Bell--Katz solenoid, the drift-form reflects the magnetic field induced from external sources: the cylindrical light beam for Bonnor, the toroidal surface current for Lynden-Bell--Katz.)  A fairly efficient loop $c$ for accruing $\int_c \omega$ is a rectangular one going (in $(r,\phi,z)$ co\"ordinates) from, say, $(a,0,0)$ to $(Ka,0,0)$ (for some $K >  1$) to $(Ka,0,Z)$ (for some $Z > 0$) to $(a,0,Z)$ and back to $(a,0,0)$; this yields $\int_c \omega = \frac ZC\(1/(1 + \frac C{B\ln K})\)$ and $L(c) = \frac ZC\((1 + \frac{2C}{B\ln K})/(1 + \frac C{B\ln K})\) + 2\int_a^{Ka} \frac{dr}{\sqrt{B\ln r + C}}$.  For fixed $K$ this gives $\int_c \omega/L(c)$ approaching $1/(1 + \frac{2C}{B\ln K})$ as $Z$ goes to $\infty$, so letting $K$ go to $\infty$ shows us $\text{\wt}(\{\omega\}  ) = 1$: category (5), spatially incomplete.

We can also explore what happens in a Bonnor cosmic string: Take the universal cover $\tilde M_{\text{Bon}}$ of $M_{\text{Bon}}$, then mod out by $\Bbb Z$-action $m\cdot(t,r,\phi,z) = (t + m\lambda, r,\phi + m\gamma, z)$, producing $M'_{\text{Bon}}$.  We have $\text{wt}(\beta_{M'_{\text{Bon}}}) = \max\{1,|\lambda|/(a\gamma )\}$ (this can be seen by examining curves $c_{n,K,Z}$ which are the same as the rectangular curves above, except the fourth side---at $r = a$---is replaced by a helix revolving $n$ times around the axis; then $\beta_{M'_{\text{Bon}}}\<c_{n,K,Z}\>/L(c_{n,K,Z})$ is monotonic increasing in $n/Z$ and approaches $\lambda/(a\gamma)$ as $n/Z \to \infty$): category (1) ($|\lambda| > a\gamma$) or (4) ($|\lambda| = a\gamma$), causally bounded and spatially incomplete (the reason why it's not category (5) is that $L_\omega(c_{n,K,Z})$ remains bounded as $n/Z \to \infty$) or (6) ($|\lambda| < a\gamma$).

(If we try to compactify Bonnor space in the $z$-direction with a time shift---i.e., mod out by the $Z$-action $m\cdot(t,r,\phi,z) = (t + m\lambda, r,\phi,z + m)$ for some $\lambda \neq 0$---then the resulting space is category (1), as loops $c_r$ of arbitrarily small length in the orbit space (from $[r,0,0]$ to $[r,0,1]$) accrue $\lambda$ in $\int_{c_r} \omega$.)

Coming back to the Lynden-Bell--Katz solenoid---call it $M_{LK}$---let us consider what information we have from the Bonnor metric in the interior of a torus circling the $z$-axis (with $B = 8I$, $I$ being the current density rolling around the torus):  As we cannot find loops of arbitrarily large change in $z$, we find that, if this were the whole of the spacetime, $\text{wt}(\beta_{M_{LK}})$ would be $< 1$.  But we still must examine the exterior metric.

In the torus exterior we have a Weyl metric
$$ds^2 = - e^{-2\psi(r,z)}\,dt^2 + e^{2\psi(r,z)}\(e^{2k(r,z)}(dr^2 + dz^2) + r^2\,d\phi^2\) $$
which is evidently a static metric---and we would have a standard static spacetime, if this were extended to the interior of the torus as well.  Thus, there is no further contribution to $\text{wt}(\beta_{M_{LK}})$ from the exterior geometry, and $M_{LK}$ is category (6).

If we build a cosmic string $M'_{LK}$ out of $M_{\text LK}$ by removing a cylinder of small radius around the $z$-axis---with a time-shift of $\lambda > 0$ and a total angular circumference of $\gamma$, as per usual---then the Weyl metric yields a weight for $\beta_{M'_{LK}}$ of $(|\lambda|/\gamma)(1/\inf(e^{2\psi}r))$.  But we also must take into account weight from the Bonnor metric; that is at least as much as comes from the usual interior of the torus, and also at least $(|\lambda|/\gamma)(1/(8I\ln(r_{\text{min}}) + C))$, where $r_{\text{min}}$ is the inner radius of the torus.

\endexample

\enddocument